\documentclass[12pt]{article}
\usepackage{amsmath,amsfonts}
\usepackage{hyperref}
\usepackage{graphicx}
\usepackage{xcolor}
\usepackage[numbers,sort&compress]{natbib}

\usepackage{amsthm}
\usepackage{amssymb}
\usepackage[matrix,arrow]{xy}
\usepackage{epsfig}
\usepackage{slashed}

\makeatletter\@addtoreset{equation}{section}\makeatother

\newcommand{\al}{\alpha}
\newcommand{\be}{\beta}
\newcommand{\te}{\theta}
\newcommand{\la}{\lambda}

\newcommand{\onov}[1]{\frac{1}{#1}}
\newcommand{\mat}[1]{\left(\begin{matrix} #1 \end{matrix}\right)}

\newcommand{\lag}{\mathcal{L}}

\newcommand{\beq}{\begin{equation}}
\newcommand{\eeq}{\end{equation}}
\newcommand{\bea}{\begin{eqnarray}}
\newcommand{\eea}{\end{eqnarray}}
\newcommand{\half}{\frac{1}{2}}

\newcommand{\vev}[1]{{\left< {#1} \right>}}

\newcommand{\mm}{\mathbb{M}}
\newcommand{\eq}[2][ ]{\begin{equation}\label{#1}{\begin{split}#2\end{split}}\end{equation}}
\newcommand{\eql}[2]{\begin{equation}\label{#1}{\begin{split}#2\end{split}}\end{equation}}

\newcommand{\hm}{\hat\mu}
\newcommand{\ha}{\hat a}
\newcommand{\hx}{\hat{\xi}}

\newcommand{\Tr}{{\rm Tr\,}}

\newcommand{\cD}{{\mathcal D}}

\newcommand{\cN}{{\mathcal N}}

\newcommand{\cO}{{\mathcal O}}

\renewcommand{\title}[1]{\vbox{\center\LARGE{#1}}\vspace{5mm}}
\renewcommand{\author}[1]{\vbox{\center\large#1}\vspace{5mm}}

\begin{document}
	\bibliographystyle{utphys}
	
	\begin{titlepage}
		\vspace{8mm}
		\begin{center}
			\title{\bf New Vortex-String Worldsheet Theories from Supersymmetric Localization}
			\vspace{8mm}
			{\large \bf Efrat Gerchkovitz}\footnote{efrat.gerchkovitz@weizmann.ac.il}
			{\bf and}
			{\large \bf Avner Karasik}\footnote{avner.karasik@weizmann.ac.il}\\\vspace{2mm}{\large\textit{ Department of Particle Physics and Astrophysics,\\Weizmann Institute of Science, Rehovot 76100, Israel}}
		\end{center}
	
\begin{abstract}
	We use supersymmetric localization techniques to study the low-energy dynamics of BPS vortex-strings in four-dimensional $\cN=2$ theories. We focus on theories with $SU(N_c)\times U(1)$ gauge group and $N_f$ hypermultiplets, all in the fundamental representation of $SU(N_c)$ but with general $U(1)$ charges. 
		Recently, we proposed a condition that determines whether the low-energy string dynamics  is captured by a two-dimensional worldsheet theory that decouples from the bulk  \cite{Gerchkovitz:2017kyi}.   
		For strings for which this decoupling applies, we propose a prescription for extracting the two-sphere partition function of the string worldsheet theory from the four-ellipsoid partition function of the parent theory.   We obtain a general formula for the worldsheet two-sphere partition function in terms of the parameters of the four-dimensional theory and identify $\cN=(2,2)$ GLSMs  that possess these partition functions in a large class of examples. In these examples, the weak coupling regime of the four-dimensional theory is mapped to the weak coupling regime of the worldsheet theory. In addition, we study the classical string zero-modes in flat space and obtain predictions for the worldsheet spectra, which agree with the low-energy spectra of the GLSMs obtained in the localization analysis.    For $N_f=2N_c=4$, we discuss the map between string worldsheet theories under four-dimensional $S$-duality and use our prescription to study  examples in which the weak coupling regime of the four-dimensional theory is mapped  to the strong coupling regime of the worldsheet theory.

\end{abstract}

\end{titlepage}

\tableofcontents
\section{Introduction and Summary}

2d sigma-models have been long known to serve as useful toy models for 4d gauge dynamics. The similarities between the 2d and 4d dynamics, in some cases, can be demonstrated not only at the qualitative, but also at the quantitative level. 
One example is the observed  matching between the quantum BPS spectra of 2d $\cN=(2,2)$ supersymmetric  ${\mathbb{CP}}^{N-1}$ sigma model and  4d $\cN=2$ $SU(N)$ gauge theory with $N$ fundamental hypermultiplets \cite{Dorey:1998yh}. This observation  received a physical explanation in later works  \citep{Hanany:2003hp,Auzzi:2003fs,Shifman:2004dr,Hanany:2004ea}, which showed that by gauging a $U(1)$ flavor symmetry in the 4d theory mentioned above and introducing a Fayet-Iliopoulos (FI) term, one obtains a theory that supports BPS vortex-strings.  The string with the minimal winding number is part of a  ${\mathbb{CP}}^{N-1}$ moduli space of string solutions. The low-energy effective theory that lives on the worldsheet of the string is the $\cN=(2,2)$ ${\mathbb{CP}}^{N-1}$ sigma-model.

Since their discovery,  BPS vortex-strings in $\cN=2$ SQCD have been studied extensively. The study of the string moduli spaces and the string worldsheet theories, which preserve $\cN=(2,2)$ supersymmetry on the worldsheet, has been generalized by adding flavors, by considering strings with general winding numbers and by considering other gauge groups. The methods one uses to study these strings range from brane constructions to explicit field theory derivations. For a partial list of references, see \citep{Auzzi:2003fs,Hanany:2003hp,Hanany:2004ea,Shifman:2004dr,Shifman:2006kd,Eto:2005yh,Auzzi:2005gr,Eto:2009zz,Ferretti:2007rp,Dorey:1999zk,Eto:2007yv,Eto:2006cx}, and the reviews \cite{Shifman:2007ce,Tong:2005un,Tong:2008qd,Eto:2006pg}.

In the past decade, the quantum and non-perturbative understanding of theories with extended supersymmetry has progressed  significantly due to the technique of supersymmetric localization. In particular,  exact formulas are now available for squashed sphere partition functions of  $4d$ $\cN=2$ Lagrangian theories \cite{Pestun:2007rz,Hama:2012bg}, as well as of $2d$ $\cN=(2,2)$ theories \cite{Doroud:2012xw,Benini:2012ui}, as long as the theories preserve appropriate $R$-symmetries. Fortunately, in the vortex-string set-up, both the parent 4d theory and the 2d worldsheet theory can be studied using localization techniques. This is due to the fact that the 4d FI parameter, while breaking the $SU(2)_R$
symmetry, preserves its Cartan, which is needed in order to preserve supersymmetry on the sphere. Similarly, the worldsheet theory inherits a vectorlike $R$-symmetry from the 4d theory. This allows placing the theory on the sphere and computing its sphere partition function using supersymmetric localization.  In this work, we will exploit the exact sphere partition function results to study the worldsheet theories of BPS vortex-strings, at the non-perturbative level, in new examples.
For previous works that use localization results to study BPS vortex-strings, see \cite{Chen:2015fta,Pan:2015hza,Chen:2011sj,Fujimori:2015zaa}. In addition, many closely related works use supersymmetric localization to study surface defects coupled to 4d theories. In particular, see the recent publications \cite{Gomis:2016ljm,Pan:2016fbl} that share some similarities with the analysis that we present below.

We will generalize the well-studied vortex-strings, which are obtained when the $U(1)$ flavor symmetry in which all the hypermultiplets have the same $U(1)$ charge is gauged, by gauging, instead,  a general $U(1)$ flavor symmetry. In other words, we will study vortex-string configurations in a 4d $\cN=2$ supersymmetric theory, with $SU(N_c)\times U(1)$ gauge group and $N_f$ hypermultiplets in the fundamental representation of $SU(N_c)$, and with charges $c_1,...,c_{N_f}\in\mathbb{Z}$ under the $U(1)$ factor of the gauge group. The parameters of this theory are two complexified gauge couplings, one for the $SU(N_c)$ factor of the gauge group and one for the $U(1)$ factor of the gauge group, and an FI parameter $\xi$. In addition, we will introduce small, non-degenerate, hypermultiplet masses.

We will be interested in baryonic strings. These strings are labeled by the winding number $K$ and by a choice of a baryonic vacuum, in which scalars belonging to $N_c$ out of the $N_f$ hypermultiplets obtain  vacuum expectation values  (VEVs).  The vacuum equations require that the sum of the $U(1)$ charges of these $N_c$ hypermultiplets, which will be denoted by $C$, is non-vanishing. 
The string configuration carries magnetic flux $\Phi={2\pi K}/{C}$. Such fluxes are allowed due to a breaking of the $SU(N_c)\times U(1)$ gauge group to a residual  $\mathbb{Z}_C$ subgroup  in the baryonic vacuum.  

For general $U(1)$ charges, not all the strings will give rise to a 2d effective theory on their worldsheet. The point is that when $N_f>N_c$
the bulk theory is not gapped and interactions between the bulk light fields and the worldsheet light fields are generally present. In \cite{Gerchkovitz:2017kyi} we studied these interactions and argued that they are suppressed at low-energies if and only if 
\begin{equation}
\frac{({c_i-c_j})\Phi}{2\pi}=\frac{({c_i-c_j})K}{C} \in\mathbb{Z} \;\;\text{for all}\; i,j\;.\;\;\;\;\;\;\;\;\text{(Decoupling Criterion)} \label{decoupling criterion}
\end{equation}
If $N_f=N_c$,  the bulk theory is gapped. In this case, the low-energy dynamics is captured by a two-dimensional worldsheet theory, regardless of whether \eqref{decoupling criterion} is satisfied or not.

When the condition \eqref{decoupling criterion} is satisfied, the low-energy dynamics in the string background is described by a 2d worldsheet theory, decoupled from a 4d bulk theory. 
The goal of this work will be to study the worldsheet theories that one obtains when the decoupling condition is satisfied (or when $N_f=N_c$). We do this by combining localization techniques on the four-ellipsoid with classical zero-mode analysis in flat space.

In order to apply localization techniques, we place the theory on the four-ellipsoid 
\begin{equation}
\frac{x_0^2}{r^2}+\frac{x_1^2+x_2^2}{l^2}+\frac{x_3^2+x_4^2}{\tilde l^2}=1\;.
\end{equation} 
The localization formula of \cite{Hama:2012bg}  allows us to write a formal expression for the four-ellipsoid partition function, as a matrix integral over Coulomb branch coordinates. Unfortunately, this formal matrix integral is not well-defined -- the Coulomb branch integral diverges due to the $U(1)$ Landau pole. In order to define the partition function, we thus introduce an ultraviolet cut-off. That is, we cut-off the Coulomb branch integral below the scale set by the Landau pole.

For a wide range of parameters, we can close the cut-offed Coulomb branch integrals in the complex plane. In analogy to the Higgs branch localization in two dimensions  \cite{Doroud:2012xw,Benini:2012ui}, we compute the integrals by summing over the residues of the encircled poles and obtain a representation of the four-ellipsoid partition function as a discrete sum, which we interpret as a sum over string contributions. This interpretation is motivated by the Higgs branch localization of \cite{Chen:2015fta,
	Pan:2015hza}; the authors of these papers used a  modified deformation term to localize  the path integral for the four-ellipsoid partition function on saddle points that solve a BPS string-like equation, and argued  that the representation that results from such a localization procedure can be obtained from the Coulomb branch representation of \cite{Pestun:2007rz,Hama:2012bg} upon closing the contours in the complex plane.
The interpretation of the residues of the poles of the integrand as encoding the string contributions is
also motivated by the results of \cite{Gaiotto:2012xa}.\footnote{In \cite{Gaiotto:2012xa}, residues of the superconformal index at poles in a $U(1)$ flavor fugacity have been identified with the superconformal index of an infrared theory in the presence of surface defects. The surface defect is understood as the RG endpoint of a BPS vortex-string that exists when the $U(1)$ flavor symmetry is gauged.}

As will be explained in detail in sections \ref{4} and \ref{general U1 charges}, we obtain a representation of the  four-ellipsoid partition function, which has the schematic  form  
\begin{equation}
Z_{S^4_b}=\sum_{v\in\text{vacua}}e^{i\xi M_v}Z^{\text{bulk}}_{v}\sum_{K\in\mathbb{Z}}e^{-\xi\Phi_{K,v}}Z^{2d}_{v,K}\label{schematic}\;.
\end{equation}
The sums in equation \eqref{schematic} are over the vacua, labeled by $v$, and the winding number $K$. The functions $Z^{\text{bulk}}_{v}$ and $Z^{2d}_{v,K}$ are independent of the FI parameter $\xi$ and are uniquely defined as the coefficients in an expansion according to the dependence on $\xi$ and the requirement that $Z^{2d}_{v,K=0}=1$. $\Phi_{K,v}$ is proportional to the magnetic flux of a string that is labeled by the vacuum $v$ and the winding number $K$. $M_v$ is related to the masses and charges of the fields that obtain an expectation value in the vacuum labeled by $v$. Let us emphasize that the expression \eqref{schematic} is a very schematic description of the representation that we find. In particular, we did not include in this schematic expression contributions of intersecting strings or contributions of mesonic vacua and strings.  The detailed analysis and the exact expressions can be found in sections \ref{4} and \ref{general U1 charges}.

$Z^{\text{bulk}}_{v}$ and $Z^{2d}_{v,K}$  are functions of the rescaled hypermultiplet masses $\sqrt{l\tilde l}\mu_i$, $i=1,...,N_f$, the gauge couplings, the theta parameters, and the squashing parameter $b=\sqrt{{l}/{\tilde {l}}}$. These functions are the output of our prescription.  We now explain how we interpret this output.

$Z^{\text{bulk}}_{v}$ is the four-ellipsoid partition function of the light bulk fields in the vacuum $v$. We check this explicitly and compare to the spectrum in this vacuum. Exactly when the condition \eqref{decoupling criterion} is satisfied,  the expression we find for $Z^{2d}_{v,K}$ is of the form obtained in \cite{Doroud:2012xw,Benini:2012ui} for two-sphere partition functions in the Higgs branch representation. 
Thus, when \eqref{decoupling criterion} is satisfied, we interpret $Z^{2d}_{v,K}$ as the two-sphere partition function of the worldsheet theory of the $K$-string labeled by the vacuum $v$. The fact that only when \eqref{decoupling criterion} is satisfied the contributions factorize to a product of two decoupled factors -- one describing the bulk fields and the other describing the worldsheet fields --  is the  supersymmetric localization derivation of the decoupling criterion \eqref{decoupling criterion}, which was derived in \cite{Gerchkovitz:2017kyi} based on classical analysis of the interaction terms. 

The uniform $U(1)$ charge example, which we work out in detail in section \ref{4}, and in fact has already been studied using localization techniques in \cite{Chen:2015fta,Pan:2015hza,Gomis:2016ljm,Pan:2016fbl}, reproduces the worldsheet theory  that appears in the literature, and is used as a test of our prescription.  
For general $U(1)$ charges obeying \eqref{decoupling criterion}, we obtain a formula for $Z^{2d}_{v,K}$, which is expressed in terms of the parameters of the 4d theory -- see equation \eqref{ZS2Kgeneral}. 
This leaves us with the challenge of finding 
a 2d $\cN=(2,2)$ theory  with a two-sphere partition function that equals to $Z^{2d}_{v,K}$ under some 4d-2d map of parameters.

In some cases, the weak coupling regime of the 4d theory is mapped to the weak coupling regime of the  2d theory, while in other cases, it is mapped to the strong coupling regime. The identification of the worldsheet theory in the latter case is significantly more challenging, as the expressions we obtain for $Z^{2d}_{v,K}$ and the expressions for two-sphere partition functions that one finds in the localization literature, are all given as weak coupling expansions.  Moreover, classical analysis of the string zero-modes, which provides useful insights in the former case, is less trustworthy in the latter case. 

We propose a condition on the $U(1)$ charges that guarantees that one can find a description of the worldsheet dynamics that is weakly coupled when the 4d theory is weakly coupled: Label the $N_c$ hypermultiplets getting vacuum expectation values by $a=1,...,N_c$ and denote $C=\sum_{a=1}^{N_c}c_a$. The condition is that for each of the extra hypermultiplets, labeled by $i=N_c+1,..., N_f$, 
\begin{equation}
\begin{aligned}\label{weak-cond}
&\Delta_{ia}\geq 0 \;\;\;\;\;\text{for all}\;\;\; 1\leq a\leq N_c \;\;\;\text{or}\\
&\Delta_{ia}\leq -1 \;\;\text{for all} \;\;\;1\leq a\leq N_c\;,
\end{aligned}
\end{equation}
where 
\begin{equation}
\Delta_{ia}=\frac{c_i-c_a}{C}\;.
\end{equation} 
In terms of the classical zero-mode analysis, this is the condition that the 4d F-term equations are satisfied trivially and do not impose any constraints on the target space of the worldsheet theory. See \cite{Gerchkovitz:2017kyi} and section \ref{21}. One can also derive this condition from our localization analysis, as the form of the general formula for the two-sphere partition function that we find changes dramatically when this condition is not satisfied.  See section \ref{general U1 charges}.

In the class of strings defined by equation (\ref{weak-cond}) we identify $Z^{2d}_{v,K}$  with the two-sphere partition function of an $\cN=(2,2)$ gauged linear sigma model (GLSM) with a $U(K)$ gauge group.
The matter content of this theory consists of one chiral multiplet in the adjoint representation,   $N_c$ chiral multiplets in the fundamental representation and $N_f-N_c$  chiral multiplets in the anti-fundamental representation. 
In addition, the spectrum contains  neutral chiral multiplets, that are coupled to the fundamental, anti-fundamental and adjoint chiral multiplets via a superpotential, and extra decoupled chiral multiplets.  The number of the neutral and the decoupled multiplets depends on the $U(1)$ charges and on the choice of the vacuum. The FI and theta parameters of the GLSM are given in terms of the Yang-Mills coupling and the theta angle of the $SU(N_c)$ factor of the gauge group,  as
\begin{equation}
\xi_{2d}=\frac{4\pi }{g^2}\;,\;\;\;\te_{2d}=\te_{4d}+(K-1)\pi\;.
\end{equation}

The worldsheet theory inherits two $U(1)$ $R$-symmetries from the 4d theory; one is related to the 4d  $R$-symmetry and the other to rotations in the transverse plane to the string. From our localization analysis, we find the masses and the two types of $R$-charges of all the 2d fields in terms of the 4d parameters. In section \ref{examples} we test the proposal that the worldsheet theory is given by the low-energy limit of the GLSM we found, and the 4d-2d dictionary that we obtained, against expectations that are based on classical analysis of the string zero-modes in flat space.

It is generally hard to study the worldsheet theories when the 4d-2d dictionary maps the weak coupling regime to the strong coupling regime.  However, in the 
$N_f=2N_c=4$ case, one can relate strongly coupled worldsheet theories to weakly coupled worldsheet theories using four-dimensional $S$-duality. 
$\cN=2$ supersymmetric $SU(2)$ gauge theory with four fundamental massless hypermultiplets has an $SO(8)$ flavor symmetry. The group $SO(8)$ has an $\bf S_3$ outer automorphism group, which  acts as the $S$-duality group of the massive theory \cite{Seiberg:1994aj}. 
The six elements of $\bf S_3$ can be generated using the transformations 
\begin{equation}
\begin{aligned}
T:\;\;&q\to \frac{q}{q-1}\;,  \;\;\;\;\;\;\;\;    S:\;\;&&q\to 1-q\;,\\
&\mu_1\to\mu_1\;,&&\mu_1\to\frac{1}{2}(\mu_1+\mu_2+\mu_3+\mu_4)\;,\\
&\mu_2\to\mu_2\;,&&\mu_2\to\frac{1}{2}(\mu_1+\mu_2-\mu_3-\mu_4)\;,\\
&\mu_3\to\mu_3\;,&&\mu_3\to\frac{1}{2}(\mu_1-\mu_2+\mu_3-\mu_4)\;,\\
&\mu_4\to-\mu_4&&\mu_4\to\frac{1}{2}(\mu_1-\mu_2-\mu_3+\mu_4)\;,
\label{triality-trans}
\end{aligned}
\end{equation}
where $q=e^{2\pi i \tau_{su(2)}}$ and $\mu_1,...,\mu_4$ are the hypermultiplet masses.

In order to allow string configurations, we need to gauge a $U(1)$ flavor symmetry. The $U(1)$ charges of the hypermultiplets, $c_1,...,c_4$, map under the $\bf S_3$ transformations in the same way the masses map in equation (\ref{triality-trans}).  
This provides us with a prediction for the worldsheet theories on strings that are related by these transformations to strings that satisfy condition (\ref{weak-cond}); the worldsheet theory has to be the same, but the 4d-2d map of the parameters is modified according to the map \eqref{triality-trans}.   In particular, applying the $S$-transformation to strings that satisfy (\ref{weak-cond}), one obtains 4d-2d dictionaries that map the weak coupling regime to the strong coupling regime.  
Indeed, in section \ref{Nc=2} we show, for winding number $K=1$, that  the expression we find for $Z^{2d}_{v,K}$ gives the expected partition functions for charge assignments that are related by triality to the equal charge case.  This check is enabled due to  hypergeometric function identities.

Let us end the introduction with a comment about the limitations of our analysis. The two-sphere partition function does not capture all the information on the theory. As a result, one cannot uniquely identify the worldsheet theory based on this observable. In the equal charge case, for example, the output of our prescription agrees with the Hanany-Tong model \cite{Hanany:2004ea,Hanany:2003hp}, which is believed to reproduce correctly various observables that are protected by supersymmetry.  However, it is known, for example, that the target space metric of the Hanany-Tong model is not the correct target space metric for winding number $K>1$~\cite{Manton:2002wb}.  

While the exact worldsheet theory cannot be determined uniquely from its sphere partition function, a large amount of non-trivial information can be extracted from this observable.  To enrich the sphere partition function, we include generic values for all the possible parameters. As a result, the output of our localization  computation depends on many parameters -- the hypermultiplet masses and $U(1)$ charges, the ellipsoid squashing parameter, the $SU(N_c)$ gauge coupling and the $SU(N_c)$ theta angle. This makes the agreement between the  expressions we compare highly non-trivial. Moreover, whenever possible, we compare the spectrum of the worldsheet theory we obtain to expectations based on classical zero-mode analysis.

The outline of the paper is as follows. In section \ref{general gauging} we review the classical vacuum and string equations and some of the properties of the string moduli space. Most of the content of this section already appeared in \cite{Gerchkovitz:2017kyi}. In section \ref{examples} we list our proposals for the worldsheet theories on strings that satisfy conditions \eqref{decoupling criterion} and \eqref{weak-cond} and explain the spectrum that we find from the classical point of view. In section \ref{4} we derive the Higgs branch representation of the four-ellipsoid partition function in the $c_i=1$ case and explain how to read from this expression the two-sphere partition functions of the worldsheet theories. In section \ref{general U1 charges} we generalize the derivation of section \ref{4} to theories with arbitrary $U(1)$ charges. We give a closed formula for the two-sphere partition function of the worldsheet theory that is applicable in all the cases in which the string decouples from the bulk. In the cases that satisfy condition \eqref{weak-cond}, we identify the two-sphere partition function with the two-sphere partition function of the GLSMs presented in section \ref{examples}. Section \ref{sec:Triality} is devoted to the triality -- the S-duality in the $N_f=2N_c=4$ case. We explain the predictions for the worldsheet theories that one can derive from the triality  and use our localization prescription to verify these predictions for  strings that are related by triality to the minimal strings in the equal charge case.   
 Technical details are collected in four appendices.

\section{The Classical String Equations and the String Moduli Space}\label{general gauging}

In this section, we will review the classical vacuum and string equations and some of the properties of the string moduli space.  This analysis is done in flat spacetime. Most of the content of this section already appeared in \cite{Gerchkovitz:2017kyi}. 

\subsection{The Classical Vacuum and String Equations}

Our starting point is 
 a 4d $\cN=2$ supersymmetric gauge theory with $SU(N_c)\times U(1)$ gauge group and  $N_f$ hypermultiplets in the fundamental representation of $SU(N_c)$. We will assume that  $N_c\leq N_f\leq 2N_c$.  The charges of the hypermultiplets under the $U(1)$ factor of the gauge group will be denoted by  $c_1,c_2,...,c_{N_f}\in\mathbb{Z}$. We will denote the complexified gauge couplings of the $SU(N_c)$ and $U(1)$
 factors of the gauge group by
 \eql{tau}{\tau_{u(1)}=\frac{\te_{u(1)}}{2\pi}+\frac{4\pi i}{e^2}\;,\;\;\;  \tau_{su(N)}=\frac{\te}{2\pi}+\frac{4\pi i}{g^2}\;.}
 To force the vacuum into the Higgs branch, we will introduce a Fayet-Ilioupulus parameter $\xi>0$.  In addition, we will introduce  non-degenerate hypermultiplet masses $\mu_i\ll e\sqrt{\xi},g\sqrt{\xi}$.\footnote{Masses of charged hypermultiplets can be shifted arbitrarily by redefinitions of the $U(1)$ gauge multiplet scalar. What we need to assume is actually $\left|\mu_i-\mu_a+\Delta_{ai}\sum_{b=1}^{N_c}{\mu_b}\right| \ll e\sqrt{\xi},g\sqrt{\xi}$.  }

The bosonic part of the action is given by
\eql{generalaction}{S&=\int d^4x\left[ \onov{4g^2}\left(F_{\mu\nu}^\al\right)^2+\onov{4e^2}\left(F_{\mu\nu}'\right)^2+\onov{g^2}|\cD_\mu a^\al|^2+\onov{e^2}|\partial_\mu a'|^2+\sum_i|\cD_\mu q_i|^2+\sum_i|\cD_\mu \tilde{q}^i|^2-V+\lag_{\te}    \right]\ , \\V&=\frac{g^2}{2}\left(\onov{g^2}f^{\al\be\gamma}a_\be^{\dagger}a_\gamma+\sum_iq^{i\ \dagger}\la^\al q_i-\sum_i\tilde{q}^i\la^\al \tilde{q}_i^{\ \dagger}\right)^2+\frac{e^2}{8}\left(\sum_ic_iq^{i\dagger}q_i-\sum_ic_i\tilde{q}_i\tilde{q}^{i\dagger}-N_c\xi\right)^2\\
	&\;\;\;+2g^2\Big|\sum_i\tilde{q}^i\la^\al q_i\Big|^2+\frac{e^2}{2}\Big|\sum_ic_i\tilde{q}^iq_i\Big|^2+\sum_i\Big|(c_ia'+\la^\al a^\al+\mu_i)q_i\Big|^2+\sum_i\Big|(c_ia'+\la^\al a^\al+\mu_i)\tilde{q}_i^\dagger\Big|^2\;,\\
	\lag_{\te}&=\frac{\te}{32\pi^2}F_{\mu\nu}^\al\tilde{F}^{\al\mu\nu}+\frac{\te_{u(1)}}{32\pi^2}F'_{\mu\nu}\tilde{F}'^{\mu\nu} \ , 
	}
where
\eq{ \cD_\mu q_i&=(\partial_\mu-ic_iA_\mu'-i\la^\al A_\mu^\al)q_i\ ,\ \cD_\mu \tilde{q}^\dagger_i=(\partial_\mu-ic_iA_\mu'-i\la^\al A_\mu^\al)\tilde{q}^\dagger_i\ ,\ \cD_\mu a^\al=(\partial_\mu\delta^{\al\gamma}-if^{\al\be\gamma}A_\be)a_\gamma\ .}
In the equation above, the scalar and the gauge field of the $U(1)$ vector multiplet are denoted by $a'$ and $A_\mu'$ respectively. Similarly,  the scalars and gauge fields of the $SU(N_c)$ vector multiplet are denoted by $a^\al$ and  $A_\mu^\al$, $\al=1,...,N_c^2-1$. The generators and structure constants of $SU(N_c)$ are denoted by $\la^\al$ and $f^{\al\be\gamma}$. $q_i^a$ and $\tilde{q}^{i}_a$ are the hypermultiplets scalars, where $i=1,..., N_f$ is the flavor index and the   $a=1,...,N_c$ is the color index. The color index was suppressed in  \eqref{generalaction}. 
In some of the equations below, the hypermultiplet scalars will be organized as matrices, \eq{\textbf{q}=\mat{q_1^1&q_2^1&\cdots&q_{N_f}^1\\q_1^2&q_2^2&\cdots&q_{N_f}^2\\\vdots&\vdots&\ddots&\vdots\\q_1^{N_c}&q_2^{N_c}&\cdots&q_{N_f}^{N_c}}\ ,\ \tilde{\textbf{q}}=\mat{\tilde{q}_1^1&\tilde{q}_2^1&\cdots&\tilde{q}_{N_c}^1\\\tilde{q}_1^2&\tilde{q}_2^2&\cdots&\tilde{q}_{N_c}^2\\\vdots&\vdots&\ddots&\vdots\\\tilde{q}_1^{N_f}&\tilde{q}_2^{N_f}&\cdots&\tilde{q}_{N_c}^{N_f}}\ .}

In the absence of the FI parameter, the theory enjoys a non-anomalous $SU(2)_R$ $R$-symmetry, which is broken by the FI parameter to its Cartan. 
The global symmetry of the theory is $U(1)^{N_f-1}$. If some of the $U(1)$ charges $c_i$ are  equal, the theory enjoys an approximate non-abelian  global symmetry which is broken by the small hypermultiplet masses.

The vacua of the theory are labeled by $SU(N_c)$ invariant operators that are charged under the $U(1)$ factor of the gauge group.  
For generic $c_i$, both charged mesons and charged baryons exist. This leads to two types of vacua -- mesonic and baryonic.

The mesonic vacua are given, up to gauge transformations, by
\eql{mes-vac}{q_i^1=\tilde{q}_1^j=\sqrt{\frac{N_c\xi}{c_i-c_j}}\; ,\;\;\;c_i>c_j\;.}
In addition, the adjoint scalars need to satisfy the constraints
\eql{mes-vac-a}{c_ia'+\la_{11}^\al a^\al+\mu_i=c_ja'+\la_{11}^\al a^\al+\mu_j=0\ .}Thus, the mesonic vacua have $N_c-2$ flat Coulomb branch directions.
The vacuum \eqref{mes-vac} breaks the gauge symmetry to $\mathbb{Z}_{|c_i-c_j|}\times SU(N_c-1)$. We will not discuss mesonic vacua and strings in this work, with one exception -- in the $N_f=2N_c=4$ example that we study in section \ref{locS-dual}, we will study also mesonic strings that are related by $S$-duality to baryonic strings in a different theory.

We will focus on the baryonic vacua and the corresponding strings. 
The baryonic vacua are labeled by choices of $N_c$ out of the $N_f$ hypermultiplets such that the sum of their charges is not zero.  Without loss of generality, we will choose the first $N_c$ hypermultiplets and assume that $C\equiv\sum_{i=1}^{N_c}c_i>0$ (flipping the sign of $C$ amounts to replacing $q\leftrightarrow\tilde{q}$).  

Up to gauge transformations, the vacuum corresponding to this choice of hypermultiplets is given by
\eql{generalvac}{\begin{aligned}
		&\textbf{q}=\mat{v\mathbb{I}_{N_c\times N_c}&0_{(N_f-N_c)\times N_c}}\; ,\;\;\;v^2=\frac{N_c\xi}{C}\; \;,\\  &c_b\delta_{ab}a'+\la^\al_{ab}a^\al+\mu_b\delta_{ab}=0\;, \;\;\;\;\;\;\;\;  a,b=1,...,N_c\;,
	\end{aligned}}
	where all the other fields vanish.
	The gauge symmetry is  broken in the vacuum  to a $\mathbb{Z}_C$ subgroup that acts on the scalars as
	\eql{zc}{q_i^a\rightarrow e^{2\pi in \Delta_{ia}}q_i^a\;,\;\;\;n=1,2,...,C\;.}	The masses of the $q_{i}^a$ excitations in the vacuum (\ref{generalvac}) are given by
	\eql{mass-in-vac}{\mu_{ia}^2=\left(\mu_i-\mu_a+\Delta_{ai}\sum_{b=1}^{N_c}{\mu_b}\right)^2\;,}
	where
	\begin{equation}
	\Delta_{ai}={\frac{c_a-c_i}{C}}\;.
	\end{equation}

		The vacuum preserves  a combination of the  $U(1)\subset SU(2)_R$ that is preserved in the presence of the FI term and an $SU(N_c)\times U(1)$ transformation.  The charges of  the hypermultiplet scalars under this combination are given by  \begin{equation}
		R(q_{i}^a)=N_c\Delta_{ai} \;,\;\;\;R(\tilde{q}_a^i)=2+N_c\Delta_{ia}\;.\label{R-charges-in-vac}
		\end{equation}  Topologically stable strings are thus characterized by the homotopy group $\pi_1(U(1)\times SU(N_c)/\mathbb{Z}_C)$, which allows fluxes quantized as \eq{\Phi=\lim_{r\to\infty}\int d\phi A'_\phi=\frac{2\pi K}{C}\;, \;\;\; K\in\mathbb{Z}\;.}

The spectrum in the vacuum consists of $2N_c(N_f-N_c)$ light hypermultiplets, with masses given by equation \eqref{mass-in-vac} for $i=N_c+1,...,N_f$, $a=1,...,N_c$. 
		  In addition, the $N_c^2$ gauge multiplets combine with the $N_c^2$ hypermultiplets labeled by ${i\leq N_c}$ via the Higgs mechanism to create $N_c^2$ long massive vector multiplets, with masses $m_{W}\sim e\sqrt{\xi}, g\sqrt{\xi}$.

Let us now derive the string equations. We will take the string to lie along the $x^3$ direction. Assuming that the configuration does not depend on the worldsheet coordinates $x^0$ and $x^3$,  that  $A_0$, $A_3$ and the off-diagonal elements of $a$ vanish, and that $a$ and $a'$ are constant, the tension can be written as  	\eql{tension-general}{T=\int dx^1dx^2&\left[\onov{2e^2}\left(B'_3\pm \frac{e^2}{2}\left(\sum_ic_i|q_i|^2-\sum_ic_i|\tilde{q}_i|^2-N_c\xi\right)\right)^2+2g^2\Big|\sum_i\tilde{q}^i\la^\al q_i\Big|^2+\frac{e^2}{2}\Big|\sum_ic_i\tilde{q}^i q_i\Big|^2\right.\\&\left.+\onov{2g^2}\left(B_3^\al\pm g^2\sum_iq^{i\dagger}\la^\al q_i\mp g^2\sum_i\tilde{q}^i\la^\al\tilde{q}_i^\dagger\right)^2+\sum_i\Big|\cD_1q_i\pm i\cD_2q_i\Big|^2+\sum_i\Big|\cD_1\tilde{q}_i\pm i\cD_2\tilde{q}^i\Big|^2\right.\\&\left.+\sum_i\Big|(c_ia'+\la^\al a^\al+\mu_i)q_i\Big|^2+\sum_i\Big|(c_ia'+\la^\al a^\al+\mu_i)\tilde{q}_i^\dagger\Big|^2ֿ\pm  N_c\xi B'_3\right]\\&\geq \pm\int dx^1dx^2 N_c\xi  B'_3\ ,}
	where $B_3'=F_{12}'\ ,\ B_3^\al=F_{12}^\al$.
		The last term in this expression is a topological term that is proportional to the magnetic flux.  The Bogomol'nyi equations are obtained by requiring the vanishing of the other terms:
\eql{Bogomolnyiset}{&B'_3\pm \frac{e^2}{2}\left(\sum_ic_i|q_i|^2-\sum_ic_i|\tilde{q}_i|^2-N_c\xi\right)=B_3^\al\pm g^2\sum_iq^{i\dagger}\la^\al q_i\mp g^2\sum_i\tilde{q}^i\la^\al\tilde{q}_i^\dagger=0\ ,\\ &\sum_i\tilde{q}^i\la^\al q_i=\sum_ic_i\tilde{q}^i q_i=\cD_1q_i\pm i\cD_2q_i=\cD_1\tilde{q}_i\pm i\cD_2\tilde{q}^i=0\ ,\\
& (c_ia'+\la^\al a^\al+\mu_i)q_i=(c_ia'+\la^\al a^\al+\mu_i)\tilde{q}_i^\dagger=0\ .}Solutions to these equations minimize the tension in a given topological sector and give rise to ${1}/{2}$-BPS strings. 

		We will focus on configurations with positive magnetic flux in the $x_3$ direction, by picking the upper sign in \eqref{tension-general},\eqref{Bogomolnyiset}. 
One can look for string solutions that satisfy the ansatz 
\eql{basicansatz}{\begin{aligned}
		&q_i^a=q_a(r,\phi)\delta_i^a\;, \;\;\;\;\;\;\;\;   \lim_{r\to\infty}q_a(r,\phi)=ve^{ik_a\phi}   \; \;,\\  &c_b\delta_{ab}a'+\la^\al_{ab}a^\al+\mu_b\delta_{ab}\;, \;\;\;\;\;\;\;\;  a,b=1,...,N_c\;.
	\end{aligned}}
where $r,\phi$ are the polar coordinates on the $x^1-x^2$ plane and $k_a$ are non-negative integers. For such a solution to exist, the diagonal elements of the gauge field need to be turned on to satisfy \eqref{Bogomolnyiset}. All other fields vanish.
	The magnetic flux carried by a solution of the Bogomol'nyi equations with an asymptotic behavior of the form \eqref{basicansatz} is given by 
	\eql{fluxphase}{\Phi=\lim_{r\to\infty}\int A'_\phi d\phi=\frac{2\pi K}{C}}
	where $K\equiv\sum_{a=1}^{N_c}k_a$ is the topological charge of the string.

\subsection{The String Moduli} \label{21}
 To discuss the string solution, one can start with an ansatz of the form \eqref{basicansatz} and plug it into the Bogomol'nyi equations. The solution for $q_a(r,\phi)$ is guaranteed to have $k_a$ zeros \cite{Taubes:1979tm}.
The string solution has translational zero-modes, related to the locations of the zeros. In the presence of generic hypermultiplet masses these are expected to be the only exact zero-modes. 

However, we will be interested in fluctuations above the string solution with energies in the range $\mu_{ia}\lesssim E\ll e\sqrt{\xi},g\sqrt{\xi}$. Therefore, we need to consider all the deformations of the string solution that become true zero-modes at the massless limit -- these will be treated as approximate zero-modes at the $\mu_{ia}\ll e\sqrt{\xi},g\sqrt{\xi}$ limit. In other words, we consider only the first two lines of \eqref{Bogomolnyiset} as the string equations.  Fixing the vector multiplet scalars on their vacuum expectation values, the terms in the third line of \eqref{Bogomolnyiset} will  give rise to small masses for the worldsheet fields.
For the tension cost of such deformations to be finite (and small), and to avoid problems related to non-normalizability when the quasi-moduli are promoted to fields on the worldsheet, one has to modify these modes at distances above $\sim \frac{1}{\mu_{ia}}$  (as in \cite{Shifman:2006kd}, for example).  This does not alter the fact that these modes approach zero-modes of the string in the massless limit. For a more detailed discussion, see \cite{Gerchkovitz:2017kyi}.

Global symmetries of the massless theory can be used to generate new approximate solutions from a given string solution. However, not all the quasi-moduli are of this form. A crucial role is played by moduli that will be referred to as size-moduli. These are deformations of the string solution that exist only when $N_f>N_c$ and affect the size of the string in the transverse plane. 

Let us consider a string solution that has the diagonal form \eqref{basicansatz} for $i\leq N_c$.  In contrast to equation \eqref{basicansatz}, however, we now allow the  $i>N_c$ entries of $\bf q$ and $\bf\tilde q$ to be non-vanishing. In \cite{Gerchkovitz:2017kyi} we showed that the Bogomol'nyi equations and the boundary conditions imply the following relation between the $i>N_c$ and $i\leq N_c$ entries:  	\eql{sizemodesgeneralzeros}{q_{i>N_c}^{a}&=\left(\prod_{b\neq a}\frac{q_{b}(r,\phi)}{\prod_{l_b=1}^{k_b}(z-z_{l_b})}\right)^{\Delta_{ia}}\left(\frac{q_a(r,\phi)}{\prod_{l_a=1}^{k_a}(z-z_{l_a})}\right)^{1+\Delta_{ia}}\times\sum_{n=0}^{\lceil k_a+\Delta_{ia} K\rceil-1}\rho^{(n)}_{ia}z^n\;,\\
	\tilde{q}^{i>N_c}_{a}&=\left(\prod_{b\neq a}\frac{q_{b}(r,\phi)}{\prod_{l_b=1}^{k_b}(z-z_{l_b})}\right)^{-\Delta_{ia}}\left(\frac{q_a(r,\phi)}{\prod_{l_a=1}^{k_a}(z-z_{l_a})}\right)^{-\Delta_{ia}-1}\times \sum_{n=0}^{\lceil-k_a-\Delta_{ia}K\rceil-1}\tilde{\rho}^{(n)}_{ia}z^n\;,}
where $z_{l_a}$, $l_a=1,...,k_a$, are the zeros of $q_a(r,\phi)$.
 ${\rho}_{ia}^{(n)}$ and $\tilde{\rho}_{ia}^{(n)}$  are complex parameters that will be referred to as the size-moduli of the string.  
 When these moduli exist, the string solution approaches the large $r$ limit of \eqref{basicansatz} with a power-law behavior, with typical size that is determined by the size-moduli (see appendix A of \cite{Gerchkovitz:2017kyi}). 
 
 From equation (\ref{sizemodesgeneralzeros}) one can see that long-range moduli for which $q^{i>N_c}$ or $\tilde q_{i>N_c}$ decay slower than $1/r$ exist if and only if $\Delta_{ia}K\notin \mathbb{Z}$ for some $i>N_c$ and $a<N_c$. In \cite{Gerchkovitz:2017kyi} we argued that these are exactly the cases in which the string and the bulk moduli do not decouple at low energies. For a supersymmetric localization derivation of this condition, see section \ref{general U1 charges}. In this work, we wish to identify the worldsheet theories on strings that decouple from the bulk. Thus, we will  assume that $\Delta_{ia}K\in \mathbb{Z}$.
 
 	When for some flavor both $q$ and $\tilde{q}$ can be excited, the F-term equations
\eql{F}{\sum_i\tilde{q}^i\la^\al q_i=\sum_i c_i\tilde{q}^iq_i=0}
 act as extra constraints on the moduli space. For strings that satisfy the condition (\ref{weak-cond}), one cannot excite $q$ and $\tilde q$ together for the same flavor, as can be seen in equation (\ref{sizemodesgeneralzeros}). For these strings, we find a description of the worldsheet theory that is weakly coupled when the four-dimensional theory is weakly coupled. We discuss these cases in the next section.

 If $n$ of the charges $c_1,...,c_{N_c}$  are equal, the theory has a global $SU(n)$ symmetry in the massless limit, acting on the equally charged hypermultiplets. This symmetry can be used to generate new string solutions.  The moduli associated with this symmetry sit in the off-diagonal elements of $q_a^b$ with $a,b\leq N_c$ and $a\neq b$. 
However, off-diagonal components of $q_a^b$ can give rise to zero-modes even when no global symmetry is involved. Evidence for the existence of these modes for strings that satisfy the condition \eqref{weak-cond} come from our localization results, and are supported by expectations from $S$-duality (see appendix \ref{offdiagonalmodes}).  When the condition \eqref{weak-cond} is satisfied, we expect each $a,b$ pair to give rise to $k_a+k_b$ complex moduli with masses $\mu_{ab}$ given by equation \eqref{mass-in-vac}.\footnote{One can show that an excitation of $q_a^b$ should be accompanied by an excitation of $q_b^a$. The first line in \eqref{Bogomolnyiset} implies that an excitation of $q_a^b$ must be accompanied with excitation of $A_{ab}$ and $A_{ba}$. The Bogomol'nyi equation $(\cD_1+i\cD_2)q=0$ therefore requires that $q_b^a\neq 0$. }

Due to the $1/2$-BPS nature of the string solutions, the worldsheet theory is $\cN=(2,2)$ supersymmetric. In addition, the worldsheet theory inherits from the 4d theory a one parameter family of vectorlike $R$-symmetries, parametrized by $R_{2d}=\al R^{(J)}+(1-\al)R^{(R)}$ with $\al\in\mathbb{R}$. In terms of the symmetries of the 4d theory, $R^{(R)}$ is related to the $U(1)$ $R$-symmetry generator preserved by the string background. When normalized as a worldsheet $R$-symmetry generator, the charges of the hypermultiplet scalars under $R^{(R)}$ are given by 
 \eql{RRcharges}{R^{(R)}\left[q_i^a\right]=N_c\Delta_{ai}\;,\;\;\;{R^{(R)}\left[\tilde{q}_a^i\right]=2+N_c\Delta_{ia}\;.}}  $R^{(J)}$ is related to rotations in the $x^1-x^2$ plane in the following way. A generic string configuration breaks rotations symmetry in the $x^1-x^2$ plane. However, a string solution of the form \eql{rotinv}{\textbf{q}=\mat{q_1(r)e^{ik_1\phi}&0&0&\cdots&0&0&\cdots&0\\0&q_2(r)e^{ik_2\phi}&0&\cdots&0&0&\cdots&0\\0&0&q_3(r)e^{ik_3\phi}&\cdots&0&0&\cdots&
			0\\\vdots&\vdots&\vdots&\ddots&\vdots&\vdots&\ddots&\vdots\\0&0&0&\cdots&q_{N_c}(r)e^{ik_{N_c}\phi}&0&\cdots&0}\ ,\;\;\;0\leq k_i\in\mathbb{Z}\;,} preserves a combination of rotations and gauge transformations. Normalized as $R$-symmetry on the worldsheet, this transformation acts on the scalars $q_i^a$ and the angle $\phi$ as
	\eql{Jaction}{q_i^a\rightarrow e^{2ik_a\phi_0+2iK\Delta_{ia}\phi_0}q_i^a\ ,\ \phi\rightarrow \phi-2\phi_0\ .}
 The $R^{(J)}$-charges of the size-modes are thus given by  
 \eql{RJcharges}{R^{(J)}\left[\rho_{ia}^{(n)}\right]=2K\Delta_{ia}+2k_a-2n\;,\;\;\;R^{(J)}\left[\tilde{\rho}_{ia}^{(n)}\right]=-2K\Delta_{ia}-2k_a-2n\;.}

\section{Examples of Worldsheet Theories}\label{examples}

In this section we discuss the worldsheet theories on strings that satisfy, in addition to the decoupling condition (\ref{decoupling criterion}),  the following condition:  for every $N_c<i\leq N_f$, 
\begin{equation}
\begin{aligned}
&\Delta_{ia}\geq 0 \;\;\;\;\;\text{for all}\;\;\; 1\leq a\leq N_c \;\;\;\text{or}\\
&\Delta_{ia}\leq -1 \;\;\text{for all} \;\;\;1\leq a\leq N_c\;.
\end{aligned}\label{weak-condition}
\end{equation} These are the cases in which there is no flavor for which both $q_i$ and $\tilde{q}^i$ give rise to zero-modes and therefore the F-terms $\sum_i\tilde{q}^i\la^\al q_i\ ,\ \sum_i c_i \tilde{q}^i q_i$ vanish identically, without imposing any additional constraints on the target space.

For strings that satisfy the condition \eqref{weak-condition}, we identify in section \ref{general U1 charges} a two-dimensional GLSM whose two-sphere partition function matches exactly with the output of our localization analysis. The corresponding low-energy non-linear sigma model (NLSM)  provides a proposal for the worldsheet theory. The map between the 4d and 2d couplings that we obtain in these cases is such that the worldsheet GLSM  is weakly coupled when the four-dimensional theory is weakly coupled.  

In this section we will describe our proposal for the worldsheet theories and will test it against expectations that are based on classical zero-mode analysis. We will start with the special case where $c_i=1$ for all $i=1,...,N_c$.  Then, in section \ref{simpleexample}, we will generalize to the case where $c_i\geq c_a$ for all $1\leq a\leq N_c$ and  $N_c+1\leq i\leq N_f$. Finally, in section \ref{qandtildeq}, we will discuss the most general case satisfying the condition \eqref{weak-condition}.

\subsection{The Equal Charge Case}
\label{equalcharge}
The case in which $c_i=1$ for every $i=1,...,N_f$ has been studied extensively in the literature. See, for example, \citep{Auzzi:2003fs,Hanany:2003hp,Hanany:2004ea,Shifman:2004dr,Shifman:2006kd,Eto:2005yh,Auzzi:2005gr}, and the reviews \cite{Shifman:2007ce,Tong:2005un,Tong:2008qd,Eto:2006pg}.  The worldsheet theory has been studied in various methods, such as brane construction \cite{Hanany:2004ea,Hanany:2003hp}, matching of the BPS spectra \cite{Dorey:1998yh,Dorey:1999zk}, moduli matrix approach \cite{Eto:2005yh,Eto:2006pg}, as well as explicit derivations in field theory \cite{Shifman:2004dr,Shifman:2006kd,Auzzi:2003fs,Hanany:2004ea,Auzzi:2005gr} and localization techniques (see, for example, \cite{Chen:2015fta,Pan:2015hza} and section \ref{42} of this work.)

For a string solution in which the hypermultiplet scalars labeled by $a=1,...,N_c$ obtain a vacuum expectation value, Hanany and Tong  proposed that the worldsheet theory can be described as the low-energy limit of an $\cN=(2,2)$ $U(K)$ gauge theory with 
\begin{itemize}
	\item 1 massless chiral multiplet, $X$, in the adjoint representation of $U(K)$,
	\item $N_c$ chiral multiplets, $\psi_a$, $a=1,...,N_c$, in the fundamental representation of $U(K)$, 
	\item  $N_f-N_c$ chiral multiplets, $\tilde{\psi}_i$, $i=N_c+1,...,N_f$, in the anti-fundamental representation of $U(K)$.
\end{itemize} 
The FI parameter of this theory is given in terms of Yang-Mills coupling of $SU(N_c)$ factor of the gauge group via 
\begin{equation}
\xi_{2d}=\frac{4\pi }{g^2}\;.\label{gtoxi}
\end{equation}
The localization analysis of \cite{Chen:2015fta,Pan:2015hza} relates the $\theta$-parameters in the following way:
\begin{equation}
\te_{2d}=\te_{4d}+(K-1)\pi\;.\label{tetote}
\end{equation}
The twisted masses and $R$-charges are summarized in table \ref{table0}.\footnote{Note that the physical parameters are combinations of the form  $m_{\psi_a}-m_{\psi_b}$ and $m_{\psi_a}+m_{\tilde{\psi_i}}$ and not the masses listed in table \ref{table0}, which can be shifted by redefinitions of the vector multiplet scalar.}
\begin{table}[h]
	\caption{\small{The spectrum of the worldsheet GLSM in the equal charge case.}}\label{table0}
	\begin{center}
		\begin{tabular}{|c|c|c|c|c|}
			\hline
			Field&$U(K)$ Representation&Twisted Mass&$R^{(J)}$&$R^{(R)}$\\
			$X$   & adjoint                &0       &    2          &      0          \\
			$\psi_a$ & fundamental       & $\mu_a$&   $2$    & $0$    \\
			$\tilde{\psi}_i$ & 
			anti-fundamental & $-\mu_i$&  $0$ &   $0$ \\
			
			\hline
		\end{tabular}
	\end{center}
\end{table}

The NLSM fields that arise from the GLSM fundamental chiral multiplets are related to vacuum symmetries that are broken by the string solutions.  The anti-fundamental chiral multiplets consist of the size-moduli and their superpartners, while the adjoint chiral multiplet consists of the translational moduli and their superpartners. Since the latter are exact zero-modes, the adjoint chiral multiplet is massless.

\subsection{Strings with no $\tilde{q}$ Excitations} \label{simpleexample}
\subsubsection{Proposal for the Worldsheet Theory}
We now relax the equal charge requirement, and instead require that 
\eql{notildecondition}{c_i\geq c_a\;\;\; \text{for all}\;\;\; a\leq N_c<i\ .}  This is the condition that there are no $\tilde{q}$ excitations, as can be easily derived from equation \eqref{sizemodesgeneralzeros}.
In section \ref{Locnotilde}, we identify the two-sphere partition that we obtain as the output of our localization analysis with the partition function of an $\cN=(2,2)$ supersymmetric theory, with $U(K)$ gauge group and  
\begin{itemize}
	\item 1 chiral multiplet, $X$, in the adjoint representation of $U(K)$,
	\item $N_c$ chiral multiplets, $\psi_a$, $a=1,...,N_c$, in the fundamental representation of $U(K)$,
	\item  $N_f-N_c$ chiral multiplets, $\tilde{\psi}_i$, $i=N_c+1,...,N_f$, in the anti-fundamental representation of $U(K)$,
	\item 
	neutral decoupled chiral multiplets, $\eta_{i,a,r}$ , 
	$a=1,...,N_c$, $i=N_c+1,...,N_f$, $r=0,...,\Delta_{ia}K-1$.
\end{itemize} 
The twisted masses and $R$-charges of the chiral superfields  are summarized in table \ref{table1}. The FI and theta parameters are given by equations \eqref{gtoxi} and \eqref{tetote}.

\begin{table}[h]
	\caption{\small{The spectrum on the worldsheet in cases where the string admits no $\tilde{q}$ excitations.}}\label{table1}
	\begin{center}
		\begin{tabular}{|c|c|c|c|c|}
			\hline
			Field&$U(K)$ Representation&Twisted Mass&$R^{(J)}$&$R^{(R)}$\\
			$X$   & adjoint                &0       &    2          &      0          \\
			$\psi_a$ & fundamental       & $\mu_a-\frac{\sum_{b=1}^{N_c}\mu_b}{C}c_a$&   $2-\frac{2Kc_a}{C}$    & $\frac{N_cc_a}{C}$    \\
			$\tilde{\psi}_i$ & anti-fundamental& $-\mu_i+\frac{\sum_{b=1}^{N_c}\mu_b}{C}c_i$&  $\frac{2Kc_i}{C}$ &   $-\frac{N_cc_i}{C}$ \\
			$\eta_{i,a,r}$& trivial &$\mu_{ai}$  &2(r+1)& $N_c\Delta_{ai}$\\

			\hline
		\end{tabular}
	\end{center}
\end{table}

\subsubsection{Comparison with the Classical Zero-Mode Analysis}\label{sec-comp-noqtilde}
The GLSM described above gives rise to an NLSM at low-energies. We now compare the spectrum of this NLSM with expectations that are based on the classical string zero-modes analysis of section \ref{21}.

The massless adjoint chiral multiplet, as before, consists of the translational moduli and their superpartners. The $N_c$ fundamental chiral multiplets, as in the equal charge case, are expected to be associated with excitations of off-diagonal $q_a^b$ entries, $a,b=1,..,N_c$. We will show below  that the masses and $R$-charges of the fundamental fields given in table \ref{table1} agree with this expectation. 
Unlike the equal charge case, the existence of these moduli does not follow from a breaking of a global symmetry in the general case.  

The decoupled and the anti-fundamental chiral multiplets are associated with the size modes $\rho_{ia}^{(n)}$ of equation \eqref{sizemodesgeneralzeros}. The division of the size modes into neutral and charged fields can be understood as follows. A scalar $q_{i}^a$ gives rise to $k_a+\Delta_{ia}K$ size modes. The decoupled modes are the ones that exist for any choice of the partition $\{k_a\}$. These are the $\Delta_{ia}K$ size modes $\rho_{ia}^{(n)}$ with $n=k_a+\Delta_{ia}K-1-r$ and $r=0,...,\Delta_{ia}K-1$. Indeed the masses, $R^{(R)}$- and $R^{(J)}$-charges of these modes are given by $\mu_{ai}$, $N_c\Delta_{ai}$ and $2(r+1)$  respectively, in agreement with the values for $\eta_{i,a,r}$ in table \ref{table1}, that were extracted from the localization analysis of section \ref{general U1 charges}. 

The anti-fundamental chiral multiplets are expected to be associated with the size-modes $\rho_{ia}^{(n)}$ of equation \eqref{sizemodesgeneralzeros}, with $n=0,...,k_a-1$. Since there are exactly $k_a$ of these for every $i,a$,  they are in one to one correspondence with the zeros of $q_{a}^{a}$, in agreement with the expectation that they transform in the anti-fundamental of $U(K)$. 
In order to understand the masses and the $R$-charges of the charged fields,  $\psi_a$ and $\tilde{\psi}_i$ 
(we use the same notation for the multiplet and its bottom component), we need to look at the physical masses and $R$-charges around the vacuum of the worldsheet theory that corresponds to a string solution of the form (\ref{rotinv}).\footnote{The vacua of the worldsheet theory are in one to one correspondence with $K$-string solutions. In the equal charge case, it was shown that monopoles in the 4d theory connect strings with different partitions of $K$ \cite{Tong:2003pz}. These monopoles map to kinks in the worldsheet theory that interpolate between the corresponding vacua \cite{Dorey:1998yh,Dorey:1999zk}.  It is easy to check that the agreement between the monopole and kink masses holds also in the cases described in this section. } 

As a warm-up, let us start with the $K=1$ case. One of the $\psi_a$ fields must get a vacuum expectation value due to the D-term constraint
\eq{\sum_{a=1}^{N_c}|\psi_a|^2-\sum_{i=N_c+1}^{N_f}|\tilde{\psi}_i|^2=\xi_{2d}\ .}
Consider the vacuum in which $\psi_{a'}$ gets a VEV for some $a'$. This vacuum corresponds to the string with windings $k_a=\delta_{a,a'}$. To cancel the contribution from the mass term to the potential, the gauge multiplet scalar will get a VEV which is equal to the twisted mass of ${\psi_{a'}}$. As a result, the masses of the other charged fields are shifted in this vacuum, 
\eq{m_{\psi_a}=\mu_{aa'}\ ,\ m_{\tilde{\psi}_i}=\mu_{a'i}\ .}

Similarly, the $R$-charges of the charged fields are shifted in the vacuum with respect to the charges given in table \ref{table1}; the $R$-transformation is accompanied by a $U(1)$ gauge transformation such that the total transformation leaves $\psi_{a'}$ invariant. 
Thus, the $R$-charges of $\psi_a$ and $\tilde{\psi}_i$ in this vacuum are
\eq{R^{(J)}_{\psi_a}=2\Delta_{a'a}\ ,\ R^{(J)}_{\tilde{\psi}_i}=2+2\Delta_{ia'}\ ,\ R^{(R)}_{\psi_a}=N_c\Delta_{aa'}\ ,\ R^{(R)}_{\tilde{\psi}_i}=N_c\Delta_{a'i}\ .} 

Comparing with equations \eqref{mass-in-vac}, \eqref{RRcharges} and \eqref{RJcharges}, we see that the masses and $R^{(R)}$-charges of $\psi_a$ and $\tilde{\psi}_i$ coincide with the masses and $R^{(R)}$-charges of the four-dimensional fields $q_{a}^{a'}$ and $q_i^{a'}$ respectively. $R^{(J)}_{\tilde{\psi}_i}$ coincide, as expected, with the charges of the size-modes $\rho_{ia'}^{(n=0)}$. 

The analysis of the $K>1$ vacua and spectrum is more involved. The string configurations of the type \eqref{rotinv} correspond to solutions of the GLSM $D$-term equations 
\eql{2dDterm}{\sum_{a=1}^{N_c}\psi_a^m\psi^{\dagger}_{a,n}+[X,X^{\dagger}]^m_{\ n}-\sum_{i=N_c+1}^{N_f}\tilde{\psi}_i^{m\ \dagger}\tilde{\psi}_{i,n}=\xi_{2d}\delta^m_{\ n}\;,
}  together with the vanishing of the mass terms, where the indices $m,n$ in (\ref{2dDterm}) are $U(K)$ color indices. In these vacua,  $\tilde{\psi}_i=0$. To solve for the vacuum expectation values of the other fields, it will be useful to divide the $K\times K$ matrices $X$ and $\psi_a\psi^\dagger_a$ into $N_c$ blocks of sizes $k_b\times k_b$, $b=1,...,{N_c}$.
We will label the entries of the b'th block by $\al{(b)},\ \be{(b)}=1,...,k_b$. Up to gauge transformations, the solution is given by
\eql{Klargevacua}{\psi_a^{\al{(b)}}=\sqrt{k_a\xi_{2d}}\delta^{\al{(a)},1}\delta^{ab}\ ,\ X^{\al{(a)}\be{(a)}}=\sqrt{(k_a-\be{(a)})\xi_{2d}}\delta^{\al{(a)},\be{(a)}+1}\ .}
It is straightforward to verify that this solution solves equation \eqref{2dDterm}. The vanishing of the mass terms is satisfied by giving the gauge multiplet scalar the vacuum expectation value \eq{\sigma^{\al{(a)}\be{(a)}}=\left(\mu_a-\frac{\sum_{b=1}^{N_c}\mu_b}{C}c_a\right)\delta^{\al{(a)}\be{(a)}}\ .}
As expected, we find that the GLSM vacua are labeled by integer partitions of $K$. 

The $R$-symmetry preserved in such a vacuum is such that the transformation $e^{i\left(\al R^{(J)}+(1-\al)R^{(R)}\right)\omega_R}$ is accompanied by a $U(K)$ gauge transformation
\eq{\psi_a\to U\psi_a\ ,\ \tilde{\psi}_i\to \tilde{\psi}_iU^\dagger\ ,\ X\to U X U^\dagger}
with 
\eq{U^{\be{(a)},\gamma{(a)}}=\delta^{\be{(a)},\gamma{(a)}}\exp\left(-i\left[\al(2\be{(a)}-2Kc_a/C)+(1-\al)\frac{N_cc_a}{C}\right]\omega_R\right)\ .}

As a result of the discussion above, the masses and the $R$-charges of the charged fields are shifted such that in the vacuum labeled by the partition $\{k_a\}$ we obtain
\begin{equation}
\begin{aligned}
&\text{for\;}\psi^{\alpha{(b)}}_a\;:\;\;\;m=\mu_{ab}\;,\;\;\;R^{(R)}=N_c\Delta_{ab}\;,\;\;\;R^{(J)}=2K\Delta_{ba}+2(1-\alpha{(b)})\;,\\
&\text{for\;}\tilde\psi^{\alpha{(a)}}_i\;:\;\;\;m=\mu_{ai}\;,\;\;\;R^{(R)}=N_c\Delta_{ai}\;,\;\;\;R^{(J)}=2K\Delta_{ia}+2\alpha{(a)}\;.
\end{aligned}
\end{equation}
Comparing with equations \eqref{mass-in-vac}, \eqref{RRcharges} and \eqref{RJcharges} we see that the $\tilde \psi_i$ spectrum matches exactly  with the spectrum of the size modes $\rho^{(n)}_{ia}$ with $n=0,...,k_a-1$. The $\psi_a$ mass and $R^{(R)}$ spectrum, as advertised in the previous section, agrees with the spectrum of $k_a+k_b$ moduli corresponding to excitations of $q_{a}^b$ (accompanied by $q_{b}^a$ excitations) for each $a,b$ pair.

\subsection{Strings with $q$ and $\tilde{q}$ Excitations}
\label{qandtildeq}
\subsubsection{Proposal for the Worldsheet Theory}
We now discuss the most general case for which we have a proposal for a worldsheet theory that is weakly coupled when the four-dimensional theory is weakly coupled. The only restriction on the $U(1)$ charges is now that the F-term constraints given in equation \eqref{F} are satisfied trivially and therefore for every $i>N_c$ there are either modes coming from $q_i$ or modes coming from $\tilde{q}^i$, but not both. Without loss of generality, we will assume that for $N_c<i\leq N_q$, there are only $q_i$ modes and for $N_q< i\leq N_f$ there are only $\tilde{q}^i$ modes, for some $N_c\leq N_q\leq N_f$. In terms of the $U(1)$ charges, this assumption means that
\eql{u1chargesbothqtildeq}{&\Delta_{ia}\geq 0\ \;\text{for all}\;\ 1\leq a\leq N_c\ ,\ N_c+1\leq i\leq N_q\ ,\\
	-&\Delta_{ia}\geq 1\ \;\text{for all}\;\ 1\leq a\leq N_c\ ,\ N_q+1\leq i\leq N_f\ .}

In section \ref{localization-q-tilde-q}, we identify our proposal for the two-sphere partition function on the string worldsheet with the two-sphere partition function of an $\cN=(2,2)$ supersymmetric $U(K)$ gauge theory with
\begin{itemize}
	\item 1 chiral multiplet, $X$, in the adjoint representation of $U(K)$,
	\item $N_c$ chiral multiplets, $\psi_a$, $a=1,...,N_c$, in the fundamental representation of $U(K)$,
	\item  $N_f-N_c$ chiral multiplets, $\tilde{\psi}_i$, $i=N_c+1,...,N_f$, in the anti-fundamental representation of $U(K)$,
	\item neutral decoupled chiral multiplets, $\eta_{i,a,r}$ ,  $a=1,...,N_c$, $i=N_c+1,...,N_q$, $r=0,...,\Delta_{ia}K-1$,
	\item neutral decoupled chiral multiplets, $\tilde{\eta}_{i,a,r}$,  $a=1,...,N_c$, $i=N_q+1,...,N_f$, $r=0,...,-\left(1+\Delta_{ia}\right)K-1$, 
	\item  neutral chiral multiplets, $\chi_{i,a,r}$, $i=N_q+1,...,N_f$, $a=1,...,N_c$, $r=0,...,K-1$.            
\end{itemize}
The twisted masses and $R$-charges of these multiplets are summarized in table \ref{table2}. The FI and theta parameters are given by equations \eqref{gtoxi} and \eqref{tetote}.

\begin{table}[h]
	
	\caption{\small{The spectrum on the worldsheet for strings satisfying \eqref{u1chargesbothqtildeq}  }}\label{table2}
	\begin{center}
		\begin{tabular}{|c|c|c|c|c|}
			\hline
			Field&$U(K)$ Representation&Twisted Mass&$R^{(J)}$&$R^{(R)}$\\
			$X$  & adjoint                 &0       &    2          &      0          \\
			$\psi_a$ &fundamental       & $\mu_a-\frac{\sum_{b=1}^{N_c}\mu_b}{C}c_a$&   $2-\frac{2Kc_a}{C}$    & $\frac{N_cc_a}{C}$    \\
			$\tilde{\psi}_i$ & anti-fundamental& $-\mu_i+\frac{\sum_{b=1}^{N_c}\mu_b}{C}c_i$&  $\frac{2Kc_i}{C}$ &   $-\frac{N_cc_i}{C}$ \\
			$\eta_{i,a,r}$& trivial& $\mu_{ai}$  &2(r+1)& $N_c\Delta_{ai}$\\
			$\tilde{\eta}_{i,a,r}$& trivial&$\mu_{ia}$ &2(r+1)& $2+N_c\Delta_{ia}$\\
			$\chi_{i,a,r}$& trivial&$\mu_{ia}$  & $ -2r+2K\Delta_{ai}$& $2+N_c\Delta_{ia}$\\
			
			\hline
		\end{tabular}
	\end{center}

\end{table}

The $\chi$ multiplets interact with the charged multiplets via the superpotential
\eql{superpotential}{W=\sum_{r=0}^{K-1}\sum_{a=1}^{N_c}\sum_{i=N_q+1}^{N_f}\al_r\,\chi_{i,a,r}\tilde{\psi}_iX^{r}\psi_a} where the color index is suppressed. The coefficients $\al_r$ cannot be fixed from the localization analysis since the two-sphere partition functions is independent of superpotential couplings \cite{Doroud:2012xw,Benini:2012ui}.\footnote{This is true if the massive superalgebra on the sphere preserves a vectorlike $R$-symmetry. If it preserves an axial $R$-symmetry, the sphere partition function will not depend on twisted superpotential couplings.} The superpotential is allowed due to the relation
\eq{m_{\chi_{i,a, I}}+m_{\tilde{\psi}_i}+m_{\psi_a}+rm_X=0\ ,\ R_{\chi_{i,a, I}}+R_{\tilde{\psi}_i}+R_{\psi_a}+rR_X=2\;,} which is satisfied for both  $R^{(J)}$ and $R^{(R)}$.

\subsubsection{Comparison with the Classical Zero-Mode Analysis}\label{sec-comp-qtildeq}

We now describe how the proposal for the worldsheet GLSM fits with the zero-mode analysis. To do this, we will study the spectrum around the GLSM vacuum \eqref{Klargevacua} and compare to the classical spectrum of zero modes around the string solution \eqref{rotinv}.
The analysis of the first $N_q$ hypermultiplets is exactly the same as in the previous section. The only difference comes from the last $N_f-N_q$ hypermultiplets. For each one of them, the scalar $\tilde{q}^i_a$ gives rise to $\Delta_{ai}K-k_a$ complex modes: $\tilde\rho_{ia}^{(n)}$, $n=0,...,\Delta_{ai}K-k_a-1$ (see equation \eqref{sizemodesgeneralzeros}).  The $\tilde\rho_{ia}^{(n)}$ size-modes for $n=\Delta_{ai}K-k_a-1-r$, $r=0,...,-(1+\Delta_{ia})K-1$, are embedded in the decoupled chiral multiplets $\tilde{\eta}_{i,a,r}$. These are the modes that exist for any choice of the partition $\{k_a\}$. Indeed, the masses and $R$-charges of these size modes, given in equations \eqref{mass-in-vac}, \eqref{RRcharges} and \eqref{RJcharges}, agree with those listed in table \ref{table2} for $\tilde{\eta}_{i,a,r}$.
The neutral chiral multiplets $\chi_{i,a,r}$, with $r=k_a,..,K-1$,   correspond to the $\tilde\rho_{ia}^{(n)}$ size-modes for $n=r-k_a$. Again, the masses and $R$-charges listed for these multiplets in table \ref{table2} agrees with equations \eqref{mass-in-vac}, \eqref{RRcharges} and \eqref{RJcharges}.  The remaining fields, $\tilde{\psi}_i$ and $\chi_{i,a,r}$ for $r=0,...,k_a-1$, must vanish around the vacuum  \eqref{Klargevacua} due to the superpotential constraint equations, as we will show now.

Let us start with $K=1$. 
The superpotential in this case is \eq{W=\al\sum_{a=1}^{N_c}\sum_{i=N_q+1}^{N_f}\chi_{i,a}\tilde{\psi}_i\psi_a}
and the constraint equations derived from it are
\begin{align}
&\frac{\partial W}{\partial \chi_{i,a}}=0\;\;\;\Rightarrow\;\;\;\tilde{\psi}_i\psi_a=0\;,\label{W1}\\&\frac{\partial W}{\partial \tilde{\psi}_i}=0\;\;\;\;\;\Rightarrow\;\;\;\sum_{a=1}^{N_c}\chi_{i,a}\psi_a=0\;,\label{W2}\\&\frac{\partial W}{\partial \psi_a}=0\;\;\;\;\;\Rightarrow\;\;\;\sum_{i=N_q+1}^{N_f}\chi_{i,a}\tilde{\psi}_i=0\;.\label{W3}
\end{align} 
Due to the D-term constraint \eql{Dterm}{\sum_{a=1}^{N_c}|\psi_a|^2-\sum_{i=N_c+1}^{N_f}|\tilde{\psi}_i|^2=\xi_{2d}\ ,} at least one of the $\psi_a$'s must obtain a VEV. Therefore, equation \eqref{W1} is solved by $\tilde{\psi}_i=0$ for every $N_q+1\leq i\leq N_f$. As a result, equation \eqref{W3} is satisfied trivially while equation \eqref{W2} still acts as extra $N_f-N_q$ complex constraints on the target space \eqref{Dterm}. Consider the vacuum in which $|\psi_a|^2=\xi_{2d}\delta_{a,a'}$ for some $a'$. This vacuum corresponds to the string with winding $k_a=\delta_{a,a'}$. Equation \eqref{W2} implies that $\chi_{i,a'}=0$ for every $i$.

Moving on to general $K$, we consider the vacuum \eqref{Klargevacua}.
In this vacuum there exist an $SU(K)$ invariant operator, positively charged under the $U(1)\subset U(K)$, that obtains non-zero expectation value. This operator is obtained by contracting the $U(K)$ color indices in 
\eql{baryon}{\prod_{a=1}^{N_c}\prod_{r=0}^{k_a-1}\left(X^r\cdot \psi_{a}\right)_{n_{a,r}}} using the antisymmetric tensor.
The superpotential \eqref{superpotential} imposes (among many others) the constraints 
\eql{Wconstraint}{\frac{\partial W}{\partial \chi_{i,a,r}}=0\;\Rightarrow\; \tilde{\psi}_iX^r\psi_{a}=0\;,\;\;\; a=1,...,N_c \;,\;\;\; r=0,...,k_a-1\ .}
Because \eqref{baryon} is not zero, the only solution of \eqref{Wconstraint} is $\tilde{\psi}_i=0$. After imposing $\tilde{\psi}_i=0$, the constraint equations become
\eql{moreconstraints}{&\sum_{a=1}^{N_c}\psi_a^m\psi^{\dagger}_{a,n}+[X,X^{\dagger}]^m_{\ n}=\xi_{2d}\delta^m_{\ n}\ ,\\
	&\sum_{r=0}^{K-1}\sum_{a=1}^{N_c}\al_r\chi_{i,a,r} \left(X^{r}\cdot\psi_a\right)_n=0\ .}
The fields $\chi_{i,a,r}$ with $r=0,...,k_a-1$ vanish in this vacuum due to the second line in \eqref{moreconstraints}.

\section{Two-Sphere Worldsheet Partition Functions from the Four-Ellipsoid Partition Function }\label{4}
We now move on to the localization analysis of the worldsheet theories. We start this section by reviewing the four-ellipsoid partition function in section \ref{41}. Then, in section \ref{42}   we  demonstrate, in the $c_i=1$ example, the prescription for extracting the worldsheet two-sphere partition function from the four-ellipsoid partition function of the original theory. In the next section we will generalize the analysis to general $U(1)$ charges. 

\subsection{The Four-Ellipsoid Partition Function}\label{41}
An $\cN=2$ supersymmetric theory that preserves a $U(1)_R\subset SU(2)_R$ $R$-symmetry can be placed on the four-ellipsoid, 
\begin{equation}
\frac{x_0^2}{r^2}+\frac{x_1^2+x_2^2}{l^2}+\frac{x_3^2+x_4^2}{\tilde l^2}=1\;,\label{ellipsoid}
\end{equation}
while preserving an $su(1|1)$ superalgebra which contains one supercharge that squares into a linear combination of the two rotations and the $R$-symmetry generator \cite{Hama:2012bg},  
\begin{equation}
Q^2=\frac{1}{l}J_{1,2}+\frac{1}{\tilde l}J_{3,4}+\left(\frac{1}{ l}+\frac{1}{{\tilde l}}\right)R\;.
\end{equation}

The four-ellipsoid partition function for such a theory has been computed by Hama and Hosomichi using supersymmetric localization \cite{Hama:2012bg}. This partition function depends on the dimensionless squashing parameter, $b^2\equiv{l}/{\tilde l}$, on the complexified gauge couplings, and on the dimensionless masses and FI parameters, $\hat\mu \equiv\sqrt{l\tilde l} \mu$, $\hx\equiv\sqrt{l\tilde l}\xi$.\footnote{From now on, we follow the conventions of \cite{Hama:2012bg} for the FI parameter.  In these conventions $\xi$ has mass dimension $1$. In the case of the round sphere, this  parameter is related to the flat space FI parameter by rescaling with the radius. Throughout this work, when a dimension one parameter appears with a hat, it means that it is measured in units of $\sqrt{l\tilde l}$.}   

The four-ellipsoid partition function is given by \cite{Hama:2012bg} 
\begin{equation}
Z_{S^4_b}=\int  [d\hat a]\, e^{-S_{\text{cl}}}\,Z^{\text{vec}}_{\text{1-loop}}\,Z^{\text{hyper}}_{\text{1-loop}}\,|Z_{\text{inst}}|^2\;,
\end{equation}
where the integral is over the Cartan subalgebra of the gauge group (the integration variables are the rescaled Coulomb branch coordinates,  $\ha=\sqrt{l\tilde l }a$).
$S_{\text{cl}}$ is the classical value of the Yang-Mills and Fayet-Illioupulus actions in the localization saddle points,
\begin{equation}
\begin{aligned}
&S_{\text{cl}}=S_{\text{YM}}+S_{\text{FI}}\;,\\
&S_{\text{YM}}=\sum_i\frac{8\pi^2}{g_i^2}\Tr{\hat a_i^2}\;,\\
&S_{\text{FI}}=-16i\pi^2\sum_{I}\hx_I\ha_{I}\;,
\end{aligned}
\end{equation}
where $g_i$ is the Yang-Mills coupling for the gauge factor $G_i$ and $\xi_I$ and $a_I$ are the FI parameters and Coulomb branch parameters for the $U(1)$ factors of the gauge group. 
The one-loop determinants for the vector multiplets and hypermultiplets are given by
\begin{equation}
\begin{aligned}
&Z^{\text{vec}}_{\text{1-loop}}=\prod_i\prod_{\alpha\in \Delta^+_i}\Upsilon_b\left(i\ha_i\cdot\alpha\right)\Upsilon_b\left(-i\ha_i\cdot\alpha\right)\;,
\\&Z^{\text{hyper}}_{\text{1-loop}}=\prod_{R}\prod_{f=1}^{n_{R}}\prod_{w\in{P(R)}}\Upsilon_b\left(i\ha\cdot w+iֿ\hm_{f} +\frac{Q}{2}\right)^{-1}\;,
\end{aligned}
\end{equation} where here $i$ runs over the simple factors of the gauge group and $\Delta^+_i$ is the set of positive roots of the simple factor labeled by $i$. The product over $R$ is a product over the irreducible representations of  the gauge group, $n_R$ is the number of hypermultiplets in the representation $R$, and $\mu_f$ denotes the hypermultiplet masses.  $P(R)$ is the set of weights for the representation $R$. The function $\Upsilon_b(x)$ is a holomorphic function which is uniquely defined by  the shift relation \begin{equation}
\Upsilon_b\left(x+b\right)=\frac{\Gamma\left(bx\right)}{\Gamma\left(1-bx\right)}b^{1-2bx}\Upsilon_b\left(x\right)\;,\label{shift}
\end{equation}
and the conditions
\begin{equation}
\begin{aligned}
&\Upsilon_b(x)=\Upsilon_{b^{-1}}(x)\;,\\
&\Upsilon_b\left(\frac{Q}{2}\right)=1\;,
\end{aligned}  
\end{equation}
where $Q=b+b^{-1}$. $\Upsilon_b(x)$ has zeros at 
\begin{equation}
\begin{aligned}
&x+mb+nb^{-1}=0\;,\;\;\;\;\;\;\;\;\;\;m,n\in\mathbb{N}\;,\\&Q-x+mb+nb^{-1}=0\;,\;\;\;m,n\in\mathbb{N}\;.
\end{aligned}
\end{equation}
$Z_{\text{inst}}$ is Nekrasov's instanton partition function in the $\Omega$-background \cite{Nekrasov:2002qd}, with the equivariant parameters $\epsilon_1=l^{-1}$, $\epsilon_2=\tilde l^{-1}$. 

The asymptotic behavior of the function $\Upsilon_b(x)$ in the large $|x|$ limit is of the form $\log\Upsilon_b(x)=\half x(x-Q)\log\left(x(Q-x)\right)+\left(\frac{3}{2}-\gamma\right)x(Q-x)+\cO(\log x)$, where $\gamma$ is the Euler-Mascheroni constant. As a result, 
the convergence of the integrals over the Coulomb branch parameters of a $G_i$ factor in the gauge group depends on the sign of  the $\beta$-function for $g_i$.\footnote{The asymptotic behavior of the $\Upsilon_b$-function implies that the convergence depends on the sign of 
\begin{equation}
2\sum_{\alpha_\in{\Delta_i^+}}(\alpha\cdot \ha_i)^2-\sum_{R}n_R\sum_{w\in P(R)}(w\cdot \ha_i)^2=\left(C(\text{adj})-\sum_{R}n_RC(R)\right)\ha_i^2\;
\end{equation} for simple factors, and that the integral over Coulomb branch parameters for $U(1)$ factors is always divergent.  }
To allow for string configurations,  we need to gauge a $U(1)$ flavor symmetry. This
will make the integral ill-defined due to the $U(1)$ Landau pole. The theory with a gauged $U(1)$, however, does make sense as a low-energy effective theory, defined at energy scales much smaller than the one set by the Landau pole. We will therefore introduce a cut-off for the Coulomb branch integral, $\hat{\Lambda}\ll e^{\frac{1}{e^2}}$, where $e$ is the $U(1)$ gauge coupling.

\subsection{Worldsheet Sphere Partition Functions from the Four-Ellipsoid Partition Function in the Equal Charge Case}\label{42}
In this subsection we will demonstrate, in the case in which $c_i=1$ for all $i$, how we extract the two-sphere partition function of the string worldsheet theory from the four-ellipsoid partition function of the full four-dimensional theory. The main result of this  section appeared already in \citep{Chen:2015fta} (see also \cite{Pan:2015hza,Pan:2016fbl}). We repeat the full derivation below and add discussions on some of the subtleties in the derivation.  In the next section we will generalize the derivation for general $U(1)$ charges. 

We consider an  $SU(N_c)$ gauge theory with $N_c\leq N_f\leq 2N_c$ hypermultiplets in the fundamental representation. We will gauge the $U(1)$ flavor symmetry under which all the hypermultiplets have charge $1$, and introduce an FI parameter $\xi>0$. The ellipsoid  partition function in this case reads
\begin{equation}
\begin{aligned}
Z_{S_b^4}=&\left(\prod_{a=1}^{N_c}\int_{-\hat{\Lambda}}^{\hat{\Lambda}} d\hat a_a\right)\; e^{-\frac{8\pi^2}{g^2}\frac{1}{N_c}\sum_{a<b} (\hat a_a-\hat a_b)^2}e^{-\frac{8\pi^2}{e^2}\frac{1}{N_c}\left(\sum_{a=1}^{N_c} \hat a_a\right)^2}e^{16i\pi^2\hat{\xi}\sum_{a=1}^{N_c}\hat a_a}\\&\frac{\prod_{a=1}^{N_c}\prod_{b\neq a} \Upsilon_b\left(i(\hat a_a-\hat a_b)\right)}{\prod_{a=1}^{N_c}\prod_{j=1}^{N_f} \Upsilon_b\left(i(\hat a_a+\hat\mu_j)+\frac{Q}{2})\right)}\;|Z_{\text{inst}}|^2\;, \label{pi-baryonic}
\end{aligned}
\end{equation}
where $g$ is the Yang-Mills coupling for the $SU(N_c)$ factor of the gauge group  and $e$ is the Yang-Mills coupling for the $U(1)$ factor.    $\mu_1,..,\mu_{N_f}$ are the hypermultiplet masses.
As explained above, we will consider the matrix integral (\ref{pi-baryonic}) with  a cut-off $\hat{\Lambda}\ll e^{\frac{1}{e^2}}$. We will further assume that $e\,,g\,,N_c\ll\hat{\Lambda}$, and that \begin{equation}
\frac{N_c\hat{\Lambda}}{e^2}\,,\frac{N_c\hat{\Lambda}}{g^2}\ll\hat{\xi}\ll\frac{\hat{\Lambda}^{2}}{e^2}\,b^{-1},\frac{\hat{\Lambda}^{2}}{g^2}\,b^{-1}\;.ֿ\label{assumption}
\end{equation}

The assumption $\frac{N_c\hat{\Lambda}}{e^2}\,,\frac{N_c\hat{\Lambda}}{g^2}\ll\hat{\xi}$ allows us to close the integration contours from above, by adding integrations over  $\hat a_a=\hat{\Lambda}e^{i\varphi_a}$, $0<\varphi_a<\pi$. The effect of this modification of the contour is an addition of an $\cO\left(\hat\Lambda e^{-\frac{\hat{\Lambda}^2}{e^2}}\right)+\cO\left(\hat\Lambda e^{-\frac{\hat{\Lambda}^2}{g^2}}\right)$ contribution to the cut-offed matrix integral.\footnote{For small values of $\varphi_a$ the smallness of the integrand is ensured by the classical Yang-Mills terms. For finite $\varphi_a$ the  smallness of the integrand is due to the FI term and the condition $\frac{N_c\hat{\Lambda}}{e^2}\,,\frac{N_c\hat{\Lambda}}{g^2}\ll\hat{\xi}$.} We neglect such terms in our computation.ֿ\footnote{In the limit where $\cO\left(\hat\Lambda e^{-\frac{\hat{\Lambda}^2}{e^2}}\right)+\cO\left(\hat\Lambda e^{-\frac{\hat{\Lambda}^2}{g^2}}\right)$ are neglected, the precise value of the cut-off is not important, because the integral from $\Lambda$ to $(1+c)\Lambda$, $c=\cO(1)$, is $\cO\left(\hat\Lambda e^{-\frac{\hat{\Lambda}^2}{e^2}}\right)+\cO\left(\hat\Lambda e^{-\frac{\hat{\Lambda}^2}{g^2}}\right)$. }
The assumption $\hat{\xi}\ll\frac{\hat{\Lambda}^2}{e^2}\,b^{-1},\frac{\hat{\Lambda}^2}{g^2}\,b^{-1}$ was added to ensure that the contributions that we want to study are much larger than the contributions that we neglect. More precisely, the condition should be $\hat{\xi}K\ll\frac{\hat{\Lambda}^2}{e^2}\,b^{-1},\frac{\hat{\Lambda}^2}{g^2}\,b^{-1}$, where $K$ is the winding number of the string configuration. The interpretation of this condition  is that we can only study strings whose energy is much smaller than the cut-off energy. However, any winding number can be placed inside the region of validity of our analysis by choosing   $\frac{\hat{\Lambda}^2}{e^2}\,b^{-1},\frac{\hat{\Lambda}^2}{g^2}\,b^{-1}$ large enough.

Closing the contours allows us to compute the integrals using Cauchy's theorem, by collecting the residues of the encircled poles. All the poles come from the zeros of the hypermultiplets $\Upsilon_b$-function, in the denominator of the one-loop determinant. We assume that  the hypermultiplet masses are non-degenerate  and therefore all the poles are simple poles.

First, we perform the integration over $\hat a_1$, summing over the residues of the poles 
\begin{equation}
\hat a_1=-\hat\mu_{l_1}+i\frac{Q}{2}+ik_1 b+ik'_1 b^{-1}\;, 1\leq l_1\leq N_f\;,k_1,k'_1,l_1\in \mathbb{N}\;.\label{polesa1}
\end{equation} Then, for each of the residues of the poles in  (\ref{polesa1}) we  perform the integration over $\ha_2$, summing over the residues of the poles,
\begin{equation}
\hat a_2=-\hat\mu_{l_2}+i\frac{Q}{2}+ik_2 b+ik'_2 b^{-1}\;, 1\leq l_2\neq l_1\leq N_f\;,k_2,k'_2,l_2\in \mathbb{N}\;,\label{polesa2}
\end{equation} (note that the residue of the pole (\ref{polesa1}) does not have a pole in $\hat a_2=-\hat\mu_{l_1}+i\frac{Q}{2}+ik_2 b+ik'_2 b^{-1}$), and so on.  Thus, the non-zero contributions to the matrix integral are given by evaluating the residues of the $N_c$ integrals in the simple poles
\begin{equation}
\hat a_a=-\hat\mu_{l_a}+i\frac{Q}{2}+ik_a b+ik'_a b^{-1}\;,\;\;\; a=1,...,N_c\;,
\end{equation}
where $k_a,k'_a,l_a\in{\mathbb N}$, $1\leq l_a\leq N_f$, and $l_a\neq l_b$ for $a\neq b$. The choice of $\{l_a\}$ corresponds to a choice of vacuum in which the $N_c$ hypermultiplets labeled by $\{l_a\}$ obtain a vacuum expectation value. Indeed, the real part of $\hat{a}_a$ matches the classical VEV of the adjoint scalar given in equation \eqref{generalvac}.
Note that we need to include only contributions that satisfy, $(\frac{Q}{2}+k_a b+k'_a b^{-1})^2+\hm_{l_a}^2<\hat{\Lambda}^2$. In addition, as mentioned above,   only contributions with \begin{equation}
\hx\left(\frac{N_cQ}{2}+\sum_{a=1}^{N_c}k_a b+\sum_{a=1}^{N_c}k'_a b^{-1}\right)\ll\frac{\hat{\Lambda}}{e^2}\;,\frac{\hat{\Lambda}}{g^2}\;,\end{equation} are actually significant in the limit we take. This is because the contribution labeled by $\{l_a,k_a,k'_a\}$ comes with a weight 
\begin{equation}
e^{-16\pi^2\hx\left(\frac{N_cQ}{2}+\sum_{a=1}^{N_c}k_a b+\sum_{a=1}^{N_c}k'_a b^{-1}\right)-16i\pi^2\hx\sum_{a=1}^{N_c}\hm_{l_a}}\;.\label{weight}
\end{equation}

In this work we consider only poles with $k'_a=0$.  We will use the dependence on the squashing parameter $b$ to separate them from the other contributions using the weight factors (\ref{weight}). From the point of view of the Higgs branch localization of \cite{Chen:2015fta,Pan:2015hza}, this amounts to focusing on solutions of the saddle point equations that wrap the squashed two-sphere 
\begin{equation}
\frac{x_0^2}{r^2}+\frac{x_1^2+x_2^2}{l^2}=1\;,\label{squashedS2}
\end{equation} 
inside the four-ellipsoid (\ref{ellipsoid}). Contributions for which both $K=\sum_{a}k_a$ and $K'=\sum_ak'_a$ are positive correspond to a $K$-string and a $K'$-string,  wrapping the two-spheres $\frac{x_0^2}{r^2}+\frac{x_1^2+x_2^2}{l^2}=1$ and $\frac{x_0^2}{r^2}+\frac{x_3^2+x_4^2}{\tilde l^2}=1$ respectively, and intersecting in the two-poles $x_0=\pm r$. See \cite{Chen:2015fta,Pan:2015hza,Pan:2016fbl}.

Using the shift identity
\begin{equation}
\frac{\Upsilon_b(x+nb)}{\Upsilon_b(x)}=\prod_{r=0}^{n-1}\frac{\gamma(b(x+rb))}{b^{2b(x+rb)-1}}\;,\;\;\;{\text{for}}\;n\in{\mathbb{N}}\;,\label{identity-decoupling}
\end{equation}
where  $\gamma(x)=\frac{\Gamma(x)}{\Gamma(1-x)}$,
we find a factorized result for the matrix integral,
\begin{equation}
\begin{aligned}
Z^{(\Lambda)}_{S^4_b}&=e^{-8\pi^2\hx{N_cQ}}\sum_{\{l_a\}}e^{-16i\pi^2\hx\sum_{a=1}^{N_c}\hm_{l_a}}Z_{\text{vac},\{l_a\}}
\sum_{K}e^{-16\pi^2\hx K b}\,Z_{K,\{l_a\}}+...\;,\label{decom}
\end{aligned}
\end{equation}
where 
\begin{equation}
\begin{aligned}
Z_{\text{vac},\{l_a\}}=&\left(2ֿ\pi i\,\text{Res}|_{x=0}(\Upsilon_b(ix)^{-1})\right)^{N_c}\\&e^{-\frac{8\pi^2}{g^2}\frac{1}{N_c}\sum_{a<b}(\hm_{l_b}-\hm_{l_a})^2}e^{-\frac{8\pi^2}{e^2}\frac{1}{N_c}\left(-\sum_{a=1}^{N_c}\hm_{l_a}+iN_c\frac{Q}{2}\right)^2}
\left(\prod_{a=1}^{N_c}\prod_{j\notin\{l_d\}}\Upsilon_b\Big(i(\hat\mu_{j}-\hat\mu_{l_a})\Big)\right)^{-1}\;,
\end{aligned}
\end{equation}
and
\begin{equation}
\begin{aligned}
&Z_{K,\{l_a\}}=\\& e^{\frac{8\pi^2}{N_ce^2}\left(K^2b^2+(b^2+1)KN_c+2iKb\sum_{a=1}^{N_c}\hm_{l_a}\right)}e^{-\frac{8\pi^2}{N_c g^2}\left(K^2b^2+2ibK\sum_{a=1}^{N_c}\hm_{l_a}\right)}b^{N_f K(1+b^2)+2K^2b^2+2ibK(\sum_{a=1}^{N_c}\hat{\mu}_{l_a}-\sum_{j\notin\{l_d\}}\hat{\mu}_j)}\\&\sum_{\{k_a\}\in\Pi(K)}\Bigg(\left({e^{\frac{8\pi^2}{g^2}}b^{N_f-2N_c}}\right)^{b^2\sum_{a=1}^{N_c}k_a^2+2ib\sum_{a=1}^{N_c}k_a\hat\mu_{l_a}}\Bigg[\prod_{a=1}^{N_c}\prod_{r=0}^{k_a-1}\frac{\prod_{b=1}^{N_c}\gamma\Big(ib(\hat\mu_{l_a}-\hat\mu_{l_b})+(r-k_b)b^2\Big)}{\prod_{j\notin\{l_d\}} \gamma\Big(1+(r+1)b^2+ib(\hat\mu_{l_a}-\hat\mu_j)\Big)}\Bigg]\\&\;\;\;\;\;\;\;\;\;\;\;\;\;\;\;\;\;\;\;\times Z^{\text{res}}_{\text{inst}}(\mm_{ij}=\hm_i-\hm_j,\{k_a\},q)Z^{\text{res}}_{\text{inst}}(\mm_{ij}=\hm_i-\hm_j,\{k_a\},\bar q)\Bigg)\;.\label{ZK}
\end{aligned}
\end{equation}
Here, $\Pi(K)$ is the set of integer partitions of $K$, $\sum_{a=1}^{N_c}k_a=K$. In addition, we denoted \begin{equation}
q=e^{2\pi i \tau_{su(N_c)}}\;,\;\;\;\;\;\;\tau_{su(N)}=\frac{\te}{2\pi}+\frac{4\pi i}{g^2}\;.
\end{equation} The expression for the instanton contributions to the residue, $Z^{\text{res}}_{\text{inst}}(\mm_{ij},\{k_a\},q)$,  can be found in appendix \ref{app-inst}.\footnote{The expression in appendix \ref{app-inst} is written for $l_a=a$.}   The dots in \eqref{decom} stand for contributions with $K'\neq 0$ and ${\cO\left(\hat\Lambda e^{-\frac{\hat{\Lambda}^2}{e^2}}\right)+\cO\left(\hat\Lambda e^{-\frac{\hat{\Lambda}^2}{g^2}}\right)}$ corrections.

In (\ref{decom})  we expanded the ellipsoid partition function according to the $\xi$ dependence. 
The interpretation of the different terms in (\ref{decom}) is the following. 
$Z_{\text{vac},\{l_a\}}$ computes the $S^4_b$ partition function of the light hypermultiplets in the Higgs vacuum labeled by $\{l_a\}$.  This partition function is multiplied by a  sum over strings with topological charge $K$.  Each  of these contributions is weighted by  a $e^{-16\pi^2\hx K b}$ factor, which amounts to the contribution to the energy due to the finite tension of the string. 
Finally, the string moduli fluctuations   are encoded in  $Z_{K,\{l_a\}}$, which computes the two-sphere partition function for the  low-energy effective theory living on the string worldsheet. In the following, we will check these claims.  

$e^{-8\pi^2\hx{N_cQ}}e^{-16i\pi^2\hx\sum_{a=1}^{N_c}\hm_{l_a}}Z_{\text{vac},\{l_a\}}$ is the residue of the integrand in the $\ha_a=-\hm_{l_a}+\frac{iQ}{2}$ poles. 
In the Higgs vacuum, the vector multiplets eat $N_c^2$ hypermultiplets and become long massive vector multiplets. Indeed, in $Z_{\text{vac},\{l_a\}}$ the vector multiplet $\Upsilon_b$-functions in the numerator are canceled with $N_c^2$ hypermultiplet $\Upsilon_b$-functions.  The $N_c(N_f-N_c)$ $\Upsilon_b$-functions left in $Z_{\text{vac},\{l_a\}}$ correspond to the $N_f-N_c$ light hypermultiplets left in the Higgs vacuum, with complex masses $\mu_j-\hm_{l_a}+\frac{iQ}{2}$. Like the vacuum expectation values for the adjoint scalars, these masses are shifted with respect to the flat space masses by $\frac{iQ}{2}$. This shift is related to the fact that the $R$-symmetry preserved in the vacuum is shifted with respect to the Cartan of  $SU(2)_R$, as explained above equation \eqref{R-charges-in-vac}. See the next section for more details.

Let us now move on to the interpretation of  $Z_{K,\{l_a\}}$ as the two-sphere partition function of the worldsheet theory. 
The strings described by $Z_{K,\{l_a\}}$ wrap the squashed two-sphere whose embedding in the four-ellipsoid (\ref{ellipsoid}) is given by equation \eqref{squashedS2}.
An $\cN=(2,2)$ supersymmetric theory with a vectorlike $U(1)$ $R$-symmetry can be placed on the round two-sphere ($l=r$) while preserving an $su(2|1)$ superalgebra \cite{Doroud:2012xw,Benini:2012ui}.\footnote{Similarly, an $\cN=(2,2)$ supersymmetric theory with an axial $U(1)$ $R$-symmetry can be placed on the round sphere while preserving a different $su(2|1)$ superalgebra, that contains the axial $R$-symmetry instead of the vector $R$-symmetry.} One can squash the sphere while preserving an $su(1|1)\subset su(2|1)$ subalgebra.  The squashed two-sphere partition function does not depend on the squashing parameter $\frac{l}{r}$ \cite{Gomis:2012wy}, and is equal to the partition function of the round two-sphere with radius $l$, which was computed in \cite{Doroud:2012xw,Benini:2012ui} using localization techniques.  

For an $\cN=(2,2)$ supersymmetric theory with gauge group $U(K)$, $N_c$ chiral multiplets in the fundamental representation, $N_f-N_c$ chiral multiplets in the anti-fundamental representation and one chiral multiplet in the adjoint representation, the Higgs branch representation of the two-sphere partition function is  \cite{Gomis:2014eya} 
\begin{equation}
\begin{aligned}
Z_{S^2,K}^{\text{SQCDA}}=\sum_{\{ k_a\}\in\Pi(K)}\Bigg[&(z\bar z)^{-i\left(\sum_{a=1}^{N_c}k_a M_a+\frac{M_X}{2}(\sum_{a=1}^{N_c}k_a^2-K)\right)}\times\\&\frac{\prod_{b=1}^{N_c}\prod_{a=1}^{N_c}\prod_{r=0}^{ k_a-1}\gamma\Big(i(M_a-M_b)+i(r- k_b)M_X\Big)}{\prod_{j=1}^{  N_f-N_c}\prod_{a=1}^{N_c}\prod_{r=0}^{k_a-1}\gamma\Big(1+iM_a+i\tilde M_j+irM_X\Big)}Z^{\text{vort}}_{\{ k_a\}}(z)Z^{\text{vort}}_{\{ k_a\}}(\bar z)\Bigg]\;,
\end{aligned}\label{ZS2}
\end{equation}
where $z=\exp\left(-2\pi\xi_{2d}+i\te_{2d}\right)$, with $\xi_{2d}$ the FI parameter of the GLSM and $\te_{2d}$ the theta-parameter. The twisted masses and $R$-charges of the chiral multiplets are combined in \eqref{ZS2} into the dimensionless complexified twisted masses, $M=l m +\frac{i}{2}R$, where $m$ denotes the twisted mass of the multiplet, $R$ its $R$-charge and $l$ is the radius of the two-sphere (which, in our case is equal to one of the radii of the ellipsoid).  In the expression above,  $M_a$ are the complexified twisted masses of the fundamental chiral multiplets, $\tilde M_j$ are the complexified twisted masses of the anti-fundamental chiral multiplets and $M_X$ is the  complexified twisted mass of the adjoint chiral multiplet.  The function $Z^{\text{vort}}_{\{k_a\}}$, describing the non-perturbative contributions, can be found in appendix \ref{app-inst}.

The dictionary between the four-dimensional and two-dimensional parameters that can be obtained by comparing  (\ref{ZK}) with (\ref{ZS2}), as we will explain below, is
\begin{equation}\label{dictionary0}
\begin{aligned}
&z=(-1)^{K-1}b^{2N_c-N_f}q\;,\\
&M_a-M_b=b\left(\hat{\mu}_{l_a}-\hat{\mu}_{l_b}\right)\;,\\
&M_a+\tilde M_j=b\left(\hat{\mu}_{l_a}-\hat{\mu}_{n_j}\right)-ib^2\;,\\
&M_X=-ib^2\;,
\end{aligned}
\end{equation}
where $\{n_1,n_2,...,n_{N_f-N_c}\}=\{1,..,N_f\} \setminus\{l_a|a=1,...,N_c\}$.
In terms of the masses and $R$-charges, the dictionary reads
\begin{equation}
\begin{aligned}
&m_a-m_b={\mu}_{l_a}-{\mu}_{l_b}\;,\;\;\;\;R_a-R_b=0\;,\\
&m_a+\tilde m_j={\mu}_{l_a}-{\mu}_{n_j}\;,\;\;\;R_a+\tilde R_{n_j}=-2b^2\;,\\
&m_X=0\;,\;\;\;\;\;\;\;\;\;\;\;\;\;\;\;\;\;\;\;\;\;\;\,R_X=-2b^2\;.
\end{aligned}
\end{equation}

Before showing agreement between (\ref{ZK}) and (\ref{ZS2}) under the dictionary \eqref{dictionary0}, let us make a few comments about the map of the parameters.
The $su(1|1)$ superalgebra on the squashed two-sphere is inherited from the $su(1|1)$ superalgebra on the four-ellipsoid in the following way
\begin{equation}
Q^2=\frac{1}{l}J_{1,2}+\frac{1}{\tilde l}J_{3,4}+\left(\frac{1}{ l}+\frac{1}{{\tilde l}}\right)R_{4d}=\frac{1}{l}\left(J_{1,2}-b^2J_{4,3}+(1+b^2)R_{4d}\right)=\frac{1}{l}\left(J_{1,2}+R_{2d}\right)\;.
\end{equation}
In other words, the $R$-symmetry that appears in the $su(1|1)$ algebra on the two-sphere is  a combination the four-dimensional $R$-symmetry and rotations in the transverse plane with coefficients $b^2+1$ and $-b^2$ respectively. We denote the corresponding worldsheet $R$-symmetries by $R^{(R)}$ and $R^{(J)}$. The dependence on $b^2$ allows us to extract the $R^{(R)}$- and $R^{(J)}$-charges of the chiral multiplets from the dictionaries we find. In section \ref{examples} we compared the $R^{(R)}$- and $R^{(J)}$-charges of the worldsheet fields that we read from the localization analysis with the expected charges based on a classical zero-mode analysis.

The $b^{2N_f-N_c}$ factor in the map (\ref{dictionary0}) can be understood in the following way.  The couplings $g^2$ and $\xi_{2d}$ that appear in the localization formulas are given at the scales $\frac{1}{\sqrt{l\tilde l}}$ and $\frac{1}{l}$ respectively.  
Thus, the correct way to read the map we obtained is
\begin{equation}
\frac{4\pi}{g^2}\left(\frac{1}{\sqrt{l\tilde l}}\right)-\xi_{2d}\left(\frac{1}{l}\right)=\frac{1}{2\pi}(2N_c-N_f)\log b\;.
\end{equation}
When the couplings are measured at the same scale, the $\log b$ correction disappears, since
\begin{equation}
\frac{4\pi}{g^2}\left(\mu\right)-\xi_{2d}\left(\mu\right)=\left[\frac{4\pi}{g^2}\left(\frac{1}{\sqrt{l\tilde l}}\right)+\frac{1}{2\pi}(2N_c-N_f)\log(\mu \sqrt{l\tilde l})\right]-\left[\xi_{2d}\left(\frac{1}{ l}\right)+\frac{1}{2\pi}(2N_c-N_f)\log(\mu l)\right]=0\;.
\end{equation}

Using the dictionary (\ref{dictionary0}), equation (\ref{ZK}) is mapped to equation (\ref{ZS2}), up to a multiplicative factor  
\begin{equation}
\begin{aligned}
\frac{Z_{K,\{l_a\}}}{Z_{S^2,K}^{\text{SQCDA}}}=&e^{-4\pi i\xi_{2d}\frac{K}{N_c}\sum_{a=1}^{N_c} M_a}\,e^{-2\pi K\xi_{2d}b^2\left(-1+\frac{K}{N_c}\right)}\\&b^{N_f K(1-b^2)+2Kb^2N_c+2K^2b^2+2iK\left(\sum_{a=1}^{N_c}M_a+\sum_{j=1}^{N_f-N_c}\tilde M_j\right)+\frac{N_f-2N_c}{N_c}\left(K^2b^2+2iK\sum_{a=1}^{N_c}M_a\right)}\\&e^{\frac{8\pi^2}{N_ce^2}\left(K^2b^2+(b^2+1)KN_c+2iKb\sum_{a=1}^{N_c}\hm_{l_a}\right)}\label{counterterm}\;.
\end{aligned}
\end{equation}
For the comparison of the non-perturbative contributions, see appendix \ref{app-inst}.
We would like to argue that the multiplicative factor (\ref{counterterm}) is not universal, in the sense that it can be removed with an appropriate choice of regularization scheme.

Two two-sphere partition functions that were obtained in two different regularization schemes  may differ by local counterterms. 
A useful way to classify the allowed counterterms, assuming that the two regularization schemes preserve the $su(2|1)$ superalgebra on the sphere,\footnote{Note that any $su(2|1)$ supersymmetric counterterm is also $su(1|1)$ supersymmetric.} is to promote the couplings to fields transforming in representations of supergravity. 
The regularization ambiguities of the sphere free-energy, $\log Z_{S^2}$, are then given by  supergravity Lagrangians composed out of these fields, evaluated at the sphere constant background. 

The complexified FI parameter, $t=i\xi_{2d}+\frac{\te_{2d}}{2\pi}$, sits in the bottom component of a background twisted chiral multiplet with $R$-charge zero. A supergravity Lagrangian that evaluates to $f(t)+\bar f(\bar t)$ in the sphere background was constructed in \cite{Gerchkovitz:2014gta}, thus showing that $Z_{S^2}$ has ambiguity of the form
\begin{equation}
Z_{S^2}\sim f(z)\bar f(\bar z)Z_{S^2}\;,\label{ambiguity1}
\end{equation}
for any holomorphic function $f$.

The masses $m_a$ and $\tilde m_j$ also need to be promoted to supergravity multiplets.  The sum of the masses $\sum_{a=1}^{N_c}{m_a}$  sits in a real  scalar component of a $U(1)$ vector multiplet, which corresponds to weakly gauging the flavor symmetry under which all the fundamental fields have  charge 1 and all the anti-fundamental fields have charge zero.  The same supergravity Lagrangian that produces the FI terms on the sphere can be used to write a local counterterm that, when evaluated in the sphere supersymmetric background, is proportional to $il\xi_{2d}\vev{\sigma}$, where  $\vev{\sigma}$ is any combination of the masses  that corresponds to a conserved $U(1)$ flavor symmetry. Thus, the two-sphere partition function has ambiguity of the form 
\begin{equation}
Z_{S^2}\sim \left(z\bar z\right)^{il\vev\sigma} Z_{S^2} \label{ambiguity2}\;. 
\end{equation}
Taking $\vev{\sigma}\propto\sum_{a=1}^{N_c}{m_a}$,  one can absorb the $\exp\left({-4\pi i\xi_{2d}\frac{K}{N_c}l\sum_{a=1}^{N_c} m_a}\right)$ term in the first line of (\ref{counterterm}).  See also the discussion in \cite{Gomis:2014eya}. 

The dependence of (\ref{counterterm}) on $b^2$ needs to be realized through a supersymmetric counterterm that contains the $R$-charges of the fields. Thus, the $R$-charges also need to be 
embedded in a supergravity representation. However, this analysis appears to be more subtle.  We leave this problem for the future; we will not try to explain the dependence  of equation \eqref{counterterm} on $b^2$ and the imaginary parts of the complexified masses. 

Finally, the last term in  (\ref{counterterm}),
\begin{equation}
e^{\frac{8\pi^2}{N_ce^2}\left(K^2b^2+(b^2+1)KN_c+2iKb\sum_{a=1}^{N_c}\hm_{l_a}\right)}\;,\label{triv}
\end{equation}
depends on two parameters of the 4d theory, $e^{-2}$ and $\sum_{a=1}^{N_c}\mu_{l_a}$, that have no interpretation in the 2d theory. To argue that (\ref{triv}) is a trivial contribution to the two-sphere partition function, we formally map these parameters to two additional parameters -- an FI term and a mass parameter that correspond to a $U(1)$ symmetry   in a decoupled theory -- and notice that this decoupled theory is in fact trivial, as (\ref{triv}) is interpreted as a supersymmetric counterterm using (\ref{ambiguity1}) and (\ref{ambiguity2}).

To conclude, the expression (\ref{ZK}) that was extracted from the four-ellipsoid partition function computes the two-sphere partition function of the GLSM described in section \ref{equalcharge}, whose low-energy limit has been conjectured in \cite{Hanany:2003hp,Hanany:2004ea} to describe the worldsheet theory on the $K$-string.

\section{General $U(1)$ Charges}\label{general U1 charges}
\subsection{The Worldsheet Two-Sphere Partition Function for a General String}
In this section we generalize the analysis of the previous section by allowing different $U(1)$ charges for different hypermultiplets. As before, the $U(1)$ charges will be denoted by $c_i$, $i=1,...,N_f$. 

The four-ellipsoid partition function is now given by the matrix integral
\begin{equation}
Z_{S^4_b}=\int \left(\prod_{a=1}^{N_{c-1}} d(w_a\cdot\ha)\right)\,d\ha'\, e^{-\frac{16\pi^2}{g^2}\ha\cdot \ha-\frac{8\pi^2}{e^2}N_c\,\ha'^2+16i\pi^2N_c\hx \ha'}\,\frac{\prod_{a\neq b}\Upsilon_b\left(iw_a\cdot\ha-iw_b\cdot\ha\right)}{\prod_{a=1 }^{N_c}\prod_{i=1}^{N_f}\Upsilon_b\left(iw_a\cdot\ha+ic_i\ha'+i\hm_i+\frac{Q}{2}\right)}\,|Z_{\text{inst}}|^2\;,
\label{general gauging s4}\end{equation}
where $w_a$ are the weights in the fundamental representation of $SU(N_c) $ (${\sum_{a=1}^{N_c}w_a=0}$). The instanton partition function is given by equation (\ref{inst-cont}) of appendix \ref{app-inst}.

The denominator $\Upsilon_b$-functions vanish at 
\begin{align}
&i w_a\cdot \ha+ic_j\ha'+i\hm_j+\frac{Q}{2}+k_{a,j}b+k'_{a,j}b^{-1}=0\;,\;\;\;a=1,...,N_c\;,\;\;\;j=1,...,N_f\;,\;\;\;k_{a,j},k'_{a,j}\in\mathbb{N}\;,\label{poles-general-1}\\
&i w_a\cdot \ha+ic_j\ha'+i\hm_j-\frac{Q}{2}-k_{a,j}b-k'_{a,j}b^{-1}=0\;,\;\;\;a=1,...,N_c\;,\;\;\;j=1,...,N_f\;,\;\;\;k_{a,j},k'_{a,j}\in\mathbb{N}\;.\label{poles-general-2}
\end{align}

As in the previous section, we introduce a UV cut-off for the matrix integral and perform the integrals step by step. In the first step we integrate over $\ha'$, closing the contour from above and picking up the residues of the encircled poles. For each one of the $\ha'$-poles we then continue with the next integrals. For the poles of $\ha'$ corresponding to color and flavor indices $\ha$ and $j$ in the hypermultiplet one-loop determinant, we perform in the next step the integration over $w_a\cdot \ha$, closing the contour from above if $c_j>0$ or from below if $c_j<0$. (There are no poles for $\ha'$ corresponding to $c_j=0$.) We continue in a similar fashion. After the matrix integral has been computed, we expand the result according to the $\hx$ dependence and attempt to identify the bulk and the string contributions as in the previous section. 

If $N_c>2$ and not all the charges are equal, there are contributions to the four-ellipsoid partition function  for  which we are not able to close all  the integration contours. These are  obtained by picking  poles with the same color index in the first two integrations, one of the form \eqref{poles-general-1} and one of the form \eqref{poles-general-2}. After the first two integrations, the FI parameter will not multiply any of the $N_c-2$ integration variables we are left with, and we will not be able to close the contours.  We will therefore stay with an $N_c-2$ dimensional matrix integral. These contributions correspond to the mesonic vacua and the mesonic strings mentioned in  section \ref{general gauging}, which have $N_c-2$ flat Coulomb branch directions. For $N_c=2$, the mesonic vacua and strings are essentially the same as the baryonic vacua and strings due to the fact that the fundamental representation of $SU(2)$ is pseudo-real. We will discuss mesonic strings
in an $N_c=2$ example in section \ref{locS-dual}.

The other contributions correspond to the baryonic vacua and strings. For these contributions, we are able to close all the contours and obtain the contribution as a residue of $N_c$ poles corresponding to $N_c$ different   color indices.  
The residue of a pole with  flavor index $j$ will not have poles in the remaining integration variables with  the same flavor index.  Therefore, the non-vanishing contributions pick $N_c$ out of the $N_f$ hypermultiplets. One can show that it is enough to consider only  poles of the form \eqref{poles-general-1} or only poles of the form \eqref{poles-general-2}, as mixed combinations of poles will cancel among themselves. We find that we need to consider poles of the form   
\begin{equation}
i w_a\cdot \ha+ic_{l_a}\ha'+i\hm_{l_a}+\frac{Q}{2}+k_{a}b+k'_{a}b^{-1}=0\;,\;\;\;a=1,...,N_c\;,\;\;\;\{l_1,...,l_{N_f}\}\subset\{1,...,N_f\}\;,\;\;\;k_{a},k'_{a}\in\mathbb{N}\;,\label{poles-general-a}
\end{equation}
with $\sum_{a=1}^{N_c}{c_{l_a}}>0$ and poles of the form 
\begin{equation}
i w_a\cdot \ha+ic_{l_a}\ha'+i\hm_{l_a}-\frac{Q}{2}-k_{a}b-k'_{a}b^{-1}=0\;,\;\;\;a=1,...,N_c\;,\;\;\;\{l_1,...,l_{N_f}\}\subset\{1,...,N_f\}\;,\;\;\;k_{a},k'_{a}\in\mathbb{N}\;,\label{poles-general-b}
\end{equation}with $\sum_{a=1}^{N_c}{c_{l_a}}<0$. 

Thus, we obtain the four-ellipsoid partition function as a sum over mesonic and baryonic contributions,
\begin{equation}
\begin{aligned}
Z^{(\Lambda)}_{S^4_b}=& \sum_{\{l_a\}}Z^{B}_{\text{vac},\{l_a\}}e^{-8\pi^2\hx{N_c^2Q}/|C_{\{l_a\}}|}e^{-16i\pi^2N_c\hx{\sum_{a=1}^{N_c}\hm_{l_a}}/C_{\{l_a\}}}
\sum_{K,K'}e^{-16\pi^2N_c\hx( K b+ K' b^{-1})/|C_{\{l_a\}}|}\,Z^{B}_{K,K'\{l_a\}}\\&+\sum_{i,j\;, c_i\neq c_j}e^{-16i\pi^2N_c\hx\frac{\hm_i-\hm_j}{c_i-c_j}}
\sum_{K, K'}e^{-\frac{16\pi^2N_c\hx}{|c_i-c_j|}( K b+ K' b^{-1})}\,Z^{M}_{K,K',i,j}\;,\label{decom-general}
\end{aligned}
\end{equation}
where $C_{\{l_a\}}=\sum_{a=1}^{N_c}c_{l_a}$.
The first line represents the baryonic vacua and strings and the second line represents the contributions of the mesonic vacua and strings. 
$Z^{B}_{\text{vac},\{l_a\}}$,  $Z^{B}_{K,K'\{l_a\}}$ and $Z^{M}_{K,K',i,j}$  are independent of $\hx$. 

We will focus on the baryonic contributions. To simplify the notation, let us take $\{l_a\}=\{1,2,..,N_c\}$, and assume that $C=C_{\{l_a\}=\{1,2,..,N_c\}}=\sum_{a=1}^{N_c}c_a>0$. The vacuum contributions are then given by
\begin{equation}
\begin{aligned}\label{zbvac}
Z^B_{\text{vac},\{1,2,...,N_c\}}=&N_c!\left(2ֿ\pi i\,\text{Res}|_{x=0}(\Upsilon_b(ix)^{-1})\right)^{N_c}\\&e^{\frac{8\pi^2}{g^2}\sum_{a=1}^{N_c}\left(-i\hm_a+i\frac{c_a}{C}\sum_{b=1}^{N_c}\hm_b+\left(\frac{c_a}{C}N_c-1\right)\frac{Q}{2}\right)^2}e^{-\frac{8\pi^2}{e^2}\frac{N_c}{C^2}\left(-\sum_{a=1}^{N_c}\hm_{a}+iN_c\frac{Q}{2}\right)^2}
\\
&\prod_{a=1}^{N_c}\prod_{i=N_c+1}^{N_f} \Upsilon_b\left(i(\hm_i-\hm_a)-i\Delta_{ia}\sum_{b=1}^{N_c}\hm_b-\Delta_{ia}N_c\frac{Q}{2}\right)^{-1}\;.
\end{aligned}
\end{equation}

As before, $Z^B_{\text{vac},\{1,...,N_c\}}$ is interpreted as the four-ellipsoid partition function of the light hypermultiplets in the vacuum  \eqref{generalvac}. Indeed, the arguments of the $\Upsilon_b$-functions imply a complexified mass spectrum on the ellipsoid; $\mu_i-\mu_a+\Delta_{ai}\sum_{b=1}^{N_c}\mu_b+\frac{i}{2}(\frac{1}{l}+\frac{1}{\tilde l})(\Delta_{ia}N_c+1)$.   The real part gives the flat space mass spectrum, which agrees  exactly with equation (\ref{mass-in-vac}). The imaginary shift in the mass is  proportional to the shift in the $R$-charge of the multiplet in the vacuum -- see the discussion above equation \eqref{R-charges-in-vac}.  This is related to the fact that a background $R$-current needs to be turned on in order to preserve  supersymmetry on the ellipsoid \cite{Hama:2012bg}. Equation (\ref{zbvac}) needs to be compared to the partition function of a theory that has been placed on the ellipsoid using the $R$-symmetry preserved by the vacuum \eqref{generalvac}.  

Next, the function $Z^B_{K,K',\{1,2,...,N_c\}}$ is given by
\begin{equation}
\begin{aligned}\label{general}
&Z^B_{K,K',\{1,2,...,N_c\}}=e^{-\frac{8\pi^2}{e^2}\frac{N_c}{C^2}\left[2i\left(-\sum_{a=1}^{N_c}\hm_{a}+iN_c\frac{Q}{2}\right)\left(Kb+K'b^{-1}\right)-\left(Kb+K'b^{-1}\right)^2\right]} \\
&\sum_{\{k_a\}\in\Pi(K)}\sum_{\{k'_a\}\in\Pi(K')}\Bigg\{(-1)^{\sum_{a=1}^{N_c}k_ak'_a}\left|Z_{\text{inst}}\right|^2|_{i w_a\cdot \ha+ic_{l_a}\ha'+i\hm_{l_a}+\frac{Q}{2}+k_{a}b+k'_{a}b^{-1}=0}\\&e^{\frac{8\pi^2}{g^2}\sum_{a=1}^{N_c}\left[2\left(-i\hm_a+i\frac{c_a}{C}\sum_{b=1}^{N_c}\hm_b+\left(\frac{c_a}{C}N_c-1\right)\frac{Q}{2}\right)\left(\frac{c_a}{C}(Kb+K'b^{-1})-k_ab-k'_ab^{-1}\right)+\left(\frac{c_a}{C}(Kb+K'b^{-1})-k_ab-k'_ab^{-1}\right)^2\right]}\\
&\left(\prod_{a=1}^{N_c}\prod_{r=0}^{k_a-1}\frac{\gamma\left(-(r+1)b^2\right)}{b^{-2(1+r)b^2-1}}\right)\left(\prod_{a=1}^{N_c}\prod_{s=0}^{k'_a-1}\frac{\gamma\left(-(s+1)b^{-2}\right)}{b^{2(1+s)b^{-2}+1}}\right)\left(\prod_{a=1}^{N_c}\prod_{r=0}^{k_a-1}\prod_{s=0}^{k'_a-1}\left(-(r+1)b-(s+1)b^{-1}\right)^{-2}\right)\\
&\prod_{a=1}^{N_c}\prod_{i=N_c+1 }^{N_f}\frac{ֿ\Upsilon_b\left(i(\hm_i-\hm_a)-i\Delta_{ia}\sum_{b=1}^{N_c}\hm_b-\Delta_{ia}N_c\frac{Q}{2}\right)}{\Upsilon_b\left(i(\hm_i-\hm_a)-i\Delta_{ia}\sum_{b=1}^{N_c}\hm_b-\Delta_{ia}N_c\frac{Q}{2}-k_ab-k'_ab^{-1}-\Delta_{ia}(Kb+K'b^{-1})\right)}\\
&\prod_{a=1}^{N_c}\prod_{b\neq a}\frac{\Upsilon_b\left(i(\hm_b-\hm_a)+i\Delta_{ab}\sum_{a=1}^{N_c}\hm_a+\Delta_{ab}N_c\frac{Q}{2}+(k_b-k_a)b+(k'_b-k'_a)b^{-1}+\Delta_{ab}(Kb+K'b^{-1})\right)}{\Upsilon_b\left(i(\hm_b-\hm_a)+i\Delta_{ab}\sum_{c=1}^{N_c}\hm_c+\Delta_{ab}N_c\frac{Q}{2}-k_ab-k'_ab^{-1}+\Delta_{ab}(Kb+K'b^{-1})\right)}\Bigg\}\;.
\end{aligned}
\end{equation}

If $N_f=N_c$ we can use the shift identity,
\begin{equation}
\frac{\Upsilon_b(x+nb+kb^{-1})}{\Upsilon_b(x)}=(-1)^{nk}\left(\prod_{r=0}^{n-1}\frac{\gamma(b(x+rb))}{b^{2b(x+rb)-1}}\right)\left(\prod_{s=0}^{k-1}\frac{\gamma(b^{-1}(x+sb^{-1}))}{(b^{-1})^{2b^{-1}(x+sb^{-1})-1}}\right)\left(\prod_{r=0}^{n-1}\prod_{s=0}^{k-1}(x+rb+sb^{-1})^2\right)\;,
\label{identity2}
\end{equation}
applicable for $n,k\in{\mathbb{N}}$, to rewrite $Z^B_{K,K'}$ in terms of $\Gamma$-functions, the ``building-blocks'' of the two-sphere partition function. When $N_f>N_c$, however, (\ref{identity2}) will be applicable only if  
\begin{equation}\Delta_{ij}K,\;\Delta_{ij}K'\in\mathbb{Z}\;,\;\;\; \text{for all }i,j\;.\label{decoupling-criterion}\end{equation} 

Indeed, when $N_f>N_c$ the theory is not gapped -- there are light fields in the bulk that in general interact with the string moduli. The contribution of a $K$-string satisfying the condition (\ref{decoupling-criterion}) to the four-ellipsoid partition function factorizes to contributions that look like a four-ellipsoid partition function multiplying two-sphere partition function. This suggests that when the condition \eqref{decoupling-criterion} is satisfied, the bulk and string moduli decouple at low energies. Remarkably, this is exactly the condition that we found in a classical analysis of the interaction terms in \cite{Gerchkovitz:2017kyi}. The contribution corresponding to positive $K$ and $K'$ that satisfy the condition (\ref{decoupling-criterion}) is expected  to describe the low-energy dynamics around a $K$-string and a $K'$-string wrapping the two-spheres ${x_0^2}/{r^2}+({x_1^2+x_2^2})/{l^2}=1$ and ${x_0^2}/{r^2}+({x_3^2+x_4^2})/{\tilde l^2}=1$ respectively, with interactions between the two worldsheet theories but not between the worldsheet modes and the bulk modes.

From now on, we will focus on the cases where the decoupling condition (\ref{decoupling-criterion}) is satisfied (if $N_f>N_c$). 
For simplicity, we will write the expression we find only for $K'=0$, and will split it as
$Z^B_{K,K'=0,\{1,2,...,N_c\}}=Z_{\text{overall}}\cdot Z_{S^2,K}$,
where
\begin{equation}
\begin{aligned} Z_{\text{overall}}=&b^{\frac{\sum_{i}c_i}{C}\left(2K^2b^2+2N_cK(b^2+1)+2ibK\sum_a\hm_a\right)-2ibK\sum_i\hm_i-2b\sum_{i,a}\Delta_{ia}K\left(i\mm_{ia}+\frac{1}{2}\Delta_{ia}Kb\right)}\\&e^{-\frac{8\pi^2}{e^2}\frac{N_c}{C^2}\left[2iKb\left(-\sum_{a=1}^{N_c}\hm_{a}+iN_c\frac{Q}{2}\right)-K^2b^2\right]}\\& e^{\frac{8\pi^2}{g^2}\left((b^2+1)K+\sum_{a=1}^{N_c}\left(-2i\frac{\hm_a c_a}{C}Kb+2i\frac{c_a^2}{C^2}Kb\sum_{b=1}^{N_c}\hm_b+\frac{c_a}{C}\left(\frac{c_a}{C}N_c-1\right)(b^2+1)K+\frac{c_a^2}{C^2}K^2b^2\right)\right)}ֿֿ\;,\label{zoverall}
\end{aligned}
\end{equation}
and 
\begin{equation}
\begin{aligned}
&Z_{S^2,K}=\sum_{\{k_a\}\in\Pi(K)}\Bigg[\left(e^{\frac{8\pi^2}{g^2}}b^{N_f-2N_c}\right)^{\sum_{a=1}^{N_c}2ib\mm_ak_a+\sum_{a=1}^{N_c}k_a^2b^2}\prod_{a,b=1}^{N_c}\prod_{r=0}^{k_b-1}\gamma\left(ib\mm_{ba}-k_ab^2+rb^2\right)\\
&\prod_{a=1}^{N_c}\frac{\prod_{i\geq N_c+1, \Delta_{ia}K+k_a\geq 1 }\prod_{r=0}^{\Delta_{ia}K+k_a-1}\gamma\left(ib\mm_{ia}+b^2K\Delta_{ia}-\left(r+1 \right)b^2\right)}{\prod_{i\geq N_c+1, \Delta_{ia}K+k_a\leq -1 }\prod_{r=0}^{-\Delta_{ia}K-k_a-1}\gamma\left(ib\mm_{ia}+b^2K\Delta_{ia}+rb^2\right)}\,Z^{\text{res}}_{\text{inst}}(\mm_{ij},\{k_a\},q)Z^{\text{res}}_{\text{inst}}(\mm_{ij},\{k_a\},\bar q)\Bigg]\;.\label{ZS2Kgeneral}
\end{aligned}
\end{equation}
In the expression  above we introduced the notation \begin{equation}
\begin{aligned}
&\mm_{i}(\{c_k\}, K)=\hm_i-\frac{c_i}{C}\left(\sum_{b=1}^{N_c}\hm_b-iN_c\frac{Q}{2}-iKb\right)\;,\\&\mm_{ij}=\mm_i-\mm_j\;.\label{512}
\end{aligned}
\end{equation}
The expression for the instanton contributions to the residue, $Z^{\text{res}}_{\text{inst}}(\mm_{ij},\{k_a\},q)$,  can be found in appendix \ref{app-inst}.
In \eqref{zoverall}, sums over $a$ and $b$ indices  are assumed to be from $1$ to $N_c$ and sums over an $i$ index are assumed to be from $N_c+1$ to $N_f$, unless stated otherwise.

We thus conjecture that the low-energy fluctuations around the string solution are captured by a 2d $\cN=(2,2)$ supersymmetric theory, whose sphere partition function is given, under some map of the parameters of the 4d theory to the parameters of the 2d theory, by equation (\ref{ZS2Kgeneral}). 
This conjecture is based on the assumption that under the 4d-2d map of the parameters, equation  (\ref{zoverall}) can be regarded as a regularization ambiguity of the two-sphere partition function  (see the discussion around equation \eqref{counterterm}.) 

\subsection{Identifications of Worldsheet Theories}\label{Loc-examples}
In this section, we will identify GLSMs whose two-sphere partition functions matches with equation \eqref{ZS2Kgeneral}, for all the cases where the condition \eqref{weak-cond} is satisfied. From the classical zero-modes point of view, this is the condition that the size-modes in equation \eqref{sizemodesgeneralzeros} satisfy the F-term constraints trivially. This condition can also be derived from  our localization analysis as the  condition that the number of $\gamma$-functions that appear in \eqref{ZS2Kgeneral} is the same for every choice of the partition $\{k_a\}$. The condition \eqref{weak-cond} allows us to identify worldsheet GLSMs that are weakly coupled when the 4d theory is weakly coupled. 

When the condition in equation \eqref{weak-cond} is not satisfied, the form of equation \eqref{ZS2Kgeneral} changes dramatically. In some of these cases (see for example section \ref{locS-dual}) the expression \eqref{ZS2Kgeneral} can be understood as a two-sphere partition function of a GLSM expanded around strong coupling.

\subsubsection{Strings with no $\tilde{q}$ Excitations}\label{Locnotilde}
Let us consider equation (\ref{ZS2Kgeneral}) in the case where $c_i\geq c_a$ for all $1\leq a\leq N_c$ and  $N_c+1\leq i\leq N_f$. This condition arises from the zero-mode analysis as the condition that there are no size zero-modes involving excitations of $\tilde q$. See equation \eqref{sizemodesgeneralzeros}. It is not difficult  to check that in this case 
\begin{equation}
Z_{S^2,K}\left(\mm_i[c_j],c_i\right)=Z_{S^2,K}\left(\mm_i[c_j],c_i=1\right)\prod_{a=1}^{N_c}{\prod_{i=N_{c}+1}^{N_f}\prod_{r=0}^{\Delta_{ia}K-1}\gamma\left(ib\mm_{ia}+b^2K\Delta_{ia}-\left(r+1 \right)b^2\right)}\;.
\label{ZS2example}
\end{equation}
\   Thus, we find that  $Z_{S^2}$ is a product of two decoupled factors. The first factor in (\ref{ZS2example}) is the two-sphere partition function of an  $\cN=(2,2)$ supersymmetric theory with gauge group $U(K)$, $N_c$ chiral multiplets in the fundamental representation, $N_f-N_c$ chiral multiplets in the anti-fundamental representation and  one chiral multiplet in the adjoint representation. The parameters of this theory are given in terms of the parameters of the four-dimensional theory by 
\begin{equation}
\begin{aligned}
&z=(-1)^{K-1}b^{2N_c-N_f}q\\
&m_a-m_b={\mu}_a-{\mu}_b+\Delta_{ba}\sum_{c=1}^{N_c}\mu_c\;,\;\;\;\;\;\;R_a-R_b=\Delta_{ab}N_c(b^2+1)+2\Delta_{ab}Kb^2\;,\\
&m_a+\tilde m_{j}={\mu}_a-{\mu}_j+\Delta_{ja}\sum_{c=1}^{N_c}\mu_c\;,\;\;\;\;\;\;R_a+\tilde R_{j}=\Delta_{aj}N_c(b^2+1)+2\Delta_{aj}Kb^2-2b^2\;,\\
&m_X=0\;,\;\;\;\;\;\;\;\;\;\;\;\;\;\;\;\;\;\;\;\;\;\;\;\;\;\;\;\;\;\;\;\;\;\;\;\;\;\;\;\;\;\;\;R_X=-2b^2\;,\label{514}
\end{aligned}
\end{equation}
where $m_a$, $R_a$, $a=1,...,N_c$, are the twisted masses and $R$-charges of the fundamental chiral multiplets,  $\tilde m_j$, $\tilde R_j$, $j=N_{c}+1,...,N_f$, are the twisted masses and $R$-charges of the anti-fundamental chiral multiplets, and $m_X$ and $R_X$ are the twisted mass and $R$-charge of the adjoint chiral multiplet.

The second factor in (\ref{ZS2example}) describes $K(\frac{N_c}{C}\sum_{i=N_c+1}^{N_f}c_i+N_c-N_f)$ free chiral multiplets, with twisted masses and $R$-charges
\begin{equation}
\begin{aligned}
&m_{i,a,r}=\mu_a-\mu_i+\Delta_{ia}\sum_{b=1}^{N_c}\mu_b\;,\\
&R_{i,a,r}=\Delta_{ai}N_c(b^2+1)-2(r+1)b^2\;,
\end{aligned} \label{515}
\end{equation} 
for $a=1,..,N_c$, $i=N_c+1,..,N_f$, $r=0,...,\Delta_{ia}K-1$. 

For a comparison between the low-energy spectrum of this theory and the classical spectrum of string zero-modes, see section \ref{sec-comp-noqtilde}.

\subsubsection{Strings with $q$ and $\tilde{q}$ Excitations}\label{localization-q-tilde-q}
We now consider a more general case, in which
\eq{&\Delta_{ia}\geq 0\ \text{ for all }\ 1\leq a\leq N_c\ ,\ N_c+1\leq i\leq N_q\ ,\\
	-&\Delta_{ia}\geq 1\ \text{ for all }\ 1\leq a\leq N_c\ ,\ N_q+1\leq i\leq N_f\ .} In terms of the classical size-modes given in  equation \eqref{sizemodesgeneralzeros}, this is equivalent to requiring that  there are no $\tilde q^i$ excitations for $N_c+1\leq i\leq N_q$ and no $q_i$ excitations for $N_
q+1\leq i\leq N_f$.  
In this case, equation \eqref{ZS2Kgeneral} reads 
\begin{equation}
\begin{aligned}
&Z_{S^2,K}=\sum_{\{k_a\}\in\Pi(K)}\Bigg[\left(e^{\frac{8\pi^2}{g^2}}b^{N_f-2N_c}\right)^{\sum_{a=1}^{N_c}2ib\mm_ak_a+\sum_{a=1}^{N_c}k_a^2b^2}\prod_{a,b=1}^{N_c}\prod_{r=0}^{k_b-1}\gamma\left(ib\mm_{ba}-k_ab^2+rb^2\right)\\
&\prod_{a=1}^{N_c}\frac{\prod_{i= N_{c}+1}^{N_q} \prod_{r=0}^{\Delta_{ia}K+k_a-1}\gamma\left(ib\mm_{ia}+b^2K\Delta_{ia}-\left(r+1 \right)b^2\right)}{\prod_{i= N_q+1}^{N_f} \prod_{r=0}^{-\Delta_{ia}K-k_a-1}\gamma\left(ib\mm_{ia}+b^2K\Delta_{ia}+rb^2\right)}\,Z^{\text{res}}_{\text{inst}}(\mm_{ij},\{k_a\},q)Z^{\text{res}}_{\text{inst}}(\mm_{ij},\{k_a\},\bar q)\Bigg]\;.\label{zs2weak}
\end{aligned}
\end{equation}

Our ansatz is that the worldsheet theory, in this case, is the low-energy limit of an $\cN=(2,2)$ GLSM with a $U(K)$ gauge group and the following matter content: 
\begin{itemize}
	\item 1 adjoint chiral multiplet, $X$, with complexified twisted mass $M_X$,
	\item $N_c$ fundamental chiral multiplets, $\psi_a$, $a=1,...,N_c$,  with complexified twisted masses $M_a$, 
	\item $N_{f}-N_c$ anti-fundamental chiral multiplets,  $\tilde{\psi}_j$, $j=N_c+1,..,N_f$, with complexified twisted masses $\tilde{M}_j$,
	\item  neutral chiral multiplets,  $\chi_{j,a,r}$, for $a=1,...,N_c$, $j=N_q+1,..,N_f$, $r=0,...,K-1$, with complexified twisted masses  $i-M_a-\tilde{M}_j-rM_X$,
	\item  neutral decoupled chiral multiplets, $\eta_{i,a,r}$, for $a=1,..,N_c$, $i=N_c+1,..., N_q$, $r=0,...,\Delta_{ia}K-1$, with complexified twisted masses $ M_{i,a,r}$,  .  \item
	neutral decoupled chiral multiplets $\tilde \eta_{i,a,r}$, for $a=1,...,N_c$, $i=N_q+1,...,N_f$, $r=0,...,-K-\Delta_{ia}K-1$, with complexified twisted masses $\tilde M_{i,a,r}$. 
\end{itemize}
The  relation between the complexified twisted masses allows a superpotential of the form 
\eq{W=\sum_{r=0}^{K-1}\sum_{a=1}^{N_c}\sum_{j=N_q+1}^{N_f}\al_r\chi_{j,a,r}\tilde{\psi}_{j}X^{r}\psi_{a}\ .}

The two-sphere partition function does not depend on the superpotential couplings $\al_r$, and therefore can be written as $Z_{\text{decoupled}}\cdot Z_{\chi} \cdot Z_{\text{charged}}$, where 
\eql{zdecoupled}{Z_{\text{decoupled}}=\prod_{i=N_c+1}^{N_q}\prod_{a=1}^{N_c}\prod_{r=0}^{\Delta_{ia}K-1}\gamma\left(-i M_{i,a,r}\right)\prod_{i=N_q+1}^{N_f}\prod_{a=1}^{N_c}\prod_{r=0}^{-K-\Delta_{ia}K-1}\gamma\left(-i\tilde M_{i,a,r}\right)} describes the decoupled fields $\eta$ and  $\tilde{\eta}$,
\eq{Z_{\chi}=\prod_{r=0}^{K-1}\prod_{a=1}^{N_c}\prod_{j=N_q+1}^{N_f}\gamma\left(1+iM_a+i\tilde{M}_j+irM_X\right)} describes the neutral $\chi$-fields and $Z_{\text{charged}}$ is given by equation (\ref{ZS2}),  which we repeat here for the convenience of the reader,
\begin{equation}
\begin{aligned}
Z_{\text{charged}}=&\sum_{\{ k_a\}\in\Pi(K)}\Bigg[{(z\bar z)}^{-i\left(\sum_{a=1}^{N_c}k_a M_a+\frac{M_X}{2}(\sum_{a=1}^{N_c}k_a^2-K)\right)}\frac{\prod_{a,b=1}^{N_c}\prod_{r=0}^{ k_a-1}\gamma\Big(i(M_a-M_b)+i(r- k_b)M_X\Big)}{\prod_{j=N_c+1}^{ N_f}\prod_{a=1}^{N_c}\prod_{r=0}^{k_a-1}\gamma\Big(1+iM_a+i\tilde M_j+irM_X\Big)}\\&\;\;\;\;\;\;\;\;\;\;\;\;\;\;\;\;\;\; Z^{\text{vort}}_{\{ k_a\}}(z)Z^{\text{vort}}_{\{ k_a\}}(\bar z)\Bigg]\;,
\end{aligned}\label{zcharged}
\end{equation}
where $Z^{\text{vort}}_{\{ k_a\}}(z)$ is given by equation \eqref{zvort}.

Due to the relation between the masses, some of the terms cancel when we multiply $Z_\chi$ with $Z_{\text{charged}}$ and we are left with
\begin{equation}
\begin{aligned}
&Z_{\chi}\cdot Z_{\text{charged}}=\\
&\sum_{\{ k_a\}\in\Pi(K)}{(z\bar z)}^{-i\left(\sum_{a=1}^{N_c}k_a M_a+\frac{M_X}{2}(\sum_{a=1}^{N_c}k_a^2-K)\right)}\frac{\prod_{a,b=1}^{N_c}\prod_{r=0}^{ k_a-1}\gamma\Big(i(M_a-M_b)+i(r- k_b)M_X\Big)}{\prod_{j=N_c+1}^{ N_q}\prod_{a=1}^{N_c}\prod_{r=0}^{k_a-1}\gamma\Big(1+iM_a+i\tilde M_j+irM_X\Big)}\\&\prod_{a=1}^{N_c}\prod_{j=N_q+1}^{N_f}\prod_{r=1}^{K-k_a}\gamma\left(1+iM_a+i\tilde{M}_j+i(K-r)M_X\right)Z^{\text{vort}}_{\{ k_a\}}(z)Z^{\text{vort}}_{\{ k_a\}}(\bar z)\;.\label{zchizcharged}
\end{aligned}
\end{equation}

To compare equation (\ref{zs2weak}) with the product of equations \eqref{zchizcharged} and \eqref{zdecoupled} it is convenient to rewrite equation (\ref{zs2weak}) as 
\begin{equation}
\begin{aligned}
&Z_{S^2,K}=
\sum_{\{k_a\}\in\Pi(K)}\Bigg[\left(e^{\frac{8\pi^2}{g^2}}b^{N_f-2N_c}\right)^{\sum_{a=1}^{N_c}2ib\mm_ak_a+\sum_{a=1}^{N_c}k_a^2b^2}\prod_{a,b=1}^{N_c}\prod_{r=0}^{k_b-1}\gamma\left(ib\mm_{ba}-k_ab^2+rb^2\right)\\&\prod_{a=1}^{N_c}\prod_{i= N_c+1}^{N_q} \prod_{r=0}^{k_a-1}\gamma\left(ib\mm_{ia}-\left(r+1 \right)b^2\right)\prod_{a=1}^{N_c}\prod_{i= N_q+1}^{N_f} \prod_{r=1}^{K-k_a}\gamma\left(1+ib\mm_{ai}+(K-r+1)b^2\right)\\&Z^{\text{res}}_{\text{inst}}(\mm_{ij},\{k_a\},q)Z^{\text{res}}_{\text{inst}}(\mm_{ij},\{k_a\},\bar q)\Bigg]\label{ZS2Kweak}\times\\
&\prod_{a=1}^{N_c}\prod_{i= N_c+1}^{N_q} \prod_{r=0}^{\Delta_{ia}K-1}\gamma\left(ib\mm_{ia}+b^2K\Delta_{ia}-\left(r+1 \right)b^2\right)\prod_{a=1}^{N_c}\prod_{i= N_q+1}^{N_f} \prod_{r=0}^{-K-\Delta_{ia}K-1}\gamma\left(1+ib\mm_{ai}-(r+K\Delta_{ia})b^2\right)\;.
\end{aligned}
\end{equation}
The first three lines of equation \eqref{ZS2Kweak} are mapped to equation \eqref{zchizcharged}\footnote{Up to an overall  factor, which we interpret as a supersymmetric counterterm.} and the last line is mapped to equation \eqref{zdecoupled}, under the dictionary 
\begin{equation}
\begin{aligned}
&z=(-1)^{K-1}b^{2N_c-N_f}q\\
&M_a-M_b=b\mm_{ab}\;,\\
&M_a+\tilde M_i=b\mm_{ai}-ib^2\;,\\
&M_X=-ib^2\;,\\
& M_{i,a,r}= b\mm_{ai}+i(K\Delta_{ia}-r-1)b^2 	\;,\\
&\tilde M_{i,a,r}= i+b\mm_{ia}+i(K\Delta_{ai}-r)b^2   \;.
\end{aligned}
\end{equation}
The $2d$ masses and $R$-charges are therefore given by
\begin{equation}
\begin{aligned}
&m_a-m_b={\mu}_a-{\mu}_b+\Delta_{ba}\sum_{c=1}^{N_c}\mu_c\;,\;\;\;\;\;\;R_a-R_b=\Delta_{ab}N_c(b^2+1)+2\Delta_{ab}Kb^2\;,\\
&m_a+\tilde m_j={\mu}_a-{\mu}_j+\Delta_{ja}\sum_{c=1}^{N_c}\mu_c\;,\;\;\;\;\;\;R_a+\tilde R_j=\Delta_{aj}N_c(b^2+1)+2\Delta_{aj}Kb^2-2b^2\;,\\
&m_X=0\;,\;\;\;\;\;\;\;\;\;\;\;\;\;\;\;\;\;\;\;\;\;\;\;\;\;\;\;\;\;\;\;\;\;\;\;\;\;\;\;\;\;\;\;R_X=-2b^2\;,\\
& m_{i,a,r}= \mu_a-\mu_i+\Delta_{ia}\sum_{b=1}^{N_c}\mu_b\;,\;\;\;\;\;\;\;\;\;\;\;R_{i,a,r}=N_c\Delta_{ai}(b^2+1)-2(r+1)b^2 	\;,\\
&\tilde m_{i,a,r}= \mu_i-\mu_a+\Delta_{ai}\sum_{b=1}^{N_c}\mu_b\;,\;\;\;\;\;\;\;\;\;\; \tilde{R}_{i,a,r}=(N_c\Delta_{ia}+2)(b^2+1)-2(r+1)b^2	\;.
\end{aligned}
\end{equation}

For a comparison between the low-energy spectrum of this theory and the classical spectrum of string zero-modes, see section \ref{sec-comp-qtildeq}.

\section{Strings in $N_c=2$, $N_f=4$ SQCD under Triality}
\label{sec:Triality}
In this section we will discuss the string configurations in the $N_c=2$, $N_f=4$ case,  with $U(1)$ charge assignments that are related by triality to the well-studied equal charge case. The triality provides predictions for the worldsheet theories, which we test against our localization results. The triality also provides  examples in which the weak coupling regime of the 4d theory is mapped to the strong coupling regime of the worldsheet theory. We will discuss such an example in detail. 

\subsection{$SO(8)$ Symmetry and Triality }
Let us start by reviewing the  triality of  $\cN=2$   $SU(2)$ gauge theory with four fundamental hypermultiplets.   The matter content of this theory can be organized in eight $\cN=1$ chiral multiplets, $Q_{i}^a\ ,\ \tilde{Q}_{a}^i$, where $i=1,..,4$ is the flavor index and $a=1,2$ is the color index. $Q$ and $\tilde{Q}$ transform, respectively, in the fundamental and anti-fundamental representations of $SU(2)$. If the hypermultiplets are massless, the theory has a classical $SO(8)$ flavor symmetry under which
\eql{so8}{Q_v^a\equiv \left(Q_{1}^a+\epsilon^{ab}\tilde{Q}^1_b\;,\;iQ_{1}^a-i\epsilon^{ab}\tilde{Q}^1_b\;,.....\;,Q_{4}^a+\epsilon^{ab}\tilde{Q}^4_b\;,\;iQ_{4}^a-i\epsilon^{ab}\tilde{Q}^4_b\right)^T\ ,}
transforms in the vector representation, which will be denoted by $8_{\text{v}}$. In the quantum level, the $SO(8)$ symmetry is enhanced to $\text{Spin}(8)$. 

Hypermultiplet masses can be introduced by weakly gauging four $U(1)$ flavor symmetries and introducing vacuum expectation values for the four gauge multiplets scalars $\vev{a_i}=\mu_i$. In the $8_\text{v}$ basis, the mass matrix can be written as\footnote{There is another basis that is very common in the literature, in which four mass parameters, $m_a,m_b,m_c,m_d$ are obtained by weakly gauging the $SU(2)_a\times SU(2)_b\times SU(2)_c\times SU(2)_d$ subgroup of $SO(8)$. The map between the two conventions is \begin{equation}
	\begin{aligned}&\mu_1=m_a+m_d,\;\;\;\;\;\;&&\mu_3=m_c+m_b\;,\\&\mu_2=-m_a+m_d,\;&&\mu_4=-m_c+m_b.\label{2nd-mass-basis}\end{aligned}\end{equation}  In this work we will always use mass basis (\ref{vev-mass}).   }
\begin{equation}
\mat{\varepsilon\mu_1& & &\\&\varepsilon\mu_2&&\\&&\varepsilon\mu_3&\\&&&\varepsilon\mu_4}\;,\label{vev-mass}\;\;\;\;\;\;\;\varepsilon=\mat{&\;\;1\\-1&}\;.
\end{equation}
If the masses are all distinct, this breaks the $SO(8)$ flavor symmetry to its maximal torus $SO(2)\times SO(2)\times SO(2)\times SO(2)$. 

The group $SO(8)$ has an $\text{\bf S}_3$ outer automorphism group, which exchanges the three 8-dimensional irreducible representations $8_\text{v}$, $8_\text{s}$ and  $8_\text{c}$, where $8_\text{s}$ and $8_\text{c}$ are the two spinorial representations. The six elements in $\text{\bf S}_3$ can be generated using two transformations,  $S$ and $T$, via $\{1,S,T,ST,TS,STS\}$. The action of $T$ on the Cartan generators is given by
\begin{equation}
\begin{aligned}
&M_{12}\to M_{12}\\
&M_{34}\to M_{34}\\
&M_{56}\to M_{56}\\
&M_{78}\to -M_{78}\;.\label{T-trans-Cartan}
\end{aligned}
\end{equation}
and the action of $S$ is given by
\begin{equation}
\begin{aligned}
&M_{12}\to\frac{1}{2} \left(M_{12}+M_{34}+M_{56}+M_{78}\right)\\
&M_{34}\to\frac{1}{2} \left(M_{12}+M_{34}-M_{56}-M_{78}\right)\\
&M_{56}\to\frac{1}{2} \left(M_{12}-M_{34}+M_{56}-M_{78}\right)\\
&M_{78}\to\frac{1}{2} \left(M_{12}-M_{34}-M_{56}+M_{78}\right)\;,\label{S-trans-Cartan}
\end{aligned}
\end{equation}
where $M_{IJ}$ correspond to the $SO(8)$ generators $(M_{IJ})_{KL}=-i\left(\delta_{IK}\delta_{JL}-\delta_{IL}\delta_{JK}\right)$.

The $\text{\bf S}_3$ outer-automorphism of the $SO(8)$ flavor symmetry of the massless theory turns out to be the duality group of the massive theory, where the masses are assigned with the transformation properties that follow from (\ref{T-trans-Cartan}-\ref{S-trans-Cartan}) and (\ref{vev-mass}). The $SU(2)$ coupling $q\equiv e^{2\pi i\tau}$ transforms as $q\to \frac{q}{q-1}$ under $T$ and $q\to 1-q$ under $S$. Thus, under $T$
\begin{equation}
\begin{aligned}
&q\to \frac{q}{q-1}\;,\\
&\mu_1\to\mu_1\;,\\
&\mu_2\to\mu_2\;,\\
&\mu_3\to\mu_3\;,\\
&\mu_4\to-\mu_4\;,\label{T-mass-trans}
\end{aligned}
\end{equation}
and under $S$,
\begin{equation}
\begin{aligned}
&q\to 1-q\;,\\
&\mu_1\to\frac{1}{2}(\mu_1+\mu_2+\mu_3+\mu_4)\;,\\
&\mu_2\to\frac{1}{2}(\mu_1+\mu_2-\mu_3-\mu_4)\;,\\
&\mu_3\to\frac{1}{2}(\mu_1-\mu_2+\mu_3-\mu_4)\;,\\
&\mu_4\to\frac{1}{2}(\mu_1-\mu_2-\mu_3+\mu_4)\;.\label{S-mass-trans}
\end{aligned}
\end{equation}

The triality of $SU(2)$ SQCD was first discovered by Seiberg and Witten \cite{Seiberg:1994aj}, who observed that the Seiberg-Witten curve is invariant under the transformations (\ref{T-mass-trans}) and (\ref{S-mass-trans}). 
The geometric origin of this triality was understood by Gaiotto. In \cite{Gaiotto:2009we}, Gaiotto constructed a family of $SU(2)$ superconformal $\cN=2$ theories by compactifying the $\cN=(2,0)$ six-dimensional theory of type $A_{1}$  on punctured Riemann surfaces. The four-dimensional theories are then determined  by the choice of the Riemann surface and the number of punctures. The duality group of the four-dimensional theory is understood in Gaiotto's construction as the symmetry group on the set of trinion decompositions of the punctured Riemann surface -- the set of different ways to obtain this surface by sewing together three-punctured spheres.    In the case of $SU(2)$ with four fundamental hypermultiplets, the Riemann surface is the four-punctured sphere, which can be sewn from two three-punctured spheres in three ways. The group of permutations of these three sewings is the $\textbf{S}_3$ triality group. 

The triality can also be interpreted, using the AGT conjecture \cite{Alday:2009aq},  as the crossing symmetry of four-point functions in Liouville theory on the four punctured sphere. 
In \cite{Alday:2009aq} Alday, Gaiotto and Tachikawa argued that the partition function of $\cN=2$ $N_f=4$ $SU(2)$ theory in the $\Omega$-background is equal  to  conformal blocks in Liouville theory. Due to the relation between the $\Omega$-background partition function and the four-ellipsoid partition function, this implies that the four-ellipsoid partition function is equal  to the four-point function in Liouville theory. 

To discuss string configurations, we need to gauge a $U(1)$ flavor symmetry. In the equal charge case, the $U(1)$ generator is embedded in $SO(8)$ as
\eql{Tb}{T_B=\mat{\;\varepsilon\;&&&\\&\;\varepsilon\;&&\\&&\;\varepsilon\;&\\&&&\;\varepsilon\;}\ .}
Using (\ref{T-trans-Cartan}-\ref{S-trans-Cartan}) we see that under $T$, the $SO(8)$ generator  (\ref{Tb}) transforms into
\eql{TTb}{T_T=\mat{\;\varepsilon\;&&&\\&\;\varepsilon\;&&\\&&\;\varepsilon\;&\\&&&-\varepsilon\;}\  ,} and that under $S$, (\ref{Tb}) transforms into 
\eql{STb}{T_S=\mat{\;2\varepsilon\;&&&\\&\;0\;&&\\&&\;0\;&\\&&&\;0\;}\ .}

As the choice of $U(1)$ breaks $\textbf{S}_3$, the strings are configurations in theories that are not self-dual under triality. However, in the limit where the $U(1)$ gauge coupling $e$ is infinitesimally small, we expect the strings to be related by triality to strings in different theories in which the dual $U(1)$'s are gauged. In particular, if the $U(1)_T$ or $U(1)_S$ flavor symmetry is gauged, the string worldsheet theory should be the one that was obtained in the equal charge gauging (for $N_c=2, N_f=4$) and the map between the 4d and 2d parameters should be the composition of the map (\ref{dictionary0}) with the transformations (\ref{T-mass-trans}) or (\ref{S-mass-trans}), respectively.  
Below, we check this claim by studying the two-sphere partition functions that result from our analysis for the minimal ($K=1$) string configurations  in the three cases where (\ref{Tb}), (\ref{TTb}) or (\ref{STb}) are gauged. 

To avoid confusion, let us end this section with a comment about the definition of the complexified gauge coupling. It is common in the literature to distinguish between the coupling that is defined using the classical prepotential, $\tau_{UV}$, and the one defined using the quantum corrected prepotential, $\tau_{IR}$.  The early works on Seiberg-Witten theories use $\tau_{IR}$, while in  recent works that use localization techniques, $\tau_{UV}$ appears naturally.  Another source of potential confusion is that the definition of the microscopic prepotential is renormalization scheme dependent. As a result,  the relation between $\tau_{UV}$ and $\tau_{IR}$ is scheme dependent. 
When the moduli space is parametrized in terms of $\tau_{IR}$, the duality group is $SL(2,\mathbb{Z})$, acting on the coupling $\tau_{\text{IR}}$ as\footnote{The unconventional $SL(2,\mathbb{Z})$ transformation is due to the normalization of $\tau_{IR}$, which is usually normalized as $\tau'_{IR}=2\tau_{IR}=\frac{\te}{\pi}+\frac{8\pi i}{g^2}$.} 
\eql{sl2}{2\tau_{\text{IR}}\rightarrow \frac{2a\tau_{\text{IR}}+b}{2c\tau_{\text{IR}}+d}\ ,\ \ a,b,c,d\in\mathbb{Z}\,,\;\;\;\;  ad-bc=1\;.} 
In the regularization scheme used   by Nekrasov \cite{Nekrasov:2002qd}, the prepotential (in the massless limit) determines the following relation between $\tau_{UV}$ and $\tau_{IR}$
\eq{q\equiv e^{2\pi i\tau_{\text{UV}}}=\lambda\left(2\tau_{\text{IR}}\right)\ ,}
where $\lambda$ is the modular lambda-function, which is defined on the upper half plane.
$\lambda(2\tau_{IR})$ is invariant under the subgroup $\Gamma(2)\subset SL(2,\mathbb{Z})$, which is generated by $\tau_{IR}\to\tau_{IR}+1$ and $\tau_{IR}\to\frac{\tau_{IR}}{1-4\tau_{IR}}$. Under   $2\tau_{IR}\to 2\tau_{IR}+1$ and $2\tau_{IR}\to-\frac{1}{2\tau_{IR}}$, $\lambda(2\tau_{IR})$ transforms as $\lambda\to\frac{\lambda}{\lambda-1}$ and $\lambda\to1-\lambda$  respectively. Therefore, in terms of the coupling $q$, the duality group is $SL(2,\mathbb{Z})/\Gamma(2)=\textbf{S}_3$ with the $T$ and $S$ transformations acting on $q$ as described above.

\subsection{Strings in $N_c=2$, $N_f=4$ SQCD under Triality -- Localization Analysis}\label{Nc=2}
The four-ellipsoid partition function for an $SU(2)\times U(1)$ gauge theory with four hypermultiplets in the fundamental representation of $SU(2)$ can be written as
\begin{equation}
\begin{aligned}
Z_{S_b^4}=&\int d\hat a\, d\hat a'\; e^{-\frac{16\pi^2}{g^2} \ha^2}e^{-\frac{16\pi^2}{e^2}\hat a'^2}e^{32i\pi^2\hat{\xi}\hat a'}\Upsilon_b\left(2i\ha\right)\Upsilon_b\left(-2i\ha\right) \\&\left(\prod_{j=1}^{4}\Upsilon_b\left(i\ha-i(c_j\ha'+\hm_j)+\frac{Q}{2}\right)\Upsilon_b\left(-i\ha-i(c_j\ha'+\hm_j)+\frac{Q}{2}\right)\right)^{-1}|Z_{\text{inst}}(\ha,\ha',c_j,\hm_j)|^2\;, 
\end{aligned}\label{zs4nc2}
\end{equation}
where $a$ is the Coulomb branch parameter for the Cartan of $SU(2)$ and $a'$ is the Coulomb branch parameter for the $U(1)$ factor.  As before, we have denoted by $c_j$ the $U(1)$ charge of the $j$'th hypermultiplet, and by $\mu_j$ its mass. For the equal-charge $U(1)_B$ (\ref{Tb}) and its $T$- and $S$-dual $U(1)$'s, (\ref{TTb}) and (\ref{STb}), 
\begin{align}
&U(1)_B\;;\;\;\;\;\;\;c_1=c_2=c_3=c_4=1\;,\label{b} \\ 
&U(1)_T\;;\;\;\;\;\;\;c_1=c_2=c_3=1\;,\;\;\;c_4=-1\;,\label{t}\\
&U(1)_S\;;\;\;\;\;\;\;c_1=2\;, \;\;\; c_2=c_3=c_4=0\;.\label{s}
\end{align}

In this section we will apply our prescription to the minimal ($K=1$) strings in these three dual cases, and will compare the duality predictions for the worldsheet theories with the output of our analysis.

\subsubsection{The $c_i=1$ Case}\label{71}
This case is a special case of section \ref{42}. In the $N_f=2N_c=4$ case, equation (\ref{ZK})  reads
\begin{equation}
\begin{aligned}
Z_{K=1,\{1,2\}}=&\, e^{\frac{4\pi^2}{e^2}\left(3b^2+2+2ib(\hm_1+\hm_2)\right)}b^{4 +6b^2+2ib(\hm_1+\hm_2-\hm_3-\hm_4)}\gamma\left(-b^2\right)\\&\Big[(q\bar q)^{ib(\hat{\mu}_2-\hat{\mu}_1)/2-b^2/4}\frac{\gamma\left(ib(\hat{\mu}_1-\hat{\mu}_2)\right)}{\prod_{j=3,4}\gamma\left(1+ib(\hat{\mu}_1-\hat{\mu}_j)+b^2\right)}\\&_2F_1(-ib\left(\hat{\mu}_1-\hat{\mu}_3\right)-b^2,-ib\left(\hat{\mu}_1-\hat{\mu}_4\right)-b^2,1-ib\left(\hat{\mu}_1-\hat{\mu}_2\right)|q) 
\\&_2F_1(-ib\left(\hat{\mu}_1-\hat{\mu}_3\right)-b^2,-ib\left(\hat{\mu}_1-\hat{\mu}_4\right)-b^2,1-ib\left(\hat{\mu}_1-\hat{\mu}_2\right)|\bar q)+\mu_1\leftrightarrow \mu_2\Big]\;,
\end{aligned}\label{ZK12}
\end{equation}
where we have written the result  for  $\{l_1,l_2\}=\{1,2\}$ for simplicity.
The dictionary (\ref{dictionary0}) is:
\begin{equation}
\begin{aligned}
&z=q\;,\\
&M_1-M_2=b\left(\hat{\mu}_1-\hat{\mu}_2\right)\;,\\
&M_1+\tilde M_1=b\left(\hat{\mu}_1-\hat{\mu}_3\right)-ib^2\;,\\
&M_1+\tilde M_2=b\left(\hat{\mu}_1-\hat{\mu}_4\right)-ib^2\;,\\
&M_X=-ib^2\;,\label{dictionary1}
\end{aligned}
\end{equation}
where $M_{1,2}$,  $\tilde M_{1,2}$,  and $M_X$ are  dimensionless complexified twisted masses 
for the positively charged, negatively charged and adjoint chiral multiplets respectively.

The dictionary \ref{dictionary1} maps  (\ref{ZK12})  to the two-sphere partition function of the desired GLSM,\footnote{Up to the overall factor (\ref{counterterm}), which is interpreted as a regularization ambiguity of $Z_{S^2}$.}
\begin{equation}
\begin{aligned}
&Z_{S^2}=2\pi (z\bar z)^{-iM_1}\gamma(-iM_X)\frac{\gamma\left(-i\tilde M_1-iM_1\right)\gamma\left(-i\tilde M_2-iM_1\right)}{\gamma\left(1+i M_2-iM_1\right)}\times \\& _2F_1\left(-iM_1-i\tilde M_1,-iM_1-i\tilde M_2,1+iM_2-iM_1|z\right)\;_2F_1\left(-iM_1-i\tilde M_1,-iM_1-i\tilde M_2,1+iM_2-iM_1|\bar z\right)\\&  +M_1\leftrightarrow M_2\;.\label{ZS2v1}
\end{aligned}
\end{equation}

\subsubsection{$T$-dual of the $c_i=1$ Case}

Due to the fact that the fundamental representation of $SU(2)$ is pseudo-real,  the transformation $c_4\to -c_4$  is a classical symmetry of the massless theory. However, this symmetry is broken by non-perturbative contributions \cite{Seiberg:1994aj}. Therefore, the difference between the theory in which (\ref{b}) is gauged and the theory in which  (\ref{t}) is gauged, is manifested only through instanton contributions, in addition to a $\mu_4\to -\mu_4$ transformation in the perturbative contributions. 
Using the instanton analysis of appendix \ref{app-inst}, we find in this case
\begin{equation}
\begin{aligned}
Z_{K=1,\{1,2\}}=&\, e^{\frac{4\pi^2}{e^2}\left(3b^2+2+2ib(\hm_1+\hm_2)\right)}b^{4 +6b^2+2ib(\hm_1+\hm_2-\hm_3+\hm_4)}\gamma\left(-b^2\right)\\&\Big[(q\bar q)^{ib(\hat{\mu}_2-\hat{\mu}_1)/2-b^2/4}\frac{\gamma\left(ib(\hat{\mu}_1-\hat{\mu}_2)\right)}{\gamma\left(1+ib(\hat{\mu}_1-\hat{\mu}_3)+b^2\right)\gamma\left(1+ib(\hat{\mu}_1+\hat{\mu}_4)+b^2\right)}\\&_2F_1(-ib\left(\hat{\mu}_1-\hat{\mu}_3\right)-b^2,ib\left(\hat{\mu}_2+\hat{\mu}_4\right)+b^2+1,1-ib\left(\hat{\mu}_1-\hat{\mu}_2\right)|q) 
\\&_2F_1(-ib\left(\hat{\mu}_1-\hat{\mu}_3\right)-b^2,ib\left(\hat{\mu}_2+\hat{\mu}_4\right)+b^2+1,1-ib\left(\hat{\mu}_1-\hat{\mu}_2\right)|\bar q)+\mu_1\leftrightarrow \mu_2\Big]\;.
\end{aligned}
\end{equation}

Using the hypergeometric function identity \eqref{id{z}{z-1}} to rewrite this expression, we get
\begin{equation}
\begin{aligned}
Z_{K=1,\{1,2\}}=&\, e^{\frac{4\pi^2}{e^2}\left(3b^2+2+2ib(\hm_1+\hm_2)\right)}b^{4 +6b^2+2ib(\hm_1+\hm_2-\hm_3+\hm_4)}\gamma\left(-b^2\right)(q\bar q)^{-b^2/4}\left[(1-q)(1-\bar q)\right]^{b^2}\\&\Big[(q\bar q)^{ib(\hat{\mu}_2-\hat{\mu}_1)/2}\left[(1-q)(1-\bar q)\right]^{ib(\hm_1-\hm_3)}\frac{\gamma\left(ib(\hat{\mu}_1-\hat{\mu}_2)\right)}{\gamma\left(1+ib(\hat{\mu}_1-\hat{\mu}_3)+b^2\right)\gamma\left(1+ib(\hat{\mu}_1+\hat{\mu}_4)+b^2\right)}\\&_2F_1\left(-ib\left(\hat{\mu}_1-\hat{\mu}_3\right)-b^2,-ib\left(\hat{\mu}_1+\hat{\mu}_4\right)-b^2,1-ib\left(\hat{\mu}_1-\hat{\mu}_2\right)\Big|\frac{q}{q-1}\right) 
\\&_2F_1\left(-ib\left(\hat{\mu}_1-\hat{\mu}_3\right)-b^2,-ib\left(\hat{\mu}_1+\hat{\mu}_4\right)-b^2,1-ib\left(\hat{\mu}_1-\hat{\mu}_2\right)\Big|\frac{\bar q}{\bar q-1}\right) +\mu_1\leftrightarrow \mu_2\Big]\;.\label{this}
\end{aligned}
\end{equation}
The composition of the dictionary \eqref{dictionary1} and the $T$-transformation  \eqref{T-mass-trans} gives the dictionary
\begin{equation}
\begin{aligned}
&z=\frac{q}{q-1}\;,\\
&M_1-M_2=b\left(\hat{\mu}_1-\hat{\mu}_2\right)\;,\\
&M_1+\tilde M_1=b\left(\hat{\mu}_1-\hat{\mu}_3\right)-ib^2\;,\\
&M_1+\tilde M_2=b\left(\hat{\mu}_1+\hat{\mu}_4\right)-ib^2\;,\\
&M_X=-ib^2\;.\label{dictionaryt}
\end{aligned}
\end{equation}
As expected, this dictionary maps the expression in \eqref{this} to the two-sphere partition function \eqref{ZS2v1}, up to the multiplicative factor
\begin{equation}
e^{\frac{4\pi^2}{e^2}\left(3b^2+2+2ib(\hm_1+\hm_2)\right)}\left(b^2\right)^{2 +3b^2+il(m_1+m_2+\tilde m_1+\tilde m_2)}(z\bar z)^{\frac{i(M_1+M_2)}{2}-\frac{ b^2}{4}}\left[(1-z)(1-\bar z)\right]^{-il\frac{m_1+m_2}{2}-il\tilde m_1-\frac{3b^2}{4}}.
\end{equation}
 
Note that the charge assignment \eqref{t} obeys the condition \eqref{weak-cond}. The strings discussed in this section are therefore a special case of the  strings discussed in section \ref{localization-q-tilde-q}, for which we identified the worldsheet sphere partition functions as sphere partition functions of GLSMs with extra fields and a superpotential. At the level of the sphere partition function, the two GLSMs we obtain in this example are equivalent. 

\subsubsection{$S$-dual of the $c_i=1$ Case}\label{locS-dual}
Let us now move on to the strings that are $S$-dual to those of section \ref{71}. We will discuss these strings in detail in this subsection. We will include also the mesonic strings in our discussion. Thus, the general formula we derived for the two-sphere partition function, equation \eqref{ZS2Kgeneral}, which applies only for baryonic strings, is not sufficient for the analysis of this section.

When the $U(1)$ flavor symmetry (\ref{STb}) is gauged, the four-ellipsoid partition function reads 
\begin{equation}
\begin{aligned}
Z_{S_b^4}=&\int d\ha\, d\hat{a}' \; e^{-\frac{16\pi^2}{g^2} \ha^2}e^{-\frac{16\pi^2}{e^2}\hat{a}'^2}e^{32i\pi^2\hat{\xi}\hat{a}'}\Upsilon_b\left(2i\ha\right)\Upsilon_b\left(-2i\ha\right) \\&\left( \Upsilon_b\left(i(\ha-2\hat{a}')-i\hat\mu_1+\frac{Q}{2}\right)\Upsilon_b\left(-i(\ha+2\hat{a}')-i\hat\mu_1+\frac{Q}{2}\right)\right)^{-1}\\&\left(\prod_{j=2}^{4}\Upsilon_b\left(i\ha-i\hat\mu_j+\frac{Q}{2}\right)\Upsilon_b\left(-i\ha-i\hat\mu_j+\frac{Q}{2}\right)\right)^{-1}|Z_{\text{inst}}|^2\;. 
\end{aligned}
\end{equation}

As we explained in section \ref{42} we use a cut-off $\hat\Lambda$ for the matrix integral, and neglect ${\cO\left(\hat\Lambda e^{-\frac{\hat{\Lambda}^2}{e^2}}\right)+\cO\left(\hat\Lambda e^{-\frac{\hat{\Lambda}^2}{g^2}}\right)}$ contributions. First we perform the integration over $\ha'$.
The poles in the upper half $\hat{a}'$-plane are located at 
\begin{equation}
2\hat{a}'=\pm\ha -\hm_1+i\left(\frac{Q}{2}+k_1b+k'_1b^{-1}\right)\;,\;\;\; k_1,k'_1\in{\mathbb{N}}\;.
\end{equation} 
The integrand is symmetric under $a\to -a$ and therefore we can focus on the poles 
\begin{equation}
2\hat{a}'=+\ha -\hm_1+i\left(\frac{Q}{2}+k_1b+k'_1b^{-1}\right)\;,\;\;\; k_1,k'_1\in{\mathbb{N}}\label{pole+}\;,
\end{equation}
and multiply the end result by 2. The residues of the poles in (\ref{pole+}) contain a $e^{16ֿ\pi^2i\hx\ha}$ factor. Therefore, we need to close the contour of the $\ha$ integrals from above. The contour will encircle the poles
\begin{equation}
\hat{a}= \pm\hm_j+i\left(\frac{Q}{2}+k_2b+k'_2b^{-1}\right)\;,\;\;\; k_2,k'_2\in{\mathbb{N}}\;, j=2,3,4\;.
\end{equation}
The Cauchy integrations therefore result in a sum 
\begin{equation}
2(2\pi i)^2e^{-16\pi^2\hat{\xi}Q}\sum_{j=2}^{4}\left(e^{-16i\pi^2\hat{\xi}(\hat{\mu}_1-\hat{\mu}_j)}Z^{+}_{\text{vac},j}\sum_{K}e^{-16\pi^2\hx Kb}Z^{+}_{K,j}+e^{-16i\pi^2\hat{\xi}(\hat{\mu}_1+\hat{\mu}_j)}Z^{-}_{\text{vac},j}\sum_{K}e^{-16\pi^2\hx Kb}Z^{-}_{K,j}\right)+...\;,
\end{equation}
where the dots stand for contributions with $k'_1+k'_2>0$.
 $Z^{\pm}_{\text{vac},j}e^{-16\pi^2\hx Kb}Z^{\pm}_{K,j}$ is the sum of the residues of the integrand in the poles
\begin{equation}
\begin{aligned}
&2\hat{a}'=+\ha -\hm_1+i\left(\frac{Q}{2}+nb\right)\;,\;\;\; n=0,1,2,..,K\;,\\
&\hat{a}= \pm\hm_j+i\left(\frac{Q}{2}+(K-n)b\right)\;.
\end{aligned}
\end{equation}
The $Z^-$ contributions correspond to the baryonic vacua and strings and the $Z^+$ contributions correspond to the mesonic vacua and strings.

Evaluating the residue of the  $K=0$ poles we find\footnote{We use the symmetry between $\mu_2,\mu_3,\mu_4$ and write the expressions for $j=2$.}
\begin{equation}
\begin{aligned}
Z^{\pm}_{\text{vac},j=2}=&e^{-\frac{4\pi^2}{g^2} (\mp 2\hat \mu_2-iQ)^2}e^{-\frac{4\pi^2}{e^2}\left(-(\hat{\mu}_1\mp\hat{\mu}_2)+iQ \right)^2}\left(\text{Res}|_{x=0}(\Upsilon_b(x+Q)^{-1})\right)^{2}\left(\prod_{k=3,4}\Upsilon_b\left(i(\pm\hat{\mu}_2+\hat\mu_k)\right)\right)^{-1}\\
&\left(\prod_{k=3,4}\Upsilon_b\left(i(\pm\hat{\mu}_2-\hat\mu_k)\right)\right)^{-1}{Z^{\pm}_{\text{inst}}(K=0,n=0,q)}\,{Z^{\pm}_{\text{inst}}(K=0,n=0,\bar q)}\;,
\end{aligned}
\end{equation}
where $Z^{\pm}_{\text{inst}}(K,n,q)$ stands for the instanton partition function in the $\Omega$-background evaluated at
\begin{equation}
\begin{aligned}
&2\hat{a}'= -(\hm_1\mp\hm_2)+i\left({Q}+Kb\right)\;,\\
&\hat{a}= \pm\hm_2+i\left(\frac{Q}{2}+(K-n)b\right)\;.
\end{aligned}
\end{equation}

The  $S$-transformation rule (\ref{S-mass-trans}) implies that the six strings we studied in section \ref{71} should map to the strings described by $Z^{\pm}_{K,j}$ in  the following way:
\begin{align}
&Z^{B}_{K,(1,j)}(q)\to Z^{-}_{K,j}(1-q)\;,\;\;\;j=2,3,4\;,\label{claim1}\\
&Z^{B}_{K,(3,4)}(q)\to Z^{+}_{K,2}(1-q)\;,\label{claim2}\\
&Z^{B}_{K,(2,4)}(q)\to Z^{+}_{K,3}(1-q)\;,\label{claim3}\\
&Z^{B}_{K,(2,3)}(q)\to Z^{+}_{K,4}(1-q)\;.\label{claim4}
\end{align}

To find the two-sphere partition function for the minimal strings, we evaluate the sum of the residues of the $K=1$ poles. After using the shift identity (\ref{identity-decoupling}) to separate the vacuum contributions we obtain
\begin{equation}
\begin{aligned}
&Z^{-}_{K=1,j=2}={b^{4ib\hat{\mu}_2+6b^2+4}}{\gamma(-b^2)}e^{\frac{4\pi^2}{e^2}\left(2ib(\hat{\mu}_1+\hat{\mu}_2)+3b^2+2\right)}\\&\Big((q\bar q)^{(-1-2b^2-2ib\hat{\mu}_2)}\frac{\gamma(2ib\hat{\mu}_2+2b^2+1)}{\prod_{k=3,4}\gamma(ib(\hat{\mu}_2-\hat{\mu}_k)+b^2+1)\prod_{k=3,4}\gamma(ib(\hat{\mu}_2+\hat{\mu}_k)+b^2+1)}Z^{-}_{\text{inst},n=1}(q)Z^{-}_{\text{inst},n=1}(\bar q)\\&+{\gamma(-2ib\hat{\mu}_2-2b^2-1)}Z^{-}_{\text{inst},n=0}(q)Z^{-}_{\text{inst},n=0}(\bar q)\Big)\;,\label{ZK1j2}
\end{aligned}
\end{equation}
with
\begin{equation}
\begin{aligned}
Z^{-}_{\text{inst},n=1}(q)&=(1-q)^{-ib(\hm_3+\hm_4)} {_2F_1}\left(ib(\hm_2-\hm_3)+b^2+1,ib(\hm_2-\hm_4)+b^2+1,2ib\hm_2+2b^2+2|q\right)\;,\\
Z^{-}_{\text{inst},n=0}(q)&=(1-q)^{-ib(\hm_3+\hm_4)}{_2F_1}\left(-ib(\hm_2+\hm_3)-b^2,-ib(\hm_2+\hm_4)-b^2,-2ib\hm_2-2b^2|q\right)\;.
\end{aligned}
\end{equation}
For the instanton contributions we used the computation of appendix \ref{app-inst} and identity \eqref{id2}.

Our claim is that equation \eqref{ZK1j2} is the two-sphere partition function of the GLSM obtained in section \ref{71},  with the dictionary between the four-dimensional and the two-dimensional parameters given by the map
\begin{equation}\label{dictionary3}
\begin{aligned}
&z=1-q\;,\\
&M_1-M_2=b(\hat\mu_3+\hat\mu_4)\;,\\
&M_1+\tilde M_1=b(\hat\mu_2+\hat\mu_4)-ib^2\;,\\
&M_1+\tilde M_2=b(\hat\mu_2+\hat\mu_3)-ib^2\;,\\
&M_X=-ib^2\;,
\end{aligned}
\end{equation}
which is simply the composition of (\ref{dictionary1}) with the S-transformation (\ref{S-mass-trans}).
Indeed, in appendix \ref{ap-ZS2-1-q} we show that equation (\ref{ZS2v1}) can be written as
\begin{equation}
\begin{aligned}
&Z_{S^2}=\gamma(-iM_X)(1-z)^{c-a-b}(1-\bar z)^{c-a-b}(z\bar z)^{\frac{c-1}{2}}2\pi (z\bar z)^{-i(M_1+M_2)/2}\times\\&\Bigg[{_2F_1}(a,b,a+b-c+1|1-z){_2F_1}(a,b,a+b-c+1|1-\bar z)\frac{\gamma(c-a-b)\gamma(a)\gamma(b)}{\gamma(c-a)\gamma(c-b)}[(1-z)(1-\bar z)]^{a+b-c}+\\&\gamma\left(a+b-c\right){_2F}_1(c-a,c-b,c-a-b+1|1-z){_2F}_1(c-a,c-b,c-a-b+1|1-\bar z)\Bigg]\;,\label{ZS2v2}
\end{aligned}
\end{equation} 
with
\begin{equation}
\begin{aligned}
-i\tilde M_1-iM_1=a\;,\\
-i\tilde M_2-iM_1=b\;,\\
1+i(M_2-M_1)=c\;.\label{ZS2v2m}
\end{aligned}
\end{equation}

Under the map \eqref{dictionary3} expressions (\ref{ZK1j2}) and (\ref{ZS2v2}) match up to the multiplicative factor
\begin{equation}
e^{\frac{4\pi^2}{e^2}\left(2ib(\hat{\mu}_1+\hat{\mu}_2)+3b^2+2\right)}{\left(b^2\right)^{il(m_1+m_2+\tilde m_1+\tilde m_2)+3b^2+2}}\left((1-z)(1-\bar z)\right)^{-1-2b^2-il(m_1+m_2+\tilde m_1+\tilde m_2)}(z\bar z)^{iM_1}\;, 
\end{equation}
which we interpret (as in the discussion below \eqref{counterterm})  as an artifact of the regularization scheme.

To complete the check of the claims in (\ref{claim1}-\ref{claim4}) for the $K=1$ strings, we need to check that $Z^{+}_{K=1,j=2}$ maps to (\ref{ZS2v2}) under the dictionary:
\begin{equation}\label{dictionarys+}
\begin{aligned}
&z=1-q\;,\\
&M_1-M_2=b(\hat\mu_3-\hat\mu_4)\;,\\
&M_1+\tilde M_1=-b(\hat\mu_2+\hat\mu_4)-ib^2\;,\\
&M_1+\tilde M_2=b(-\hat\mu_2+\hat\mu_3)-ib^2\;,\\
&M_X=-ib^2\;,
\end{aligned}
\end{equation}
where $Z^{+}_{K=1,j=2}$ is given by
	\begin{equation}
	\begin{aligned}
	Z^{+}_{K=1,j=2}=&{b^{-4ib\hat{\mu}_2+6b^2+4}}{\gamma(-b^2)}e^{\frac{4\pi^2}{e^2}\left(2ib(\hat{\mu}_1-\hat{\mu}_2)+3b^2+2\right)}\\&\Bigg[{\gamma(2ib\hat{\mu}_2-2b^2-1)}\frac{Z^{+}_{\text{inst}}(K=1,n=1,q)Z^{+}_{\text{inst}}(K=1,n=1,\bar q)}{Z^{+}_{\text{inst}}(K=0,n=0,q)Z^{+}_{\text{inst}}(K=0,n=0,\bar q)}\,+\\&(q\bar q)^{(-1-2b^2+2ib\hat{\mu}_2)}\frac{\gamma(-2ib\hat{\mu}_2+2b^2+1)}{\prod_{k=3,4}\gamma(ib(-\hat{\mu}_2-\hat{\mu}_k)+b^2+1)\prod_{k=3,4}\gamma(ib(-\hat{\mu}_2+\hat{\mu}_k)+b^2+1)}\\&\times \frac{Z^{+}_{\text{inst}}(K=1,n=0,q)Z^{+}_{\text{inst}}(K=1,n=0,\bar q)}{Z^{+}_{\text{inst}}(K=0,n=0,q)Z^{+}_{\text{inst}}(K=0,n=0,\bar q)}\Bigg]\;,\label{Z+K1j2}
	\end{aligned}
	\end{equation}
	Thus, we need to check that
\begin{align}
&\frac{Z^{+}_{\text{inst}}(K=1,n=0,q)}{Z^{+}_{\text{inst}}(K=0,n=0,q)}=g(q,\hm)\,_2F_1\left(ib(\hat{\mu}_2+\hat{\mu}_4)-b^2,ib(\hat{\mu}_2-\hat{\mu}_3)-b^2,2ib\hat{\mu}_2-2b^2|q\right)\;,\label{cond3}\\
&\frac{Z^{+}_{\text{inst}}(K=1,n=1,q)}{Z^{+}_{\text{inst}}(K=0,n=0,q)}=g(q,\hm)\,_2F_1\left(1-ib(\hat{\mu}_2+\hat{\mu}_3)+b^2,1-ib(\hat{\mu}_2-\hat{\mu}_4)+b^2,2-2ib\hat{\mu}_2+2b^2|q\right)\;,\label{cond4}\end{align}
with the same function $g(q,\mu)$.
This is not a special case of the instanton analysis of appendix \ref{app-inst}, which applies only to the baryonic strings. We have checked that (\ref{cond3}-\ref{cond4}) hold up to two instanton level.

\subsubsection{Other Transformations}
We discussed three out of the six triality elements. To complete the analysis, let us discuss the other three elements. 

Each of the three $U(1)$ generators we discussed is invariant under one of the triality transformations; $T_B$, $T_T$ and $T_S$ are invariant, respectively, under $STS$, $S$  and $T$. This invariance  manifests itself in the $\mathbb{Z}_2$ symmetry of the two-dimensional worldsheet theory \eql{z2transformation}{t\rightarrow-t\ ,\ m_i\leftrightarrow\tilde{m}_i\ ,} where $m_i$ and $\tilde m_{i}$ are the twisted masses of the fundamental and anti-fundamental worldsheet chiral multiplets.

\section*{Acknowledgments}

We would like to thank Shlomo Razamat and Talya Vaknin for fruitful discussions. We especially thank Jaume Gomis and Zohar Komargodski for collaboration throughout a
large portion of this project and for many useful discussions. 
We thank the Perimeter
Institute for Theoretical Physics and the Galileo Galilei Institute for Theoretical Physics for hospitality during the course of this project. This research was supported in part by Perimeter Institute for Theoretical Physics. Research at Perimeter Institute is supported by the
Government of Canada through the Department of Innovation, Science and Economic Development and by the Province of Ontario through
the Ministry of Research and Innovation.
E.G. and A.K. are supported by the ERC STG grant 335182.

\appendix

\section{Non-Perturbative Contributions}\label{app-inst}

\subsection{Non-Perturbative Contributions to the Four-Ellipsoid Partition Function}
The non-perturbative contributions to equation \eqref{general gauging s4} are given in terms of Nekrasov's instanton partition function \cite{Nekrasov:2002qd}, which in this case reads
\begin{equation}
\begin{aligned}
&Z_{\text{inst}}=\sum_{\vec Y}q^{|\vec Y|}b^{(2N_c-N_f){|\vec Y|}}\prod_{a,b=1}^{N_c}Z_{ab}^{\text{vec}}\left({\vec{{\hat a}},\vec Y}\right)\prod_{i=1}^{N_f}\prod_{a=1}^{N_c}Z_{ai}^{\text{hyp}}\left({\vec{{\hat a}},\vec{{\hat \mu}},\vec Y}\right)\;,\\
&Z_{ab}^{\text{vec}}=\prod_{r,s=1}^\infty\frac{\Gamma\left(Y_{ar}-Y_{bs}+b^2(r-s-1)+ib(w_a\cdot \ha-w_b\cdot \ha) \right)}{\Gamma\left(Y_{ar}-Y_{bs}+b^2(r-s)+ib(w_a\cdot \ha-w_b\cdot \ha) \right)}\frac{\Gamma\left(
	b^2(r-s)+ib(w_a\cdot \ha-w_b\cdot \ha) \right)}{\Gamma\left(b^2(r-s-1)+ib(w_a\cdot \ha-w_b\cdot \ha) \right)}\;,\\
&Z_{ai}^{\text{hyp}}=\prod_{r=1}^\infty\frac{\Gamma\Big(Y_{ar}+b^2(r-1)+b\left(i(w_a\cdot \ha+c_i\hat{a}'+\hm_i)+\frac{Q}{2}\right) \Big)}{\Gamma\Big(b^2(r-1)+b\left(i(w_a\cdot \ha+c_i\hat{a}'+\hm_i)+\frac{Q}{2}\right) \Big)}\;,	\label{inst-cont}
\end{aligned}
\end{equation}
where $q=e^{2\pi i \tau_{su(N_c)}}$.\footnote{We ignore subtleties in the instanton partition function related to $U(1)$ factors. These subtleties are expected to affect our results  only through the overall factors multiplying the two-sphere partition functions that we identify as regularization ambiguities. }
The sum over $\vec Y$ is a sum over $N_c$-tuples of Young diagrams, $\vec Y=(Y_1,..,Y_{N_c})$, $Y_{ar}$ is the height of the $r$'th column in the diagram $Y_a$,  $Y_{ar}\geq Y_{a,r+1}$, and $|\vec Y|=\sum_{a=1}^{N_c}\sum_{r=1}^{\infty}Y_{ar}$.

We need to evaluate this for 
\begin{equation}
i w_a\cdot \ha+ic_a\ha'+i\hm_a+\frac{Q}{2}+k_ab=0\;, \;\;\;a=1,...,N_c\;.\label{polesss}
\end{equation}
After a change of variables, this computation becomes a special case of computations that appeared in  \cite{Chen:2015fta,Pan:2016fbl,Pan:2015hza}. We repeat the computation here for the convenience of the reader. Substituting \eqref{polesss} into \eqref{inst-cont} we obtain 
\begin{equation}
\begin{aligned}
&Z^{\text{res}}_{\text{inst}}(\mm_{ij},\{k_a\},q)=\sum_{\{\vec Y|Y_{a,k_a+1}=0\}}q^{|\vec Y|}b^{(2N_c-N_f){|\vec Y|}}\prod_{i=1}^{N_f}\prod_{a=1}^{N_c}\prod_{r=1}^{k_a}\frac{\Gamma\left(Y_{ar}+b^2(r-1-k_a)+ib\mm_{ia} \right)}{\Gamma\left(b^2(r-1-k_a)+ib\mm_{ia} \right)}\\
&\prod_{a,b=1}^{N_c}\prod_{r=1}^{\infty}\prod_{s=1}^{\infty}\frac{\Gamma\left(Y_{ar}-Y_{bs}+b^2(r-s-1)+ib\mm_{ba}-b^2(k_a-k_b) \right)}{\Gamma\left(Y_{ar}-Y_{bs}+b^2(r-s)+ib\mm_{ba}-b^2(k_a-k_b) \right)}\frac{\Gamma\left(
	b^2(r-s)+ib\mm_{ba}-b^2(k_a-k_b) \right)}{\Gamma\left(b^2(r-s-1)+ib\mm_{ba}-b^2(k_a-k_b) \right)}\;,	
\end{aligned}
\end{equation}
where \begin{equation}
\mm_{ia}=\hm_i-\hm_a+\Delta_{ai}\left(\sum_{b=1}^{N_c}\hm_b-iN_c\frac{Q}{2}-iKb\right)\;,
\end{equation}
and we used the fact that diagrams for which $Y_{a,k_a+1}>0$ cannot contribute since the hypermultiplet contribution for $r=k_a+1$ and  $i=a$ multiplies everything by $0$.
Simplifying this expression we get
\begin{equation}
\begin{aligned}
Z^{\text{res}}_{\text{inst}}(\mm_{ij},\{k_a\},q)=&\sum_{\{\vec Y|Y_{a,k_a+1}=0\}}q^{|\vec Y|}b^{(2N_c-N_f){|\vec Y|}}(-1)^{N_c|\vec Y|}\prod_{i=N_c+1}^{N_f}\prod_{a=1}^{N_c}\prod_{r=0}^{k_a-1}\left(-b^2(r+1)+ib\mm_{ia}\right)_{Y_{a,k_a-r}}\\
&\times\prod_{a,b=1}^{N_c}\prod_{r=0}^{k_a-1}\prod_{s=0}^{k_{b}-1}\frac{1}{\left(1-Y_{b,k_b-s}+Y_{a,k_a-r}+b^2(s-r)+ib\mm_{ba} \right)_{Y_{b,k_b-s}-Y_{b,k_b-s+1}}}\\&\times\prod_{a,b=1}^{N_c}\prod_{r=0}^{k_{a}-1}\frac{\left(1+b^2(k_b-r)+ib\mm_{ba}+Y_{a,k_a-r}-Y_{b,1} \right)_{Y_{b,1}}}{\left(1+b^2(k_b-r)+ib\mm_{ba} \right)_{Y_{a,k_a-r}}}\;.
\end{aligned}
\end{equation}
Here, $(x)_n$ is the Pochhammer symbol, defined as 
\begin{equation}
(x)_n=x(x+1)(x+2)...(x+n-1)=\frac{\Gamma(x+n)}{\Gamma(x)}\;.
\end{equation}

\subsection{Non-Perturbative Contributions to the Two-Sphere Partition Function}
The two-sphere partition function of an $\cN=(2,2)$ supersymmetric theory with gauge group $U(K)$, $N_c$ chiral multiplets in the fundamental representation, $ N_f-N_c$ chiral multiplets in the anti-fundamental representation and one chiral multiplet in the adjoint representation, is given by equation \eqref{ZS2}, where the function $Z^{\text{vort}}_{\{ k_a\}}$ is given by
\begin{equation}
\begin{aligned}
Z^{\text{vort}}_{\{ k_a\}}(z)=&\sum_{\tilde Y}\left[(-1)^{N_c+K-1}z\right]^{\sum_{a=1}^{N_c}\sum_{r=0}^{k_a-1}\tilde Y_{a,k_a-r}}\\&\times\prod_{a=1}^{N_c}\prod_{r=0}^{k_a-1}\frac{\prod_{j=1}^{N_f-N_c}\left(-i\tilde M_j-iM_a-irM_X\right)_{\tilde Y_{a,k_a-r}}}{\prod_{b=1}^{N_c}\left(1+iM_b-iM_a+(k_b-r)iM_X\right)_{\tilde Y_{a,k_a-r}}}
	\\&\times\prod_{a=1}^{N_c}\prod_{r=0}^{k_a-1}\frac{\prod_{b=1}^{ N_c}\left(1+iM_b-iM_a+(k_b-r)iM_X+\tilde Y_{a,k_a-r}-\tilde Y_{b,1}\right)_{\tilde Y_{b,1}}}{\prod_{b=1}^{N_c}\prod_{s=0}^{k_b-1}\left(1+iM_b-iM_a+(s-r)iM_X+\tilde Y_{a,k_a-r}-\tilde Y_{b,k_b-s}\right)_{\tilde Y_{b,k_b-s}-\tilde Y_{b,k_b-s+1}}}\;.\label{zvort}
\end{aligned}
\end{equation}
The sum over $\tilde Y$ is a sum over functions $\tilde Y: \{1,...,N_c\}\times\{0,...,k_a-1\}\to \mathbb{N}$, such that  $\tilde Y_{a,k_a-r}\in \mathbb{N}$, where $a=1,...,N_c$, $r=0,..,k_a-1$, satisfies $\tilde Y_{a,k_a-r}\geq \tilde Y_{a,k_a-r+1}$. As before $z=e^{-2\pi\xi_{2d}+i\te_{2d}}$, and  $M_a$,  $\tilde M_j$ and $M_X$ are the complexified twisted masses for the chiral multiplets transforming in the fundamental, anti-fundamental and adjoint representations respectively.

We therefore find that the map 
\begin{equation}
\begin{aligned}
&z=(-1)^{K-1}b^{2N_c-N_f}q\;,\\
&m_a-m_b={\mu}_a-{\mu}_b+\Delta_{ba}\sum_{c=1}^{N_c}\mu_c\;,\;\;\;\;\;\;\;\;\;\;\;R_a-R_b=\Delta_{ab}N_c(b^2+1)+2\Delta_{ab}Kb^2\;,\\
&m_a+\tilde m_{j-N_c}={\mu}_a-{\mu}_j+\Delta_{ja}\sum_{c=1}^{N_c}\mu_c\;,\;\;\;\;\;\;R_a+\tilde R_{j-N_c}=\Delta_{aj}N_c(b^2+1)+2\Delta_{aj}Kb^2-2b^2\;,\\
&m_X=0\;,\;\;\;\;\;\;\;\;\;\;\;\;\;\;\;\;\;\;\;\;\;\;\;\;\;\;\;\;\;\;\;\;\;\;\;\;\;\;\;\;\;\;\;\;\;\;\;\;R_X=-2b^2\;,\label{mapp}
\end{aligned}
\end{equation} where $a,b=1,...,N_c$, $j=N_{c}+1,...,N_f$, maps $Z^{\text{res}}_{\text{inst}}(\mm_{ij},\{k_a\},q)$ to $Z^{\text{vort}}_{\{ k_a\}}(z)$. This is true for any $U(1)$ charge assignment. For the strings discussed in sections \ref{42},\ref{Locnotilde} and \ref{localization-q-tilde-q}, this completes the identification of the outputs of our prescription as the two-sphere partition functions of the proposed worldsheet theories. In other examples, the observation above does not imply that the  parameters of the charged sector of the worldsheet theory are given in terms of the four-dimensional parameters as in (\ref{mapp}). For example, see section \ref{locS-dual} (and appendix \ref{ap-ZS2-1-q}), where we discuss a case in which the weak coupling regime maps to the strong coupling regime of the worldsheet theory and the identification of the worldsheet theory requires writing the worldsheet two-sphere partition  function as an expansion around $z=1$.

\section{Useful Identities}\label{useful}
In section \ref{Nc=2} and appendix \ref{ap-ZS2-1-q} we use the following hypergeometric and Gamma-function identities,  
\begin{align}
&_2F_1\left(a,b,c|z\right)=\frac{\Gamma(c)\Gamma(c-a-b)}{\Gamma(c-a)\Gamma(c-b)}{_2F_1}(a,b,a+b-c+1|1-z)+\label{id1-z}\\&\;\;\;\;\;\;\;\;\;\;\;\;\;\;\;\;\;\;\;\;\;\;\;\nonumber\frac{\Gamma(a+b-c)\Gamma(c)}{\Gamma(a)\Gamma(b)}        {_2F}_1(c-a,c-b,c-a-b+1|1-z)(1-z)^{c-a-b}\;,\\
&{_2F_1}(a,b,c|z)=(1-z)^{-a}{_2F_1\left(a,c-b,c\;\Big|\frac{z}{z-1}\right)}\;,\label{id{z}{z-1}}\\
&_2F_1(a,b,c|z)=(1-z)^{c-a-b}\,_2F_1(c-a,c-b,c|z)\label{id2}\;,\\&\Gamma(x)^2=\gamma(x)\frac{\pi}{\sin(\pi x)}\;.\label{Gamma^2}\end{align}
In appendix \ref{ap-ZS2-1-q} we also use the identities:
\begin{align}
&\frac{\gamma(a)\gamma(b)}{\gamma(c)}  \left(\frac{\Gamma(c)\Gamma(c-a-b)}{\Gamma(c-a)\Gamma(c-b)}\right)^2+\begin{pmatrix}a\to 1-c+a\\b\to 1-c+b\\c\to 2-c\end{pmatrix}
=\frac{\gamma(c-a-b)\gamma(a)\gamma(b)}{\gamma(c-a)\gamma(c-b)}\label{i1}\;,\\
&\frac{\gamma\left(a\right)\gamma\left(b\right)}{\gamma\left(c\right)}\left(\frac{\Gamma(a+b-c)\Gamma(c)}{\Gamma(a)\Gamma(b)}\right)^2+\begin{pmatrix}a\to 1-c+a\\b\to 1-c+b\\c\to 2-c\end{pmatrix}=\gamma(a+b-c)\label{i2}\;,\\
&\frac{\gamma\left(a\right)\gamma\left(b\right)}{\gamma\left(c\right)}\frac{\Gamma(c)\Gamma(c-a-b)}{\Gamma(c-a)\Gamma(c-b)}\frac{\Gamma(a+b-c)\Gamma(c)}{\Gamma(a)\Gamma(b)}+\begin{pmatrix}a\to 1-c+a\\b\to 1-c+b\\c\to 2-c\end{pmatrix}=0\;.\label{i3}
\end{align}

To derive \eqref{i1} one can use equation \eqref{Gamma^2} to get
\begin{equation}
\begin{aligned}
&\frac{\gamma(a)\gamma(b)}{\gamma(c)}  \left(\frac{\Gamma(c)\Gamma(c-a-b)}{\Gamma(c-a)\Gamma(c-b)}\right)^2+\begin{pmatrix}a\to 1-c+a\\b\to 1-c+b\\c\to 2-c\end{pmatrix}=\\&\frac{\gamma(c-a-b)\gamma(a)\gamma(b)}{\gamma(c-a)\gamma(c-b)}\left(\frac{\sin\left(\pi(c-a)\right)\sin\left(\pi(c-b)\right)}{\sin\left(\pi c\right)\sin\left(\pi(c-a-b)\right)}+\frac{\sin\left(\pi(1-a)\right)\sin\left(\pi(1-b)\right)}{\sin\left(\pi (2-c)\right)\sin\left(\pi(c-a-b)\right)}\right)=\\&\frac{\gamma(c-a-b)\gamma(a)\gamma(b)}{\gamma(c-a)\gamma(c-b)}\;,
\end{aligned}
\end{equation}
where we also used the fact that $\gamma(x)\gamma(1-x)=1$. In the second step we 
used  standard trigonometric identities. The derivation of (\ref{i2}-\ref{i3}) is similar.

\section{Derivation of equation (\ref{ZS2v2})}\label{ap-ZS2-1-q}
We start with the formula (\ref{ZS2v1}) for the two-sphere partition function. For simplicity we denote 
\begin{align}
-i\tilde M_1-iM_1=a\;,\\
-i\tilde M_2-iM_1=b\;,\\
1+i(M_2-M_1)=c\;,
\end{align}
such that equation (\ref{ZS2v1}) becomes
\begin{equation}
\begin{aligned}
Z_{S^2}=2\piֿ\gamma(-iM_X) (z\bar z)^{-i(M_1+M_2)/2}\Bigg((z\bar z)^{\frac{c-1}{2}}\frac{\gamma\left(a\right)\gamma\left(b\right)}{\gamma\left(c\right)} \;_2F_1\left(a,b,c|z\right)\;_2F_1\left(a,b,c|\bar z\right) +\begin{pmatrix}a\to 1-c+a\\b\to 1-c+b\\c\to 2-c\end{pmatrix}\Bigg)\;.
\end{aligned}
\end{equation}
Using identity \eqref{id1-z} we can write
$Z_{S^2}=2\pi\gamma(-iM_X) (z\bar z)^{-i(M_1+M_2)/2}\Bigg(I_1+I_2+I_3+I_4\Bigg)\;,
$
with
\begin{equation}
\begin{aligned}
I_1=&(z\bar z)^{\frac{c-1}{2}}\frac{\gamma\left(a\right)\gamma\left(b\right)}{\gamma\left(c\right)}\left(\frac{\Gamma(c)\Gamma(c-a-b)}{\Gamma(c-a)\Gamma(c-b)}\right)^2{_2F_1}(a,b,a+b-c+1|1-z){_2F_1}(a,b,a+b-c+1|1-\bar z)\\&+\begin{pmatrix}a\to 1-c+a\\b\to 1-c+b\\c\to 2-c\end{pmatrix}\;,\\I_2=&(z\bar z)^{\frac{c-1}{2}}\frac{\gamma\left(a\right)\gamma\left(b\right)}{\gamma\left(c\right)}\left(\frac{\Gamma(a+b-c)\Gamma(c)}{\Gamma(a)\Gamma(b)}\right)^2(1-z)^{c-a-b}(1-\bar z)^{c-a-b}\,\times\\&{_2F}_1(c-a,c-b,c-a-b+1|1-z){_2F}_1(c-a,c-b,c-a-b+1|1-\bar z)+\begin{pmatrix}a\to 1-c+a\\b\to 1-c+b\\c\to 2-c\end{pmatrix}\;,\\I_3=&(z\bar z)^{\frac{c-1}{2}}\frac{\gamma\left(a\right)\gamma\left(b\right)}{\gamma\left(c\right)}\frac{\Gamma(c)\Gamma(c-a-b)}{\Gamma(c-a)\Gamma(c-b)}\frac{\Gamma(a+b-c)\Gamma(c)}{\Gamma(a)\Gamma(b)}\,\times\\&{_2F_1}(a,b,a+b-c+1|1-z){_2F}_1(c-a,c-b,c-a-b+1|1-\bar z)(1-\bar z)^{c-a-b}+\begin{pmatrix}a\to 1-c+a\\b\to 1-c+b\\c\to 2-c\end{pmatrix}\;,\\I_4=&(z\bar z)^{\frac{c-1}{2}}\frac{\gamma\left(a\right)\gamma\left(b\right)}{\gamma\left(c\right)}\frac{\Gamma(c)\Gamma(c-a-b)}{\Gamma(c-a)\Gamma(c-b)}\frac{\Gamma(a+b-c)\Gamma(c)}{\Gamma(a)\Gamma(b)}\,\times\\&{_2F_1}(a,b,a+b-c+1|1-\bar z){_2F}_1(c-a,c-b,c-a-b+1|1- z)(1- z)^{c-a-b}+\begin{pmatrix}a\to 1-c+a\\b\to 1-c+b\\c\to 2-c\end{pmatrix}\;.
\end{aligned}
\end{equation}

Identity \eqref{id2} implies that  $z^{\frac{c-1}{2}}{_2F_1}(a,b,a+b-c+1|1-z)$ and $z^{\frac{c-1}{2}}{_2F_1}(c-a,c-b,c-a-b+1|1-z)$ are invariant under $\begin{pmatrix}a\to 1-c+a\\b\to 1-c+b\\c\to 2-c\end{pmatrix}$.
We can therefore write:
\begin{equation}
\begin{aligned}
I_1=&(z\bar z)^{\frac{c-1}{2}}{_2F_1}(a,b,a+b-c+1|1-z){_2F_1}(a,b,a+b-c+1|1-\bar z)\,\times\\&\left(\frac{\gamma\left(a\right)\gamma\left(b\right)}{\gamma\left(c\right)}\left(\frac{\Gamma(c)\Gamma(c-a-b)}{\Gamma(c-a)\Gamma(c-b)}\right)^2+\begin{pmatrix}a\to 1-c+a\\b\to 1-c+b\\c\to 2-c\end{pmatrix}\right)\;,\\I_2=&(z\bar z)^{\frac{c-1}{2}}\left(\frac{\gamma\left(a\right)\gamma\left(b\right)}{\gamma\left(c\right)}\left(\frac{\Gamma(a+b-c)\Gamma(c)}{\Gamma(a)\Gamma(b)}\right)^2+\begin{pmatrix}a\to 1-c+a\\b\to 1-c+b\\c\to 2-c\end{pmatrix}\right)\,\times\\&{_2F}_1(c-a,c-b,c-a-b+1|1-z){_2F}_1(c-a,c-b,c-a-b+1|1-\bar z)(1-z)^{c-a-b}(1-\bar z)^{c-a-b}\;,\\I_3=&(z\bar z)^{\frac{c-1}{2}}\left(\frac{\gamma\left(a\right)\gamma\left(b\right)}{\gamma\left(c\right)}\frac{\Gamma(c)\Gamma(c-a-b)}{\Gamma(c-a)\Gamma(c-b)}\frac{\Gamma(a+b-c)\Gamma(c)}{\Gamma(a)\Gamma(b)}+\begin{pmatrix}a\to 1-c+a\\b\to 1-c+b\\c\to 2-c\end{pmatrix}\right)\,\times\\&{_2F_1}(a,b,a+b-c+1|1-z){_2F}_1(c-a,c-b,c-a-b+1|1-\bar z)(1-\bar z)^{c-a-b}\;,\\I_4=&(z\bar z)^{\frac{c-1}{2}}\left(\frac{\gamma\left(a\right)\gamma\left(b\right)}{\gamma\left(c\right)}\frac{\Gamma(c)\Gamma(c-a-b)}{\Gamma(c-a)\Gamma(c-b)}\frac{\Gamma(a+b-c)\Gamma(c)}{\Gamma(a)\Gamma(b)}+\begin{pmatrix}a\to 1-c+a\\b\to 1-c+b\\c\to 2-c\end{pmatrix}\right)\,\times\\&{_2F_1}(a,b,a+b-c+1|1-\bar z){_2F}_1(c-a,c-b,c-a-b+1|1- z)(1- z)^{c-a-b}\;.
\end{aligned}
\end{equation}

Using identities (\ref{i1}-\ref{i3}) we 
can simplify these expressions, and write
\begin{equation}
\begin{aligned}
I_1=&(z\bar z)^{\frac{c-1}{2}}\frac{\gamma(c-a-b)\gamma(a)\gamma(b)}{\gamma(c-a)\gamma(c-b)}{_2F_1}(a,b,a+b-c+1|1-z){_2F_1}(a,b,a+b-c+1|1-\bar z)\;,\\I_2=&(z\bar z)^{\frac{c-1}{2}}\gamma\left(a+b-c\right)(1-z)^{c-a-b}(1-\bar z)^{c-a-b}\,\times\\&{_2F}_1(c-a,c-b,c-a-b+1|1-z){_2F}_1(c-a,c-b,c-a-b+1|1-\bar z)\;,\\I_3=&I_4=0\;.
\end{aligned}
\end{equation}
We therefore finally get
\begin{equation}
\begin{aligned}
&Z_{S^2}=2\pi\gamma(-iM_X) (z\bar z)^{-i(M_1+M_2)/2}(1-z)^{c-a-b}(1-\bar z)^{c-a-b}(z\bar z)^{\frac{c-1}{2}}\times\\&\Bigg[{_2F_1}(a,b,a+b-c+1|1-z){_2F_1}(a,b,a+b-c+1|1-\bar z)\frac{\gamma(c-a-b)\gamma(a)\gamma(b)}{\gamma(c-a)\gamma(c-b)}[(1-z)(1-\bar z)]^{a+b-c}+\\&\gamma\left(a+b-c\right){_2F}_1(c-a,c-b,c-a-b+1|1-z){_2F}_1(c-a,c-b,c-a-b+1|1-\bar z)\Bigg]\;.
\end{aligned}
\end{equation}

\section{Consistency Check}
\label{offdiagonalmodes}
In this appendix we restrict to $N_c=2$ and $N_f=4$, and to strings that satisfy condition \eqref{notildecondition}. The $STS$ transformation in $\text{\bf S}_3$ keeps this condition invariant. We will show that the worldsheet theories  and the spectrum of the worldsheet fields we proposed in sections ֿ\ref{simpleexample} and \ref{Locnotilde} are consistent with this transformation.

Note that the $STS$ transformation can be used to  map strings with $c_1=c_2$ to strings with $c_1\neq c_2$. This allows us to understand the moduli of the  $c_1\neq c_2$ strings (in particular, the spectrum of off-diagonal moduli) in terms of the better understood moduli of the $c_1=c_2$ strings.

The $S T S$ transformation maps $q$ to ${1}/{q}$, and acts on the masses and $U(1)$ charges as 
\begin{equation}
\begin{aligned}
\mu_i&\to\half(\mu_1+\mu_2+\mu_3+\mu_4)-\mu_{5-i}\;,\\c_i&\to\half(c_1+c_2+c_3+c_4)-c_{5-i}\;.
\end{aligned}
\end{equation} This transformation exchanges $\mm_{12}\leftrightarrow\mm_{34}$, $\mm_{13}\leftrightarrow\mm_{24}$,  and $\mm_{14}\leftrightarrow\mm_{23}$, where $\mm_{ij}$ is defined in equation \eqref{512}. According to equations (\ref{514}-\ref{515}), the effect of this transformation on the worldsheet theory is that $z$ is mapped to $1/z$,  the fundamental and anti-fundamental multiplets are interchanged, while  the spectrum of the neutral and adjoint fields remains invariant.  This agrees  exactly with the expectation based on the duality, since  
$z\to1/z$ acts as charge conjugation on the worldsheet theory. Therefore, accompanied by exchanging the fundamental and anti-fundamental representations, it is a symmetry of the worldsheet theory.

\bibliography{WorldsheetFinal2}
	\end{document}